\RequirePackage{rotating}
\documentclass[fleqn,usenatbib]{mnras}

\usepackage[utf8]{inputenc}
\usepackage{tabulary}

\usepackage[T1]{fontenc}
\usepackage{ae,aecompl}
\usepackage{xcolor}
\usepackage{multirow}
\usepackage[normalem]{ulem} 

\usepackage{graphicx}	
\usepackage{amsmath}	
\usepackage{amssymb}	
\usepackage{hyperref}
\usepackage{caption}
\usepackage{adjustbox}
\usepackage{subcaption}
\usepackage{rotating}
\usepackage{supertabular}                                             
\usepackage{lscape} 
\usepackage{newtxtext,newtxmath}

\title[Galactic abundance gradients]{Gradients of chemical abundances in the Milky Way from H~{\sc ii} regions:  distances derived from Gaia EDR3 parallaxes and temperature inhomogeneities }

\author[J. E. M\'endez-Delgado et al.]
{J. E. M\'endez-Delgado$^{1,2}$ \thanks{E-mail: jemd@iac.es},
A. Amayo$^{3}$, K. Z. Arellano-C\'ordova$^{4}$, C. Esteban$^{1,2}$   
\newauthor 
J. Garc\'ia-Rojas$^{1,2}$, L. Carigi$^{3}$ and G. Delgado-Inglada$^{3}$ \\
\\
$^{1}$Instituto de Astrof\'isica de Canarias (IAC), E-38205 La Laguna, Spain\\
$^{2}$Departamento de Astrof\'isica, Universidad de La Laguna, E-38206 La Laguna, Spain\\
$^{3}$Instituto de Astronom\'ia, Universidad Nacional Aut\'onoma de M\'exico, Ap. 70-264, 04510 CDMX, M\'exico\\
$^{4}$Department of Astronomy, The University of Texas at Austin, 2515 Speedway, Stop C1400, Austin, TX 78712, USA}

\date{Accepted XXX. Received YYY; in original form ZZZ}

\pubyear{2021}
\begin{document}
\label{firstpage}
\pagerange{\pageref{firstpage}--\pageref{lastpage}}
\maketitle

\begin{abstract}
We present a reassessment of the radial abundance gradients of  He, C, N, O, Ne, S, Cl, and Ar in the Milky Way using the deep optical spectra of 42 H~{\sc ii} regions presented in Arellano-C\'ordova et al. (2020, 2021) and M\'endez-Delgado et al. (2020) exploring the impact of: (1) new distance determinations based on Gaia EDR3 parallaxes and (2) the use of Peimbert's temperature fluctuations paradigm ($t ^ 2> 0$) for deriving ionic abundances. We find that distances based on Gaia EDR3 data are more consistent with kinematic ones based on Galactic rotation curves calibrated with radio parallaxes, which give less dispersion and uncertainties than those calibrated with spectrophotometric stellar distances. The distances based on the Gaia parallaxes --DR2 or EDR3-- eliminate the internal flattening observed in previous determinations of the Galactic gradients at smaller distances than $\sim 7$ kpc. Abundances and gradients determined assuming $ t ^ 2> 0 $ ---not only for O but also for the rest of elements--- are not affected by the abundance discrepancy problem and give elemental abundances much consistent with the solar ones for most elements. We find that our radial abundance gradient of He is consistent with the most accurate estimates of the primordial He abundance. We do not find evidence of azimuthal variations in the chemical abundances of our sample. Moreover, the small dispersion in the O gradient ---indicator of metallicity in photoionized regions--- indicate that the gas of the H~{\sc ii} regions is well mixed in the sampled areas of the Galaxy.

\end{abstract}



\section{Introduction}
\label{sec:introduction}

Chemical evolution models of galaxies require precise observational constraints as the determinations of abundance gradients. In order to obtain good determinations of these gradients, it is necessary to have accurate radial distances and precise determinations of the chemical abundances. In the Milky Way, the determination of the Galactocentric distances is particularly problematic, since both the Sun and the observed nebulae are placed within the Galactic disk. Although it is common for Galactic abundance gradient studies to focus mainly on the determination of chemical abundances, in this paper we will emphasize the effects of using the more precise distance determinations.

Distances to Galactic H~{\sc ii} regions can be determined using different techniques. Some of the most common methodologies are based on the study of radial velocities of the nebulae or photometry of their associated stars \citep[see e.~g.][and references therein]{Caplan00, Russeil03, Quireza06, Russeil07, Balser2011, moises2011, Foster15, Anderson2015, Wenger2018}. However, their results are strongly dependent on the rotation and extinction models of the Galaxy, which are still matter of debate. This problem can be avoided by using direct parallax-based distance determinations of their stellar populations \citep[e.~g.][and references therein]{Menten07,Kraus07,Hirota07,Bartkiewicz08,Reid09,Brunthaler09,Zhang09,Xu09}. However, until a few years ago, there were not enough reliable parallax measurements. Since the availability of the optical parallaxes from the Second Data Release of Gaia \citep[hereinafter Gaia DR2,][]{Gaiadr2} there are billions of reliable optical stellar parallaxes, from which distance can be determined directly. Moreover, in the recent Gaia Early Data Release 3 \citep[hereinafter EDR3,][]{Gaiadr3}, these parallaxes have substantially improved their precision. 

Regarding the determination of chemical abundances of H~{\sc ii} regions, there is a significant source of uncertainty due to the Abundance Discrepancy (AD) problem, which is the systematic difference between the derived abundances based on the intensity of collisionally excited lines (CELs) and recombination lines (RLs), where the RLs give always higher abundances. The origin of this problem is still not clear. However, \citet{garciarojasyesteban07} showed that the Abundance Discrepancy Factor (ADF)\footnote{Measured as the ratio between the abundances derived with RLs and CELs.} is rather constant for O$^{2+}$ in Galactic H~{\sc ii} regions --presenting a different behavior from what is observed in planetary nebulae--, being consistent with the predictions of the temperature fluctuations paradigm proposed by \citet{Peimbert67}.

In this work, we derive the Galactic radial abundance gradients of He, C, N, O, Ne, S, Cl, and Ar based on the 42 Galactic H~{\sc ii} regions studied by \citet{Mendez2020} and \citet{arellano2020, arellano2021}. We analyze the distances based on the Gaia EDR3 parallaxes and the probabilistic approach of \citet{Bailer-Jones2021} and derive the chemical composition assuming the temperature fluctuations ($t^2>0$) paradigm \citep[][]{Peimbert67}.

In Sec.~\ref{sec:obs_sample} we describe the sample of spectra adopted from the literature. In Sec.~\ref{sec:distances}, we analyze the distances to the Galactic H~{\sc ii} regions of our sample using the Gaia EDR3 parallaxes, as well as the kinematic distance determinations. In Sec.~\ref{sec:chemical_abundances_grad}, we analyze the determination of chemical abundances in the Galactic H~{\sc ii} regions of our sample under the paradigm of temperature fluctuations -- $t ^ 2> 0 $ -- as well as the resulting chemical abundance gradients. In Sec.~\ref{sec:disc} we discuss our results and summarize them in Sec.~\ref{sec:concl}. In the Appendix we include some data tables and figures.

\section{Observational sample}
\label{sec:obs_sample}

The spectra of the 42 Galactic H~{\sc ii} regions used in this paper are of high and intermediate spectral resolution, taken with different telescopes with diameters ranging from 4~m to 10~m. The objects which spectra are published in  \citet{Esteban:2004, Esteban:2013} and \citet{Garcia-Rojas:2004, Garcia-Rojas:2005, Garcia-Rojas:2006, Garcia-Rojas:2007}\footnote{See Table~\ref{tab:ionic_abundances_1_t2} to know in which reference the spectra of each object were published.} were observed with the UT2 (Kueyen) of the Very Large Telescope (VLT) using the Ultraviolet–Visual Echelle Spectrograph (UVES), covering the spectral range from 3570 to 10,400 \AA\ and with $R\sim$ 8000. The spectra of the H~{\sc ii} regions observed by \citet{Esteban:2017, Esteban:2018} and \citet{arellano2021} were taken with the 10.4 m Gran Telescopio Canarias (GTC) using OSIRIS (Optical System for Imaging and low-Intermediate-Resolution Integrated Spectroscopy) spectrograph in longslit mode, covering from 3600 to 7750 \AA\, with $R\sim$ 1000 or 2500. The spectra of three H~{\sc ii} regions were observed by \citet{Garcia-Rojas:2014} or \citet{Fernandez-Martin:2017} with the Intermediate dispersion Spectrograph and Imaging System (ISIS) attached to the 4.2~m William Herschel Telescope (WHT), the spectral coverage was from 3600 to 9200 \AA\ or 3200 to 10,000 \AA\, with $R\sim$ 1000. In addition, we have included  17 Galactic H~{\sc ii} regions and ring nebulae around massive stars for the distance analysis but not in the abundance one. This is because we do not have the emission line ratios of their spectra available or, in the case of ring nebulae, their He, O, and N abundances may be polluted by stellar ejecta.

\section{Revising the distances of a sample of Galactic H~{\sc ii} regions with Gaia EDR3}
\label{sec:distances}

We conduct an extensive literature search on ionizing sources of our sample of Galactic H~{\sc ii} regions in order to determine their distances. The ionizing stars are expected to be O, early B or Wolf-Rayet-type stars, so they should be among the most luminous stars contained in the boundaries or immediate surroundings of the H~{\sc ii} regions. In addition, their parallax should be more precise than those of the rest of their associated stellar population. However, some of these stars may be located in dusty environments or belong to multiple systems that may be not well spatially resolved. In these cases, we look for other stars or young stellar objects (YSOs) that belong to the H~{\sc ii} region, even if they are not their ionizing sources.

After finding the stars contained in the nebulae, we locate them in the Gaia DR2 and EDR3 catalogs using their coordinates, to finally obtain their parallax-based distances \citep{Bailer-Jones18,Bailer-Jones2021}. Besides the parallaxes, the Gaia EDR3 data contain color and magnitude measurements, allowing to estimate photometric distances. Taking advantage of this, \citet{Bailer-Jones2021}, have estimated the distance to the stars using two methods, obtaining the so-called geometric and photogeometric distances. Geometric distances are estimated from the parallaxes, considering the non-linearity of the transformation and the asymmetry of the probability distribution through a prior parameter that varies smoothly as a function of Galactic longitude and latitude, independently of assumptions of stellar physics or interstellar extinction. Geometric distances were also calculated with the Gaia DR2 parallaxes in \citet{Bailer-Jones18}. The photogeometric distances, meanwhile, use the same distance prior, in addition to a color-magnitude one based on a model of the Galactic interstellar extinction. 

In Table~\ref{tab:identi_stars} we present the identified ionizing/associated stars in each nebula, as well as their spectral type, the Gaia DR2 ID, heliocentric geometric distances derived from their parallaxes in the Gaia DR2 and EDR3 data \citep{Bailer-Jones18,Bailer-Jones2021} and the references to the identification of ionizing/associated stars (I/A) and spectral type (SpT). For each distance determination, we specify the uncertainty associated with the parallax. In general, these uncertainties decrease considerably between the Gaia DR2 and Gaia EDR3 data except in two cases: the stars MFJ~SH~2-212~2 and HD~253327, associated with the H~{\sc ii} regions Sh~2-212 and Sh~2-257, respectively. In both stars, the EDR3 parallax is noticeably worse than the DR2 one and give very different distances. \citet{Moffat79} cataloged 14 possible ionizing stars in Sh~2-212. However, \citet{Deharveng08} showed that several of these sources are evolved stars that may not belong to the same star cluster. \citet{Moffat79} and \citet{Deharveng08} agree that the main ionizing star of the region is MFJ~SH~2-212~2, which has the most uncertain parallax. However, the confirmed membership of MFJ~SH~2-212~5, MFJ~SH~2-212~7 and MFJ~SH~2-212~11 to the same stellar group as MFJ~SH~2-212~2 allow us to estimate the distance to Sh~2-212 with a high confidence. The odd value of the parallax of MFJ~SH~2-212~2 does not seem to be related to the presence of dust in the center of Sh~2-212 since most of its optical extinction is of interstellar origin \citep{Deharveng08}. It is more likely that the discrepant parallax is due to MFJ~SH~2-212~2 being a binary or multiple system. This is suggested by its {\it renormalised unit weight error} (RUWE) value of 8.48, which is larger than the expected threshold of 1.4 for single stars \citep{Gaiadr3_cat_validation}. 

The Gaia parallax of HD~253327 (in Sh~2-257) is also very different when comparing DR2 and EDR3 data. \citet{Ojha11} showed that Sh~2-255 and Sh~2-257 are associated with the same gas complex and therefore they are at the same distance. The lower uncertainty of the EDR3 parallax of the ionizing star of Sh~2-255, LS~19, indicates that its distance is the most suitable one for HD~253327. In this case, the RUWE value of 1.209 does not point to  HD~253327 being a binary or multiple system (although it does not rule it out). Additionally, the similar optical extinction coefficients of Sh~2-255 and Sh~2-257 (see Table~7 from \citet{Shaver83}) seems to rule out dust extinction as the cause of the problem.
There are several ionizing stars for some nebulae included in  Table~\ref{tab:identi_stars}. If the distance obtained for each ionizing source is consistent, we take the more precise individual parallax as the representative one for the nebula. In the cases of NGC~3603, M~16, M~17 and M~42, we adopt the distances derived by \citet{Drew19} and \citet{Binder2018} which are based on the analysis of the Gaia DR2 parallaxes of multiple stars associated with the photoionized regions. The use of several stars diminishes the statistical errors, reaching a higher precision in the distance estimates.

A large fraction of the sample of stars analyzed in this work has Gaia EDR3 parallaxes with uncertainties smaller than 10\,per cent ($e_\text{plx}/\text{plx}<0.10$), enabling us to make a reliable estimate of the distance directly from inverting the parallax \citep{Bailer-Jones2021}. In Gaia DR2, quasars observations indicate a global zeropoint bias in the parallax measurements of about $-$0.03~mas, with variations dependent on magnitude, color and position of the stars \citep{Lindegren2018}. For Gaia EDR3, \citet{Lindegren2021a,Lindegren2021b} provide a tentative expression for a better correction of the zeropoint bias including its non-trivial dependence on the ecliptic latitude, magnitude and color. For our sub-sample with the best parallaxes (with uncertainties smaller than 10\,per cent), this zeropoint varies from $-$0.055~mas to $-$0.011~mas, although the median is $-$0.030~mas with $1\sigma$ deviation of 0.010~mas. This indicates that the deviation from a constant value is rather small, minimizing the dependence of the distance with the zeropoint correction and the variables on which it depends. From the Gaia EDR3 data we can have up to 3 distance estimates for the same star which will be analyzed in Sec.~\ref{subsec:Galactocentric_dis}, along with some relevant kinematic distance estimates.

\subsection{Galactocentric Distances}
\label{subsec:Galactocentric_dis}

We adopt a solar Galactocentric distance of $R_{0}=8.2 \pm 0.1$ kpc to estimate the Galactocentric distances of the nebulae from the heliocentric parallax-based ones. That value of $R_{0}$ is the best estimate to the overall distribution of 26 recent values with tracers in the Galactic centre, bulge, disc and halo \citep{Bland-Hawthorn2016}, and it is very close to the geometric distance to the Galactic centre black hole estimated by \citet{GravityCollaboration2019} of $R_{0}=8.178 \pm 0.013  \text{\it (stat)} \pm 0.022  \text{\it (sys)}$ kpc, where {\it stat} and {\it sys} stands  for the contribution of statistical and systematic uncertainties, respectively. This same value was adopted by \citet{Mendez2020}, \citet{arellano2020,arellano2021}, so the comparison with results of those  papers is straightforward.

In Table~\ref{tab:Galactocentric_dis} we compare our parallax-based estimates of Galactocentric distances with the kinematic determinations of \citet{Russeil03, Russeil07} and \citet{Wenger19} in the nebulae in common. Although there is a large number of articles devoted to the estimation of kinematic distances, the ones of \citet{Russeil03} and \citet{Russeil07} are representative of the usual method based on the \citet{Brand93} rotation curve, calibrated through the observation of around 400 Galactic H~{\sc ii} regions with spectrophotometric distances to their associated stars \citep{Brand88}, and the IAU-defined solar motion parameters\footnote{$R_0 = 8.5$ kpc, $\theta_0 =220 \text{ km s}^{-1}$, $U_{\odot}=10.27 \text{ km s}^{-1}$, $V_{\odot}=15.32 \text{ km s}^{-1}$, $W_{\odot}=7.74 \text{ km s}^{-1}$} \citep{Kerr86}. On the other hand, the kinematic determinations of \citet{Wenger19} are based on a Monte Carlo technique adapted to the more recent rotation curve of \citet{Reid2014}, which was calibrated with around 100 parallax-distances from masers observed at radio wavelengths, with updated solar parameters\footnote{$R_0 = 8.34 \pm 0.16$ kpc, $\theta_0 =240 \pm 8 \text{ km s}^{-1}$, $U_{\odot}=10.5 \pm 1.7 \text{ km s}^{-1}$, $V_{\odot}=14.4 \pm 6.8 \text{ km s}^{-1}$, $W_{\odot}=8.9 \pm 0.9 \text{ km s}^{-1}$} \citep[although the sample has been expanded to $\approx 200$ regions in ][]{Reid2019}. The results of this method generally present smaller uncertainties than the direct use of the rotation curve of \citet{Reid2014} \citep{Wenger2018}.

In Fig.~\ref{fig:geo_EDR3vs_other_Gaia_distances} we show the comparison between the Gaia EDR3 geometric Galactocentric distances and other estimates obtained from Gaia Parallaxes. In the top panel, the consistency with the geometric Gaia DR2 distance is remarkable, besides the reduction of the uncertainties in the EDR3 values. The middle panel shows that the photogeometric values are practically the same than the geometric ones for the whole sample, with the exception of Sh~2-127, where the difference is appreciable, although both estimates are consistent within the error bars. In the more distant regions, the uncertainties in the photogeometric distances are slightly smaller due to the use of more information about the stars. However, all the objects we analyze in this paper are located at almost zero Galactic latitudes, which makes the correction for extinction less accurate \citep{Bailer-Jones2021}. Therefore, there is little reward on adopting the photogeometric distances in some particular cases, while it seems more reasonable to use the geometric EDR3 ones in the whole sample as the representative parallax-based distances. The bottom panel shows the consistency between the geometric EDR3 Galactocentric distances and those simply calculated as the inverse of the parallax for the sub-sample with parallax uncertainties smaller than 10\,per cent.

Fig.~\ref{fig:kin_compari} shows a comparison between the geometric Gaia EDR3 Galactocentric distances and the kinematic estimations by \citet{Russeil03}, \citet{Russeil07} and \citet{Wenger19}. Besides the adoption of different solar parameters, the main difference between both kinematic estimates is in the Galactic rotation curve used. As we can see in Fig.~\ref{fig:kin_compari}, at Galactocentric distances $< 13 $ kpc, the kinematic determinations from \citet{Russeil03} and \citet{Russeil07} show a moderate good agreement with our estimates based on Gaia EDR3 parallaxes, although there is a slight overestimation of the kinematic distances due to the larger solar $R_0$ adopted. More important is the underestimation of the kinematic distances in a group of regions with $R_G$ between 5 and 7 kpc, which includes Sh~2-29, Sh~2-32, M~8 and M~20. Finally, at distances greater than $\approx 13 $ kpc, the kinematic distances from \citet{Russeil03} and \citet{Russeil07}, seem to be significantly overestimated for most objects. The bottom panel of Fig.~\ref{fig:kin_compari} shows that the kinematic distances from \citet{Wenger19} are in fairly good agreement with our determinations. Although there is debate about which rotation curve of the Milky Way is the most appropriate \citep[see][]{Russeil17}, in our case, the one proposed by \citet{Reid2014} gives more consistent results. However, it should be kept in mind that our sample of Galactic H~{\sc ii} regions and ring nebulae is quite limited and only covers two quadrants of the Galactic disc.

\begin{figure}
\includegraphics[width=\columnwidth]{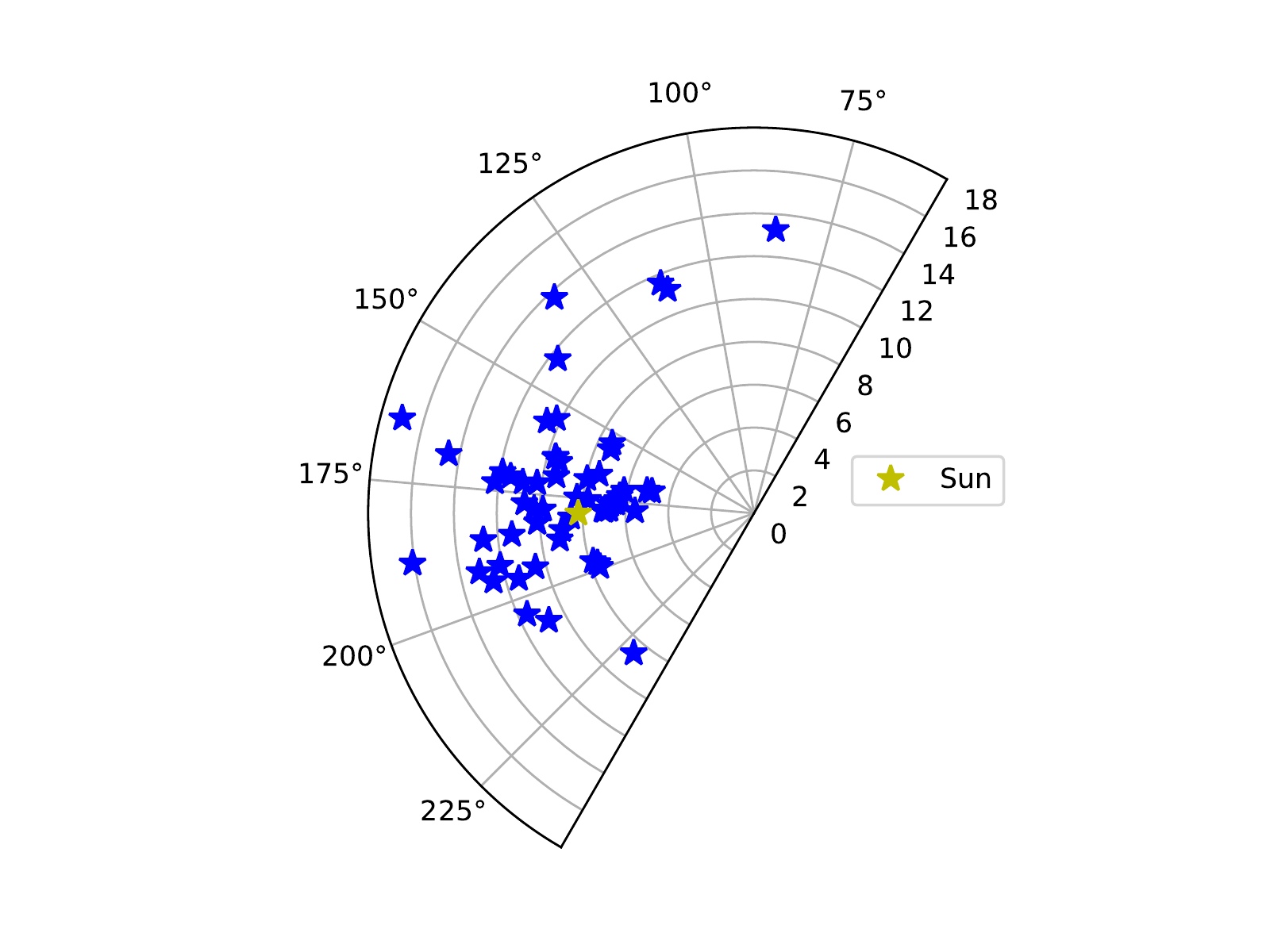}
\caption{Galactic distribution of the H~{\sc ii} regions and ring nebulae of our sample. The adopted solar Galactocentric distance is $R_0 = 8.2$ kpc. }
\label{fig:Gal_dis_fig}
\end{figure}

Considering observational evidences, we define the following criteria to adopt the representative distances of our sample of Galactic nebulae: (i)  we adopt the EDR3 geometric distance when the uncertainties of the parallax are smaller than 20\,per cent; (ii) when (i) is not satisfied, we adopt the kinematic distances of \citet{Wenger19} if their uncertainties are smaller than 20\,per cent; (iii) when criteria (i) and (ii) cannot be satisfied (the only case is Sh~2-48), we adopt the EDR3 geometric distance, however, it must be used with caution since its uncertainty is relatively high. Exceptions to the aforementioned procedure are Sh~2-209 and Sh~2-270, and both cases will be discussed in Sec.~\ref{subsec:H_exceptions}. Table~\ref{tab:rec_distances} presents the adopted heliocentric and Galactocentric distances for our sample of Galactic nebula while Fig.~\ref{fig:Gal_dis_fig} shows their Galactic distribution.

\begin{table}
\centering
\caption{Adopted heliocentric and Galactocentric distance estimates for the H~{\sc ii} regions and ring nebulae of our sample.}
\label{tab:rec_distances}
\begin{tabular}{ccccccccc}
\hline 
Nebula & $d$ (kpc) & $R_G$ (kpc) & Method & Notes \\
\hline
Sh~2-29 & $1.15 ^{+0.04} _{-0.03}$ & $7.06 \pm 0.14$ & Plx &  \\
Sh~2-32 & $1.46 \pm 0.05$ & $6.75 \pm 0.15$ & Plx &  \\
Sh~2-47 & $1.54 \pm 0.05$ & $6.73 ^{+0.15} _{-0.14}$ & Plx &  \\
Sh~2-48 & $3.59 ^{+0.70} _{-0.61}$ & $4.88 ^{+0.72} _{-0.66}$ & Plx & 1 \\
Sh~2-53 & $3.38 \pm 0.18$ & $5.10 ^{+0.24} _{-0.26}$ & Plx &  \\
Sh~2-54 & $1.91 \pm 0.05$ & $6.42 \pm 0.15$ & Plx &  \\
Sh~2-57 & $2.10 ^{+0.06} _{-0.08}$ & $6.32 \pm 0.17$ & Plx &  \\
Sh~2-61 & $2.39 ^{+0.07} _{-0.08}$ & $6.15 \pm 0.16$ & Plx &  \\
Sh~2-82 & $0.78 \pm 0.06$ & $7.76 \pm 0.13$ & Plx &  \\
Sh~2-83 & $16.12 \pm 1.23$ & $13.24 \pm 1.13$  & Kin & 2 \\
Sh~2-88 & $2.06 ^{+0.08} _{-0.07}$ & $7.44 ^{+0.12} _{-0.11}$ & Plx &  \\
Sh~2-90 & $3.36 ^{+0.39} _{-0.28}$ & $7.33 ^{+0.12} _{-0.10}$ & Plx &  \\
Sh~2-93 & $3.64 ^{+0.08} _{-0.07}$ & $7.66 ^{+0.19} _{-0.21}$  & Kin & 2 \\
Sh~2-98 & $11.24 ^{+2.59} _{-2.52}$ & $11.21 ^{+2.00} _{-1.79}$ & Plx &  \\
Sh~2-100 & $11.40 ^{+1.14} _{-1.01}$ & $11.61 \pm 0.87$ & Plx &  \\
Sh~2-127 & $10.13 ^{+1.53} _{-1.22}$ & $13.76 ^{+1.31} _{-0.87}$  & Kin & 2 \\
Sh~2-128 & $7.27 ^{+1.10} _{-0.82}$ & $11.78 ^{+0.78} _{-0.64}$  & Kin & 2 \\
Sh~2-132 & $4.53 ^{+0.29} _{-0.27}$ & $10.21 ^{+0.28} _{-0.26}$ & Plx &  \\
Sh~2-152 & $4.54 ^{+0.86} _{-0.74}$ & $10.61 ^{+0.71} _{-0.43}$  & Kin & 2 \\
Sh~2-156 & $2.56 ^{+0.22} _{-0.19}$ & $9.40 ^{+0.21} _{-0.22}$ & Plx &  \\
Sh~2-158 & $2.84 ^{+0.14} _{-0.16}$ & $9.61 \pm 0.20$ & Plx &  \\
Sh~2-175 & $2.02 \pm 0.05$ & $9.38 \pm 0.13$ & Plx &  \\
Sh~2-203 & $2.39 ^{+0.08} _{-0.09}$ & $10.22 \pm 0.17$ & Plx &  \\
Sh~2-206 & $2.96 ^{+0.17} _{-0.15}$ & $10.88 ^{+0.25} _{-0.26}$ & Plx &  \\
Sh~2-207 & $3.59 ^{+0.23} _{-0.26}$ & $11.48 \pm 0.34$ & Plx &  \\
Sh~2-208 & $4.02 ^{+0.27} _{-0.25}$ & $11.87 ^{+0.37} _{-0.34}$ & Plx &  \\ 
Sh~2-209 & $9.33 \pm 0.70$ & $17.00 \pm 0.70$ & Avg & 5\\
Sh~2-212 & $6.65 ^{+1.36} _{-1.26}$ & $14.76 \pm 1.30$  & Kin & 2 \\
Sh~2-219 & $4.16 ^{+0.32} _{-0.28}$ & $12.17 ^{+0.39} _{-0.41}$ & Plx &  \\
Sh~2-228 & $2.56 \pm 0.09$ & $10.72 \pm 0.19$ & Plx &  \\
Sh~2-235 & $1.66 \pm 0.07$ & $9.85 ^{+0.16} _{-0.17}$ & Plx &  \\
Sh~2-237 & $2.07 \pm 0.06$ & $10.26 ^{+0.16} _{-0.17}$ & Plx &  \\
Sh~2-255 & $1.96 ^{+0.12} _{-0.09}$ & $10.13 \pm 0.21$ & Plx &  \\
Sh~2-257 & $1.96 ^{+0.12} _{-0.09}$ & $10.13 \pm 0.22$ & Plx &  \\
Sh~2-266 & $4.60 ^{+0.53} _{-0.51}$ & $12.69 ^{+0.62} _{-0.64}$ & Plx &  \\
Sh~2-270 & $8.08 \pm 1.29$  & $16.10 \pm 1.40$ & Avg & 6\\
Sh~2-271 & $3.25 ^{+0.11} _{-0.12}$ & $11.34 \pm 0.22$ & Plx &  \\
Sh~2-283 & $5.37 ^{+0.58} _{-0.41}$ & $13.12 ^{+0.62} _{-0.65}$ & Plx &  \\
Sh~2-285 & $4.38 ^{+0.42} _{-0.31}$ & $12.07 ^{+0.51} _{-0.47}$ & Plx &  \\
Sh~2-288 & $5.06 ^{+0.49} _{-0.47}$ & $12.54 ^{+0.54} _{-0.52}$ & Plx &  \\
Sh~2-297 & $1.08 ^{+0.04} _{-0.05}$ & $8.98 ^{+0.14} _{-0.13}$ & Plx &  \\
Sh~2-298 & $4.12 ^{+0.54} _{-0.40}$ & $11.38 ^{+0.58} _{-0.51}$ & Plx &  \\
Sh~2-301 & $3.21 \pm 0.14$ & $10.50 \pm 0.20$ & Plx &  \\
Sh~2-308 & $1.51 ^{+0.10} _{-0.08}$ & $9.14 ^{+0.16} _{-0.17}$ & Plx &  \\
Sh~2-311 & $5.28 ^{+0.40} _{-0.35}$ & $11.60 ^{+0.37} _{-0.39}$ & Plx &  \\
IC~5146 & $0.76 \pm 0.01$ & $8.29 \pm 0.10$ & Plx &  \\
RCW~52 & $2.33 \pm 0.07$ & $7.83 \pm 0.10$ & Plx &  \\
RCW~58 & $2.70 ^{+0.13} _{-0.11}$ & $7.60 \pm 0.10$ & Plx &  \\
G2.4+1.4 & $2.65 \pm 0.15$ & $5.57 ^{+0.25} _{-0.26}$ & Plx &  \\
NGC~2579 & $5.19 ^{+0.48} _{-0.35}$ & $10.82 ^{+0.41} _{-0.44}$ & Plx &  \\
NGC~3576 & $2.48 ^{+0.11} _{-0.10}$ & $7.66 \pm 0.10$ & Plx &  \\
NGC~3603 & $7.02 \pm 0.10$ & $8.61 \pm 0.11$ &  Plx & 3 \\
NGC~6888 & $1.67 \pm 0.04$ & $7.95 \pm 0.10$ & Plx &  \\
NGC~7635 & $2.83 ^{+0.13} _{-0.10}$ & $9.64 ^{+0.17} _{-0.18}$ & Plx &  \\
M~8 & $1.22 \pm 0.04$ & $6.99 \pm 0.14$ & Plx &  \\
M~16 & $1.71 \pm 0.18$ & $6.58 ^{+0.26} _{-0.28}$ & Plx & 4 \\
M~17 & $1.82 \pm 0.16$ & $6.46 \pm 0.25$ & Plx & 4 \\
M~20 & $1.42 ^{+0.09} _{-0.08}$ & $6.79 \pm 0.18$ & Plx &  \\
M~42 & $0.41 \pm 0.01$ & $8.54 \pm 0.11$ & Plx & 4 \\
\hline
\end{tabular}
\begin{description}
\item 1: Uncertainties in the parallax larger than 20\,per cent.
\item 2: Kinematic determinations taken from \citet{Wenger19}.
\item 3, 4: Geometric determinations based on the Gaia DR2 parallaxes taken from \citet{Drew19} and \citet{Binder2018}, respectively.
\item 5, 6: Average of several spectrophotometric and/or kinematic determinations obtained by \citet{Esteban:2017} and \citet{Esteban:2018}, respectively.
\end{description}
\end{table}

 \begin{figure}
\includegraphics[width=\columnwidth]{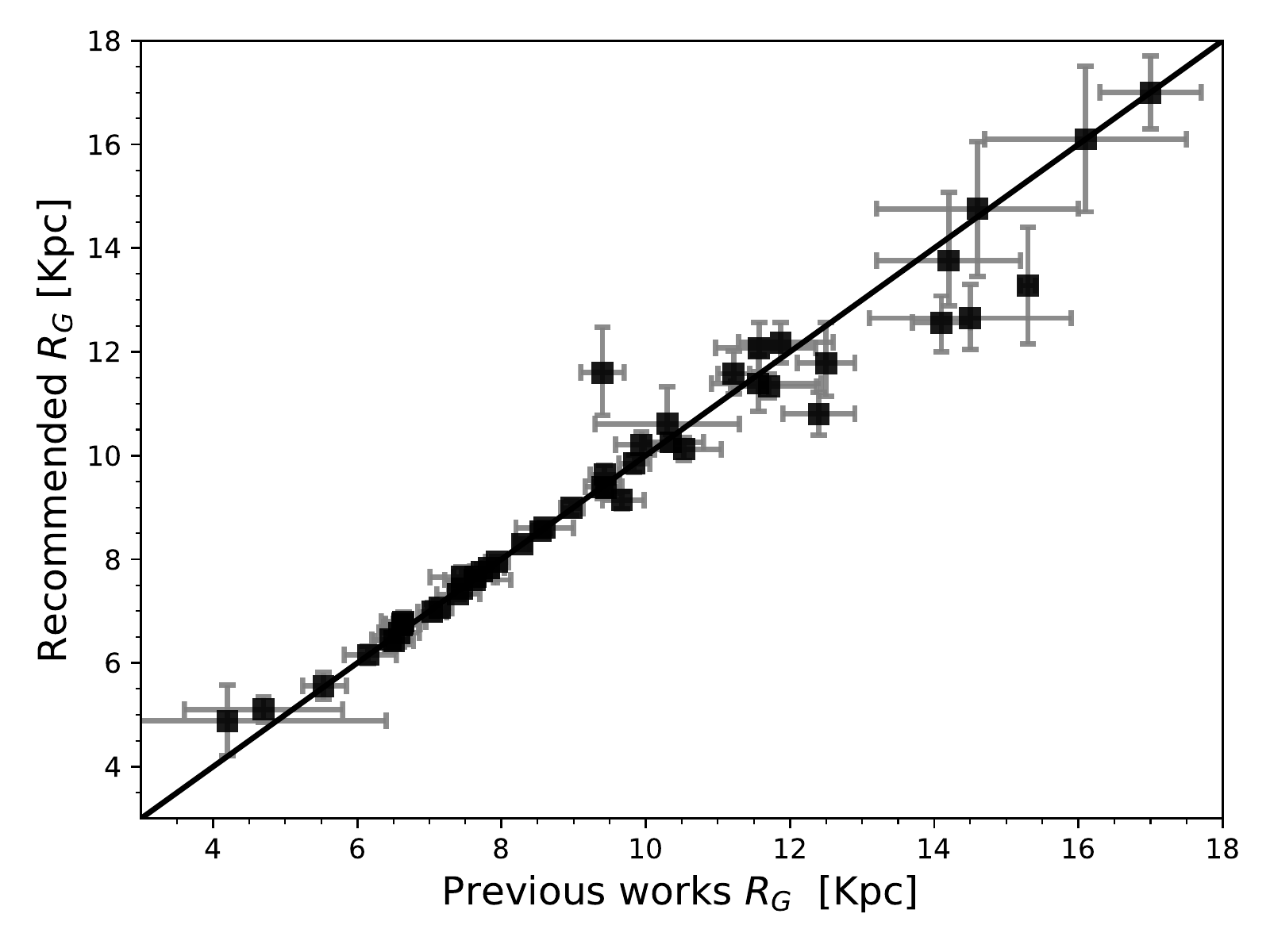}
\caption{Comparison between the updated Galactocentric distances adopted in this work with those used by  \citet{Mendez2020,arellano2020,arellano2021}.}
\label{fig:old_vs_new_distances}
\end{figure}

In Fig.~\ref{fig:old_vs_new_distances} we show the comparison between the values adopted by \citet{Mendez2020} and \citet{arellano2020,arellano2021} and our updated values given in Table~\ref{tab:rec_distances}. In general, there is a good agreement between both sets of distances. In \citet{Mendez2020} and \citet{arellano2020,arellano2021}, the ionizing/associated stars of some of the nebulae have very uncertain Gaia DR2 parallaxes. In those cases, the adopted distance was the average of different kinematic and spectrophotometric estimates. However, most of the kinematic distances used in the calculation of the aforementioned average were based on the rotation curve of \citet{Brand93}, which, as shown in Fig.~\ref{fig:kin_compari}, seems to overestimate the distance of the objects located at Galactocentric distances $\gtrsim 13$ kpc, which explains the larger previous distance determinations of several nebula located between $\approx 13$ and $\approx 14$ kpc shown in Fig.~\ref{fig:old_vs_new_distances}.

\subsection{On the distances of Sh~2-209 and Sh~2-270}
\label{subsec:H_exceptions}

From our sample of Galactic H~{\sc ii} regions, only Sh~2-209 and Sh~2-270 show differences between  parallax-based and kinematic/spectrophotometric distances that can not be explained by methodological issues. The Galactocentric distances adopted in \citet{Esteban:2017} and \citet{Esteban:2018} are $R_G=17.00\pm 0.70 \text{ kpc}$ and $R_G=16.10\pm 1.40 \text{ kpc}$ for Sh~2-209 and Sh~2-270, respectively, but the parallax-based distances derived from Gaia EDR3 are only about 65\,per cent of those values (see Table~\ref{tab:Galactocentric_dis}). Sh~2-209 and Sh~2-270 were precisely among the objects which stars have very uncertain Gaia DR2 parallaxes and for which \citet{Esteban:2017} and \citet{Esteban:2018} adopted the average value of several spectrophotometric and/or kinematic distances, as commented at the end of the previous section. 
If we assume that the Gaia EDR3 parallax-based distances are the true ones for these regions, Sh~2-209 would be located only at 0.36 kpc from Sh~2-206 in the Galactic plane (both nebulae are less than one degree apart). There are several determinations of the Galactic rotation velocity of Sh~2-209 in the local standard of rest \citep[][]{Brand93,Caplan00,Russeil03,Balser2011}, all being consistent with the most precise one by \citet{Balser2011} of $V_\text{lsr}=-51.18 \pm 0.09 \text{ km s}^{-1}$. The same occurs with Sh~2-206, where \citet{Balser2011} determined $V_\text{lsr}=-26.61 \pm 0.05 \text{ km s}^{-1}$. These numbers imply a ratio of 1.92 between the Galactic rotation velocity of Sh~2-209 with respect to that of  Sh~2-206, which is entirely inconsistent for two neighbor objects located at very similar Galactocentric distance. 

In the case of Sh~2-270, if we assume the Gaia EDR3 parallax-based distance as the true one, this region would be only 0.22 kpc apart from Sh~2-255 and Sh~2-257. \citet{Balser2011} estimated a $V_\text{lsr}=-5.05 \pm 0.09 \text{ km s}^{-1}$ for Sh~2-257 (which is located at the same distance as Sh~2-255), while \citet{Russeil03} reported a $V_\text{lsr}=25.6 \text{ km s}^{-1}$ for Sh~2-270. However, the three objects are in the general direction of the Galactic anticentre and such velocity difference is not conclusive.

\citet{Foster15} performed a spectrophotometrical analysis of the ionizing/associated stars of Sh~2-209 and Sh~2-270 included in Table~\ref{tab:identi_stars}. From their distance modulus, they found heliocentric distances of $d=10.58 \pm 0.57\text{ km s}^{-1}$ and $d=9.27 \pm 1.85\text{ km s}^{-1}$ for Sh~2-209 and Sh~2-270, respectively, which are consistent with the distances adopted by \citet{Esteban:2017} and \citet{Esteban:2018}. This rejects the misidentification of the ionizing/associated stars as the cause of the discrepancy in the distances of these objects and points out to possible errors in the reported Gaia EDR3 parallaxes. Although these errors are rather rare, there are cases -- as MFJ~SH~2-212~2 and HD~253327, commented before -- where the Gaia DR2 and EDR3 parallaxes can differ beyond the formal uncertainty bars.

As a result of this discussion, we adopt the distances reported by \citet{Esteban:2017} and \citet{Esteban:2018} for Sh~2-209 and Sh~2-270, both having convenient conservative error bars. Improved optical or radio parallaxes for the stars associated with these nebulae would help to solve the distance inconsistencies.

\section{Effects of temperature fluctuations on the chemical abundances}
\label{sec:chemical_abundances_grad}

\citet{Peimbert67} introduced the concept and the formalism of the paradigm of temperature fluctuations in gaseous nebulae. In his eqs. (9) and (12), he defines the average temperature ($T_0$) and the root mean square temperature fluctuation parameter ($t^2$), that accounts for the level of inhomogeneities or fluctuations in the spatial distribution of the electron temperature within the observed volume. According to this paradigm, when we estimate the electron temperature from an auroral-nebular ratio of CELs -- as [O~{\sc iii}] I($\lambda 4363$)/I($\lambda 5007$) --, the resulting value will be conditioned to $t^2$. The derived temperature will favor the hottest areas within the line of sight and consequently, ionic abundance determinations based on CELs will be underestimated.

On the other hand, the intensity ratios of RLs have a very slight dependence on temperature and the impact of temperature fluctuations is practically negligible. Because of this, it is possible that temperature inhomogeneities could cause or contribute to the AD problem, making abundances based on CELs systematically underestimated, while those of RLs are not. 

As mentioned in Sec.~\ref{sec:introduction}, \citet{garciarojasyesteban07} conclude that for M~8, M~16, M~17, M~20, M~42, NGC~3576, NGC~3603 and Sh~2-311, the ADF(O$^{2+}$) is rather constant around a factor of 2 and consistent with the temperature fluctuations paradigm. However, the ADF of other ions, such as O$^{+}$, does not have to be the same as ADF(O$^{2+}$). In fact, if there is a global value of $t^2>0$ in a given nebula, this will affect differently to each ion, depending on the excitation energies of the upper levels of their observed CELs as shown in eq.~(11) from \citet{Peimbert04}. Therefore, considering a constant ADF on all ions is different from considering a global constant $t^2$. 

\begin{figure}
\includegraphics[width=\columnwidth]{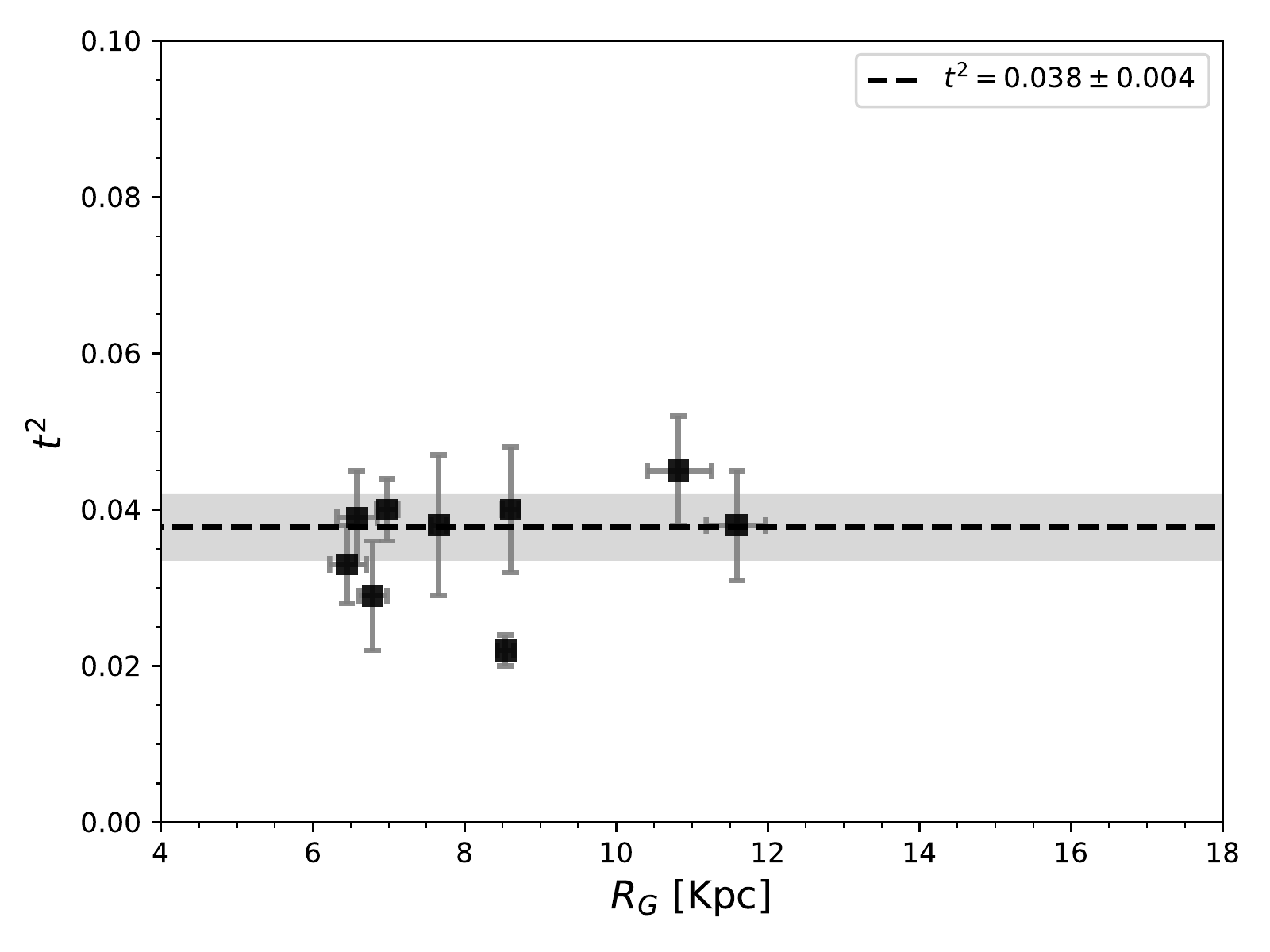}
\caption{$t ^ 2$ values determined for the Galactic H~{\sc ii} regions M~8, M~16, M~17, M~20, M~42, NGC~2579, NGC~3576,  NGC~3603 and Sh~2-311 from the literature (see text for references) with respect to their adopted Galactocentric distance. The horizontal line represents the weighted average of these values and its uncertainty.}
\label{fig:t2_medidas}
\end{figure}

There are several methods to estimate $T_0$ and $t^2$. It is possible to calculate them by comparing different temperature diagnostics of the gas that belongs to the same volume. It is also common to determine them from the ADF, although this procedure assumes that temperature fluctuations are the cause of the AD problem. From our sample, M~8, M~16, M~17, M~20, M~42, NGC~2579, NGC~3576,  NGC~3603 and Sh~2-311 have deep high spectral-resolution spectra obtained with the Very Large Telescope (VLT), with reliable estimates of $t^2$. Fig.~\ref{fig:t2_medidas} shows that the radial distribution of $t^2$ in the aforementioned sample of H~{\sc ii} regions along the Galactic disc seems to be rather flat, at least at $R_G$ values between 6.45 and 11.6 kpc. Therefore, it seems reasonable to adopt a weighted average of $t^2=0.038\pm 0.004$ as a representative-$t^2$ value for the Galactic H~{\sc ii} regions. We must say that although its uncertainty of  $t^2$ is the smallest of the whole sample, the Orion Nebula has been eliminated in the calculation of the weighted average. This is because the Orion Nebula shows the lowest value, by far, of $t^2$. Considering it to calculate the weighted mean we obtain $t^2=0.029 \pm 0.008$. A value that does not seem to represent properly the rest of the H~{\sc ii} regions included in Fig.~\ref{fig:t2_medidas}. In any case, if included, both determinations of the weighted average of $t^2$ are consistent within the $1\sigma$ error bars. The conclusions of the paper are not altered nor do they depend on the inclusion or exclusion of this point.

\subsection{Ionic and total abundances}
\label{subsec:ionic_and_total}

We use eq. (15) from \citet{Peimbert67} as well as eq. (8), eq. (9) and eq. (11) from \citet{Peimbert04} to calculate the impact of assuming $t^2=0.038\pm 0.004$ in the ionic abundances. In this process, we also use the excitation energies of the atomic levels associated with each corresponding emission line, together with the physical conditions and the ionic abundances determined by \citet{arellano2020,arellano2021} assuming $t^2=0$. 
In these previous works, the electron density ($n_{\rm e}$) was obtained using the [\ion{O}{II}] $\lambda$3729/$\lambda$3726, [\ion{S}{II}] $\lambda$6717/$\lambda$6731 and/or [\ion{Cl}{III}] $\lambda$5518/$\lambda$5538 line intensity ratios (although the [\ion{Cl}{III}] diagnostic was discarded if it led to large uncertainties).  For cases where the  derived $n_{\rm e}$ was too low ($n_{\rm e} < 100$ cm$^{-3}$), a value of $100$ cm$^{-3}$ was adopted. On the other hand, the electronic temperatures ($T_{\rm e}$) were derived using a low- and high-ionization zones scheme, using the [\ion{N}{II}] ($\lambda$6548+$\lambda$6584)/$\lambda$5755 ($T_{\rm e}$([\ion{N}{II}])) and the [\ion{O}{III}] ($\lambda$4959+$\lambda$5007)/$\lambda$4363 ($T_{\rm e}$([\ion{O}{III}])) line intensity ratios, respectively. The temperature relation proposed by \citet{Esteban2009} was used to derive either $T_{\rm e}$([\ion{N}{II}]) or $T_{\rm e}$([\ion{O}{III}]) when [\ion{O}{II}] $\lambda$4363 or [\ion{N}{II}] $\lambda$5755 was not detected.
The resulting ionic abundances derived for $t^2=0.038\pm 0.004$ are shown in Tables~\ref{tab:ionic_abundances_1_t2} and \ref{tab:ionic_abundances_2_t2}.  Table~\ref{tab:ionic_abundances_1_t2} also shows the references of the spectroscopic data as well as the values of $n_{\rm e}$ and $T_{\rm e}$ computed by \citet{arellano2020} and \citet{arellano2021}.

To obtain the total abundances of C, N, Ne, Cl, S and Ar, it is necessary to use ionization correction factors (ICFs) that take into account the contribution of the unseen ions to the total abundances of these elements. In this work, we adopt the ICFs from \citet{Amayo2021}. These ICFs have been determined from a carefully selected large set of realistic photoionization models for extragalactic H~{\sc ii} regions, covering a wide range of ionization degrees. Moreover, they allow us to include the uncertainties associated to each ICF as a function of the ionization degree, which leads to a more robust determination of the chemical abundances. When comparing  their selected sample of $\sim$1800 models (from which the ICFs where derived) to our sample of Galactic H~{\sc ii} regions, we find good consistency with the vast majority of our objects, supporting the validity of using these ICFs in this work even though they were proposed for a different kind of objects. In some objects, it is possible to obtain the total abundances of N, S and Cl by considering only their ionic abundances (without using an ICF). We find good agreement between the derived total abundances of N, S and Cl and the predictions of the ICFs. The differences are smaller than 0.08 dex, a  value smaller than the mean uncertainties obtained for the total abundances of these elements (of $\sim$0.12 dex).

In order to propagate the uncertainties in the total abundances, we generate a log-normal distribution of 400 Monte Carlo experiments with a $1\sigma$ defined by the lower and upper uncertainties of the ionic abundances (following eq. 13 from \citealt{Amayo2021}) and choose a random ICF from a second log-normal distribution, defined in the same way by the lower and upper uncertainties of each ICF. We then compute the total abundances and report the nominal values and the 16 and 84 percentiles as their lower and upper uncertainties, respectively.

In the estimates of the total abundance of Ar, we only consider the regions where both [Ar~{\sc iii}] and [Ar~{\sc iv}] emission lines are observed, as \citet{Amayo2021} recommends. This criterion reduces the number of objects where the total abundance of Ar is derived to 10 H~{\sc ii} regions. 

In some objects, it is possible to derive the total abundances of N, S and Cl without an ICF by simply adding the ionic abundances of the available ions because either the ionization degree or the depth of the spectra allow to observe emission lines of all the ions of each element that should be present in the nebula. Here, we follow the same criteria used by \citet{arellano2020} to identify these objects, although in some cases the number of H~{\sc ii} regions changes with $t^2>0$ because it can produce a slight change in the ionization degree. The resulting total abundances are shown in Table~\ref{tab:total_abundances}.

For Sh~2-83, Sh~2-100, Sh~2-128, Sh~2-209, M~17, M~42, NGC~2579, NGC~3576 and NGC~3603 \citet[][]{Mendez2020} estimated a small value of the radiation softness parameter \citep[$\text{log(}\eta) <0.9$, see][for a definition of $\eta$]{Vilchez88}. In these cases, the amount of neutral helium within the ionized gas is negligible \citep{Pagel92}. Therefore, we only consider these regions to our analysis of the total helium abundances, assuming $\text{He}/\text{H}\approx \text{He}^{+}/\text{H}^{+}$.

\subsection{The radial abundance gradients}
\label{subsec:radial_grad}

Using the abundances obtained and the updated Galactocentric distances, we compute the radial abundance gradients assuming $t^2 = 0.0$ and $t^2 > 0.0$. We have strict control of the uncertainties in both quantities: distances and total abundances (which include uncertainties from the ICFs). The fit parameters (slope and intercept) of the revised radial gradients are given in Table~\ref{tab:gradients_table}. 

\begin{figure*}
\begin{minipage}{\textwidth}
\centering
\includegraphics[width=.42\textwidth]{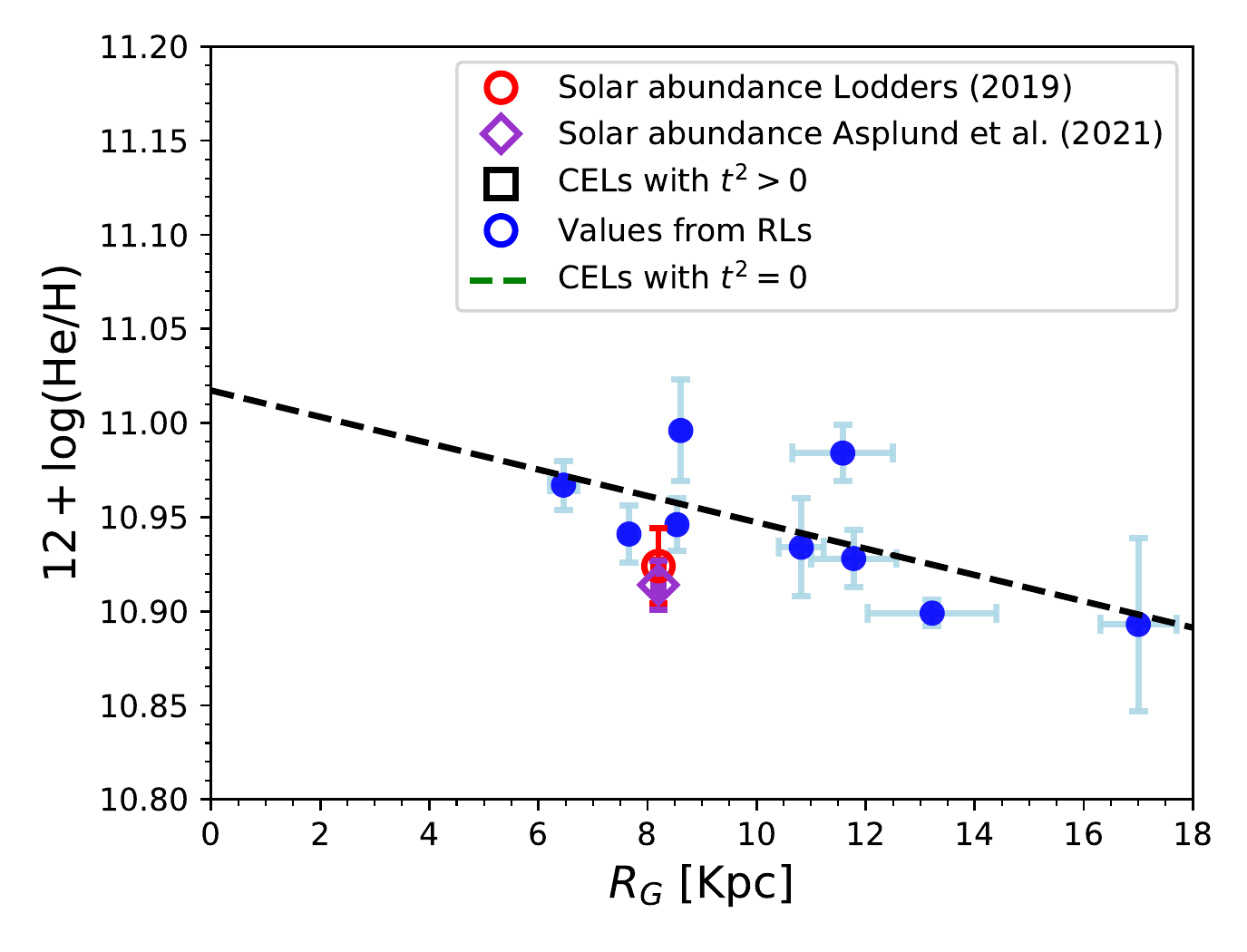}
\includegraphics[width=.42\textwidth]{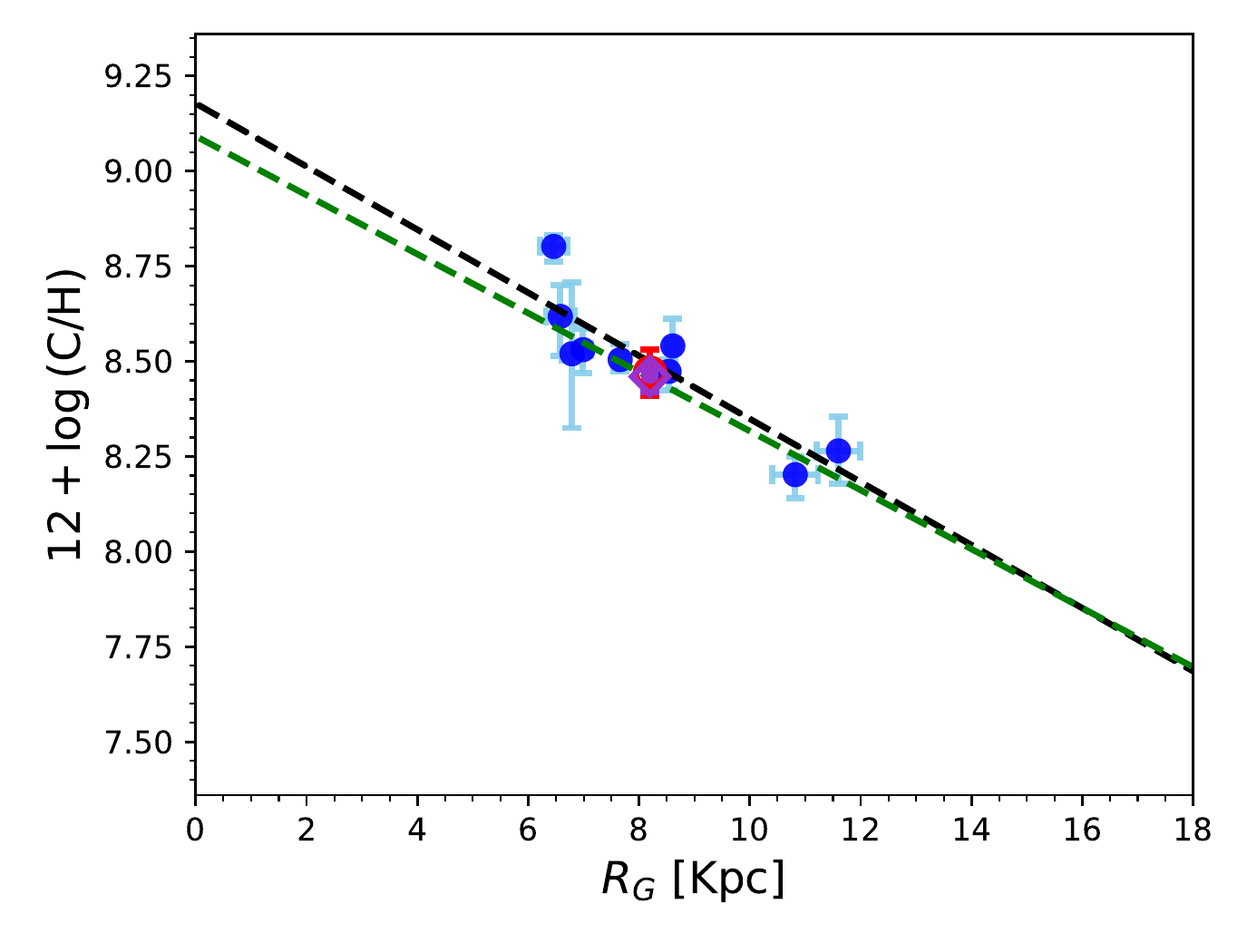}
\includegraphics[width=.42\textwidth]{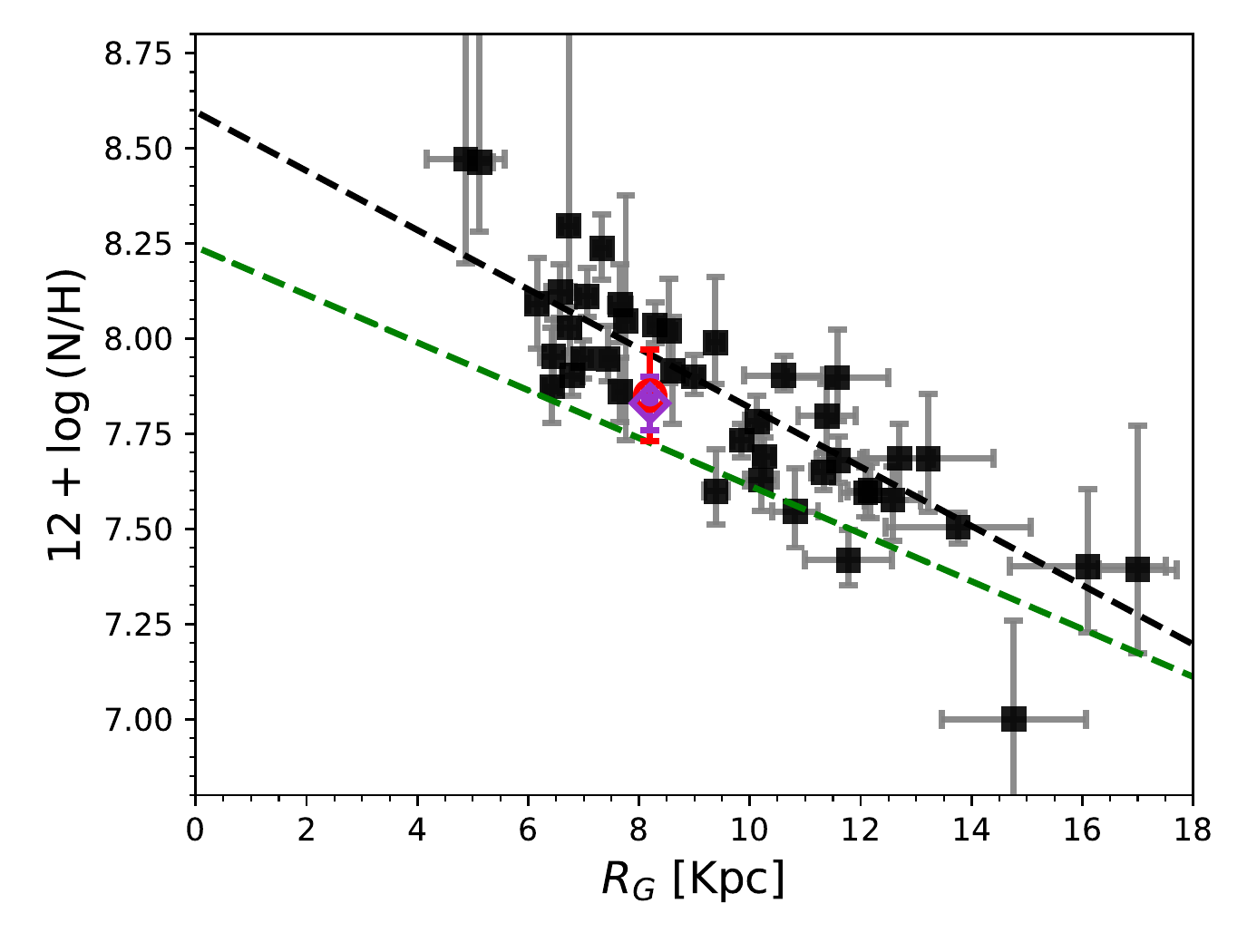}
\includegraphics[width=.42\textwidth]{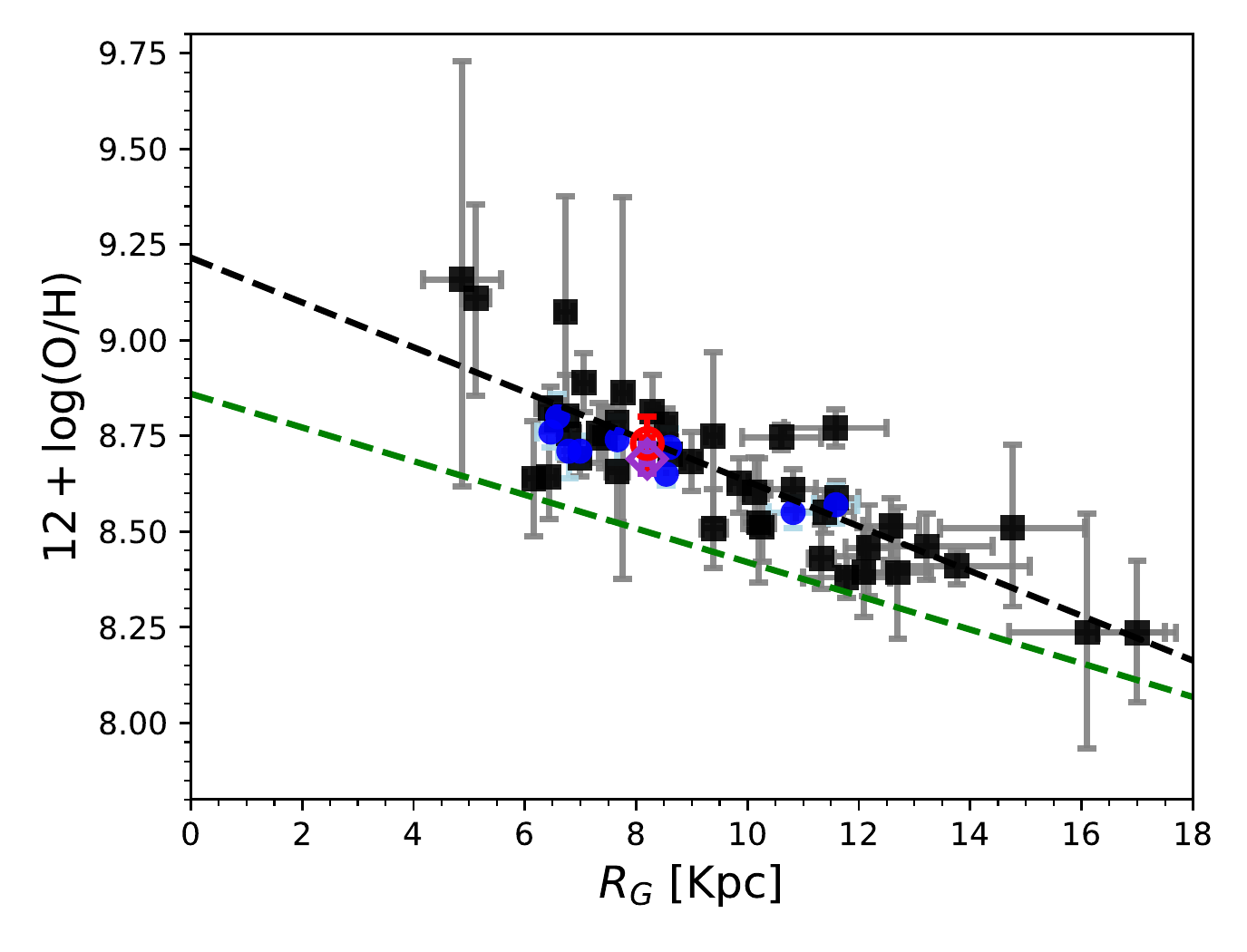}
\includegraphics[width=.42\textwidth]{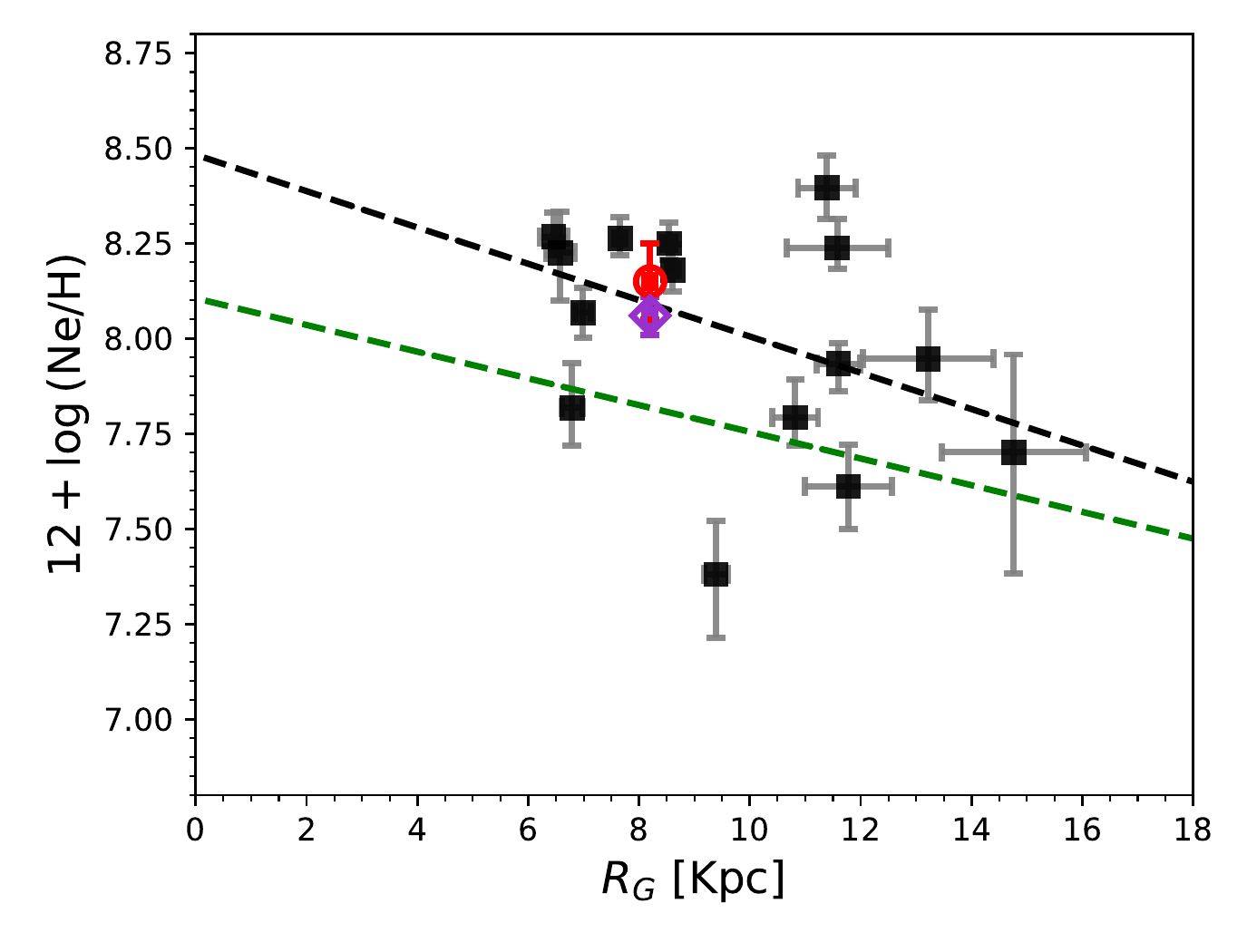}
\includegraphics[width=.42\textwidth]{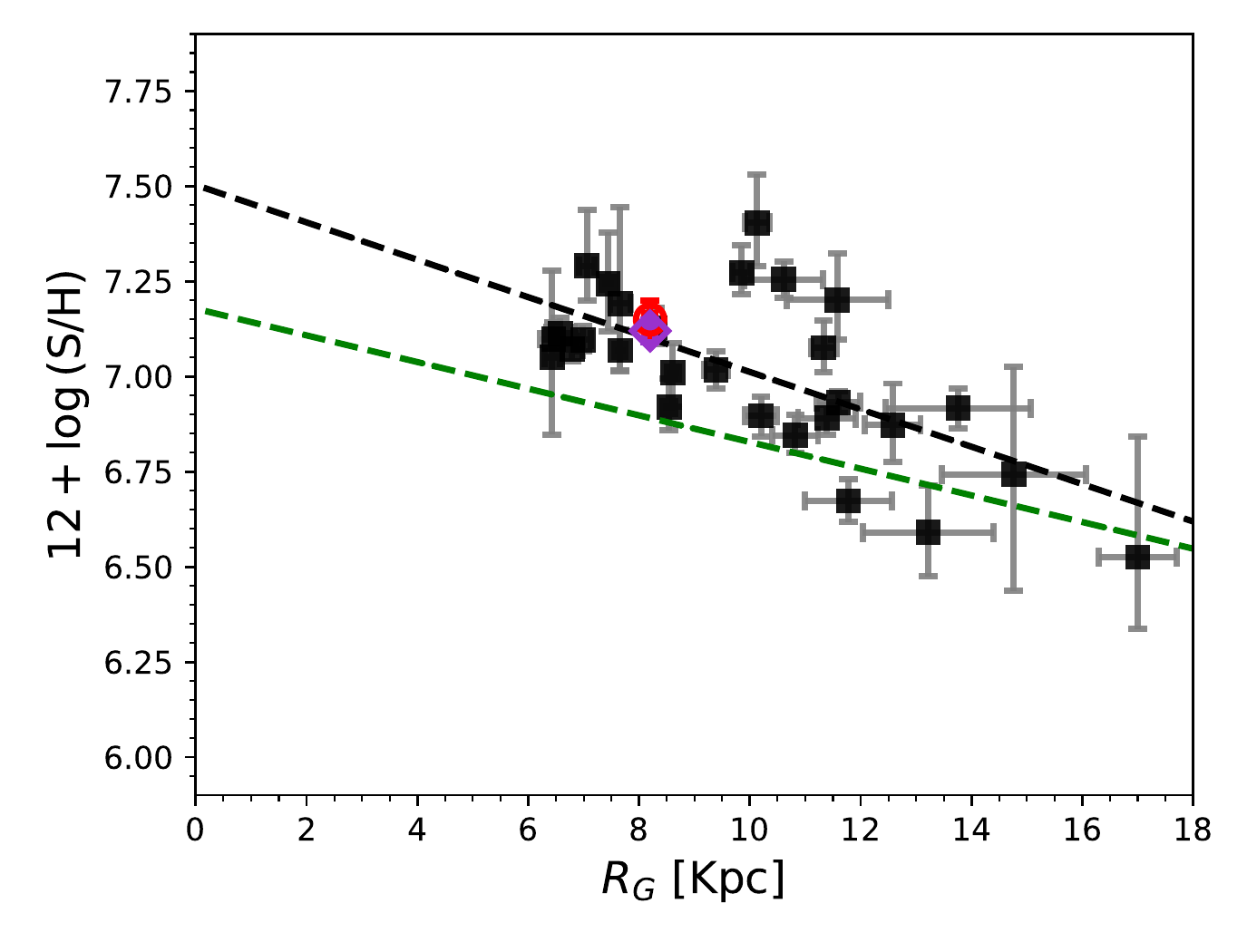}
\includegraphics[width=.42\textwidth]{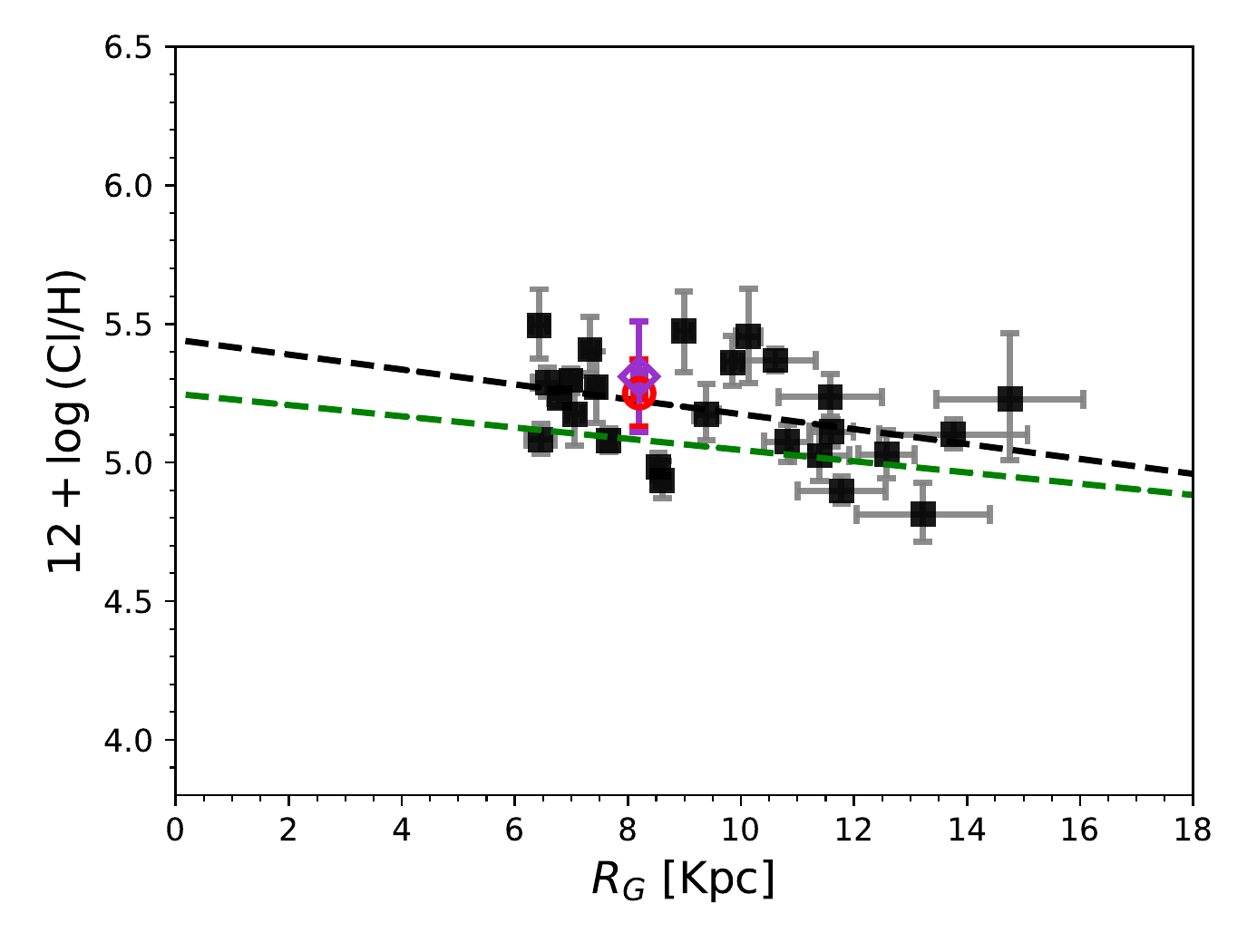}
\includegraphics[width=.42\textwidth]{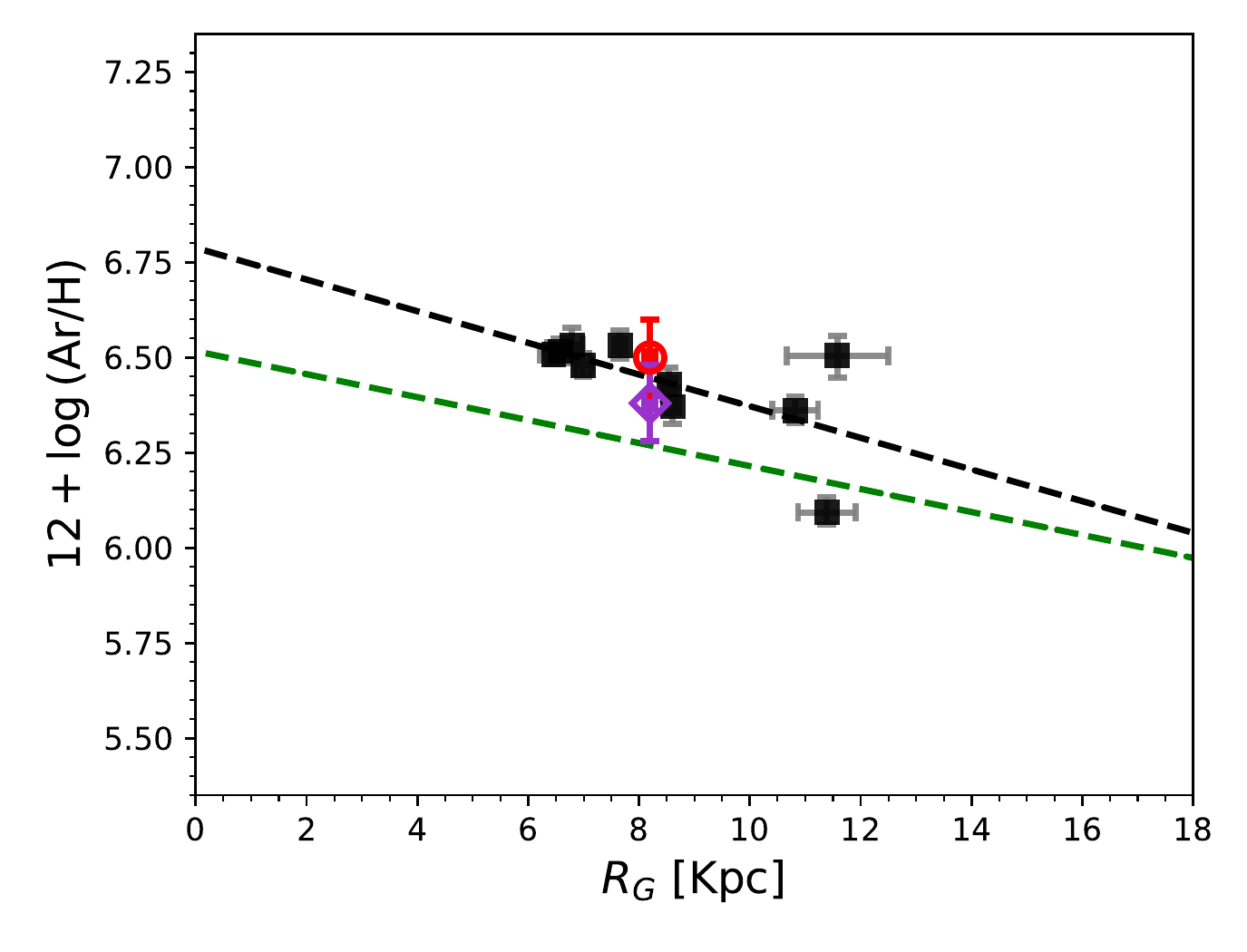}
\end{minipage}
\caption{Radial distribution of He, C, N, O, Ne, S, Cl, and Ar abundances for the sample of Galactic H~{\sc ii} regions assuming $t^2 > 0$ as a function of their revised distances. The black dashed line represents the linear fit using chemical abundances based on CELs assuming $t^2 > 0$ or RLs, while the green dashed line represents the linear fit using CELs assuming $t ^ 2 = 0$. The red circle and purple diamond indicate the solar values recommended by \citet{Lodders2019} and \citet{Asplund2021}, respectively. Blue circles correspond to chemical abundances derived from RLs.}
\label{fig:AbundGradients}
\end{figure*}

The radial gradients for He, C, N, O, Ne, S, Cl and, Ar are shown in Fig.~\ref{fig:AbundGradients}. For the abundances determined from CELs, the black dashed lines indicate the gradients obtained assuming $t^2>0$ while the green ones indicate those with $t^2=0.0$. Although the abundances of C$^{2+}$ have been determined from RLs (and therefore they are independent on $t^2$), the total abundances of C depend on the ICF(C$^{+}$), which varies slightly with $t^2$. Following \citet{arellano2020, arellano2021}, we discard several objects to compute the radial abundance gradients of N and Ne, because of their odd abundances and high uncertainties. In the case of Cl, we discarded the abundances obtained for Sh~2-219, Sh~2-237 and Sh~2-271, because of their very high and uncertain Cl/H ratios. In fact, these  objects are of very low ionization degree, implying that their associated ICFs are very high and uncertain. The red circles and the purple diamonds in the aforementioned figure represent the present-day photospheric solar abundances given by \citet{Lodders2019} and \citet{Asplund2021}, respectively.

The comparison between the chemical composition of the Sun, which was formed 4.57 Gyr ago \citep{Asplund2021}, and that of the H~{\sc ii} regions, which have been formed much more recently, may not be straightforward. Nuclear processing, radioactive decay and diffusion can change the derived chemical composition of the Sun \citep[][]{Asplund2021}. Also, most of the proto-solar abundances are expected to be lower than abundances of young objects, due to the chemical enrichment of the ISM from the Sun's birth until the present-day. According to \citet[][]{Asplund2021}, the difference between the proto-solar and photospheric abundances of helium is 0.070 dex, and of heavier elements is 0.064 dex, although the latter may be even smaller, of 0.03 dex, depending on the mixing scheme considered. On the other hand, \citealt{Carigi2019} estimate an O enrichment of $\sim$0.1 dex for the last 4.6 Gyr in our galaxy at $R_G=8$ kpc, based on chemical evolution models. Since both the impact of the solar diffusion as well as the chemical enrichment during the last $\sim$5 Gyr are similar or even smaller than the typical uncertainties in the abundance determinations in ionized nebulae and stellar abundances, we will use the present-day photospheric solar abundances in our comparisons, keeping in mind its possible associated uncertainty. 

When comparing with the solar abundances, it should be considered the possible radial migration of the Sun since its formation. Recent dynamical and chemodynamical models for the Milky Way disk have estimated a low radial migration for the Sun \citep[e.g.,][]{Klacka2012, MartinezBarbosa2015, MartinezMedina2017}.
\citeauthor{Klacka2012} found a radial-migration average of $0.23 \pm 0.68$ kpc, \citeauthor{MartinezBarbosa2015} claimed that the Sun has unlikely migrated, and \citeauthor{MartinezMedina2017} estimated an average of radial migration of $0.32 \pm 1.40$ kpc.
Hence, we consider the Sun is representative of the solar neighborhood and consequently the comparison of the chemical gradients (at $R_G \sim 8$ kpc) with the solar composition is appropriate, if the age difference between H~{\sc ii} regions and the Sun is neglected, as we mention above.

\begin{table*}
\centering
\caption{Galactic radial abundance gradients determined from the spectra of H~{\sc ii} regions assuming $t ^ 2 = 0.0$ and $t ^ 2> 0$.}
\label{tab:gradients_table}
\begin{tabular}{ccccccccc}
\hline 
&\multicolumn{2}{c}{$t^2=0.0$} & \multicolumn{2}{c}{$t^2=0.038\pm 0.004$}\\
12+log(X/H) & Slope (dex kpc$^{-1}$)         & intercept (dex)           & Slope (dex kpc$^{-1}$)         & intercept (dex) \\
\hline
He          & $-0.0072^{+0.0037} _{-0.0035}$ & $11.02 ^{+0.03} _{-0.04}$ & $-0.0072^{+0.0037} _{-0.0035}$ & $11.02 ^{+0.03} _{-0.04}$ \\
C           & $-0.077 \pm 0.019$             & $9.05 \pm 0.16$           & $-0.077 \pm 0.017$             & $9.12 \pm 0.15$\\
N           & $-0.063 \pm 0.009$             & $8.20 \pm 0.09$           & $-0.078 \pm 0.013$             & $8.60 \pm 0.14$\\
O           & $-0.044 \pm 0.009$             & $8.86 \pm 0.09$           & $-0.059 \pm 0.012$             & $9.22 \pm 0.12$\\
Ne          & $-0.035 \pm 0.018$             & $8.10 \pm 0.16$           & $-0.048 \pm 0.019$             & $8.48 \pm 0.16$\\
S           & $-0.035 \pm 0.011$             & $7.20 \pm 0.10$           & $-0.049 \pm 0.012$             & $7.50 \pm 0.12$\\
Cl          & $-0.021 \pm 0.010$             & $5.23 \pm 0.09$           & $-0.027 \pm 0.010$             & $5.44 \pm 0.09$\\
Ar          & $-0.029 \pm 0.008$             & $6.50 \pm 0.07$           & $-0.042 \pm 0.011$             & $6.79 \pm 0.09$\\
\hline
\end{tabular}
\end{table*}

\section{Discussion}
\label{sec:disc}

The use of distances based on Gaia DR2 parallaxes by \citet{Mendez2020,arellano2020,arellano2021} and our revision based on Gaia EDR3 have been an important step forward to improve the quality of the abundance gradients determinations. Comparing the results obtained assuming kinematic or parallax-based distances have raised interesting issues. In the case of \citet{Esteban:2018}, the underestimation of the Galactocentric distances of Sh~2-61, Sh~2-90, M~8, M~16, M~17, M~20 and NGC~3576, produced an apparent flattening at $R_G<8 \text{ kpc}$ in the abundance gradients of several elements. If true, as \citet{Esteban:2018} discuss, this would imply a somewhat different chemical evolution of the central parts of the Galaxy, invoking the possible action of the Galactic bar. However, this seems to be a spurious result  produced  by the underestimation of small $R_{\rm G}$ values due to the use of rotation model-dependent distance estimates. It is important to note that the latest calibrations made to the Galactic rotation curves by \citet{Reid2014,Reid2019}, produce consistent results with the parallax-based distances from Gaia EDR3 for our sample of H~{\sc ii} regions.

\subsection{The reliability of the derived gradient of helium}
\label{sec:dis_helium}

\citet{Mendez2020} found a negative slope in the radial abundance gradient of He determined from H~{\sc ii} regions of the Milky Way. Although there was marginal evidence of a negative helium gradient in some previous works \citep[see][]{Hawley78,Peimbert78, Talent79,Shaver83, Deharveng2000,Fernandez-Martin:2017}, the uncertainties of those determinations made them compatible with a flat distribution. There are several factors that make it difficult to determine the abundance gradient for this element: (i) unlike the heavier elements, the abundance of primordial He is relatively high, so its relative increase due to stellar nucleosynthesis is rather small, (ii) the known metastability of the 2$^3$S level of the He atom in its triplet configuration makes it very sensitive to self-absorptions and collisions, which generates deviations in the populations of the triplet levels expected by pure recombination, and (iii) estimating the He$^0$/H$^+$ abundance is a difficult issue, and its contribution is expected to be important for the range of ionization degree of most Galactic H~{\sc ii} regions. Considering these factors, we restrict ourselves to 9 highly-ionized H~{\sc ii} regions where $\text{He}^{+}/\text{H}^{+}\approx\text{He}/\text{H}$ due to the hardness of their ionizing sources, covering a range of Galactocentric distances between 6.45 and 17 kpc.

\begin{figure}
\includegraphics[width=\columnwidth]{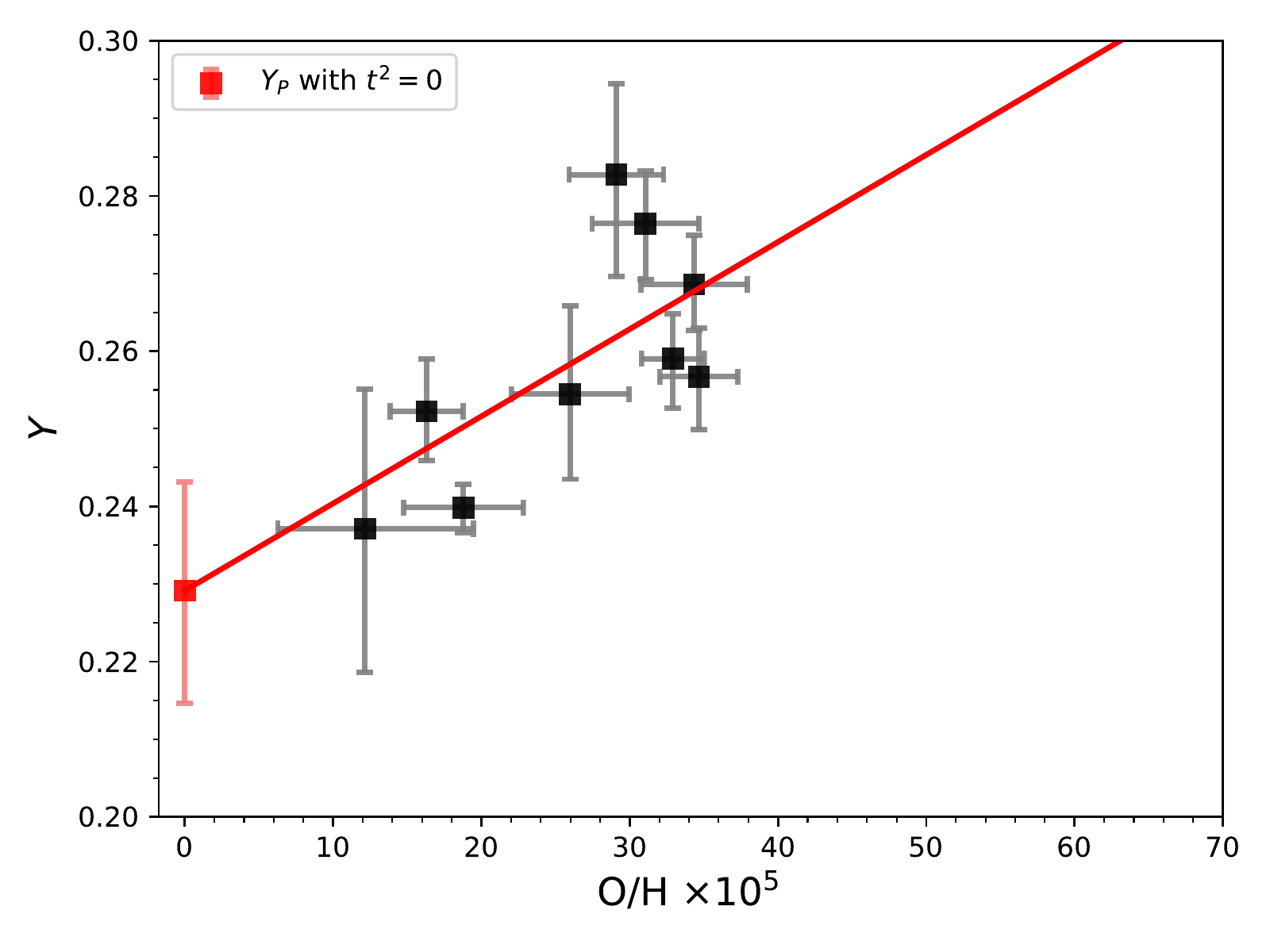}
\includegraphics[width=\columnwidth]{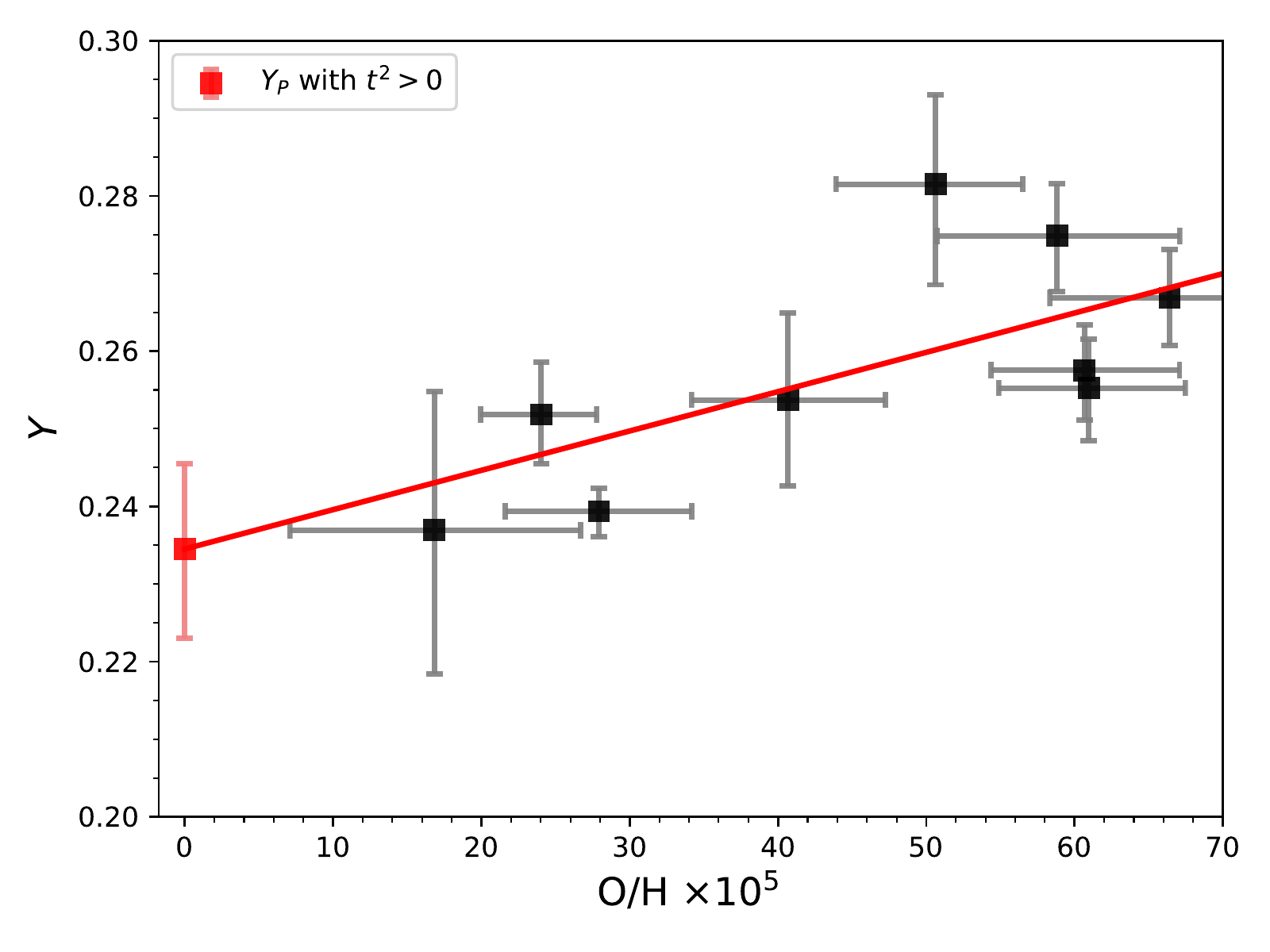}
\caption{Estimation of the fraction of primordial baryonic mass in helium $Y_p$ using the abundances of He and O in the sample of regions used in the determination of the helium gradient in the Milky Way. \textit{Top panel:} Considering $ t ^ 2 = 0$. \textit{Bottom panel:} Considering $ t ^ 2 = 0.038 \pm 0.004$.}
\label{fig:primordial_Galactic_helium}
\end{figure}

If the thus determined radial gradient of He is accurate, the relations between $Y$ (He abundance by mass) and O/H must be used to extrapolate the primordial helium value. The most recent determinations of the primordial helium abundance are consistent with a fraction of primordial baryonic mass in helium $Y_p$ ranging between 0.243 to 0.247 \citep[][]{Aver2015,Peimbert2016,fernandez2019,Hsyu2020,PlanckCollaboration2020, Kurichin2021,Valerdi2021}. In Fig.~\ref{fig:primordial_Galactic_helium} we show the determinations of $Y_p$ considering the sample of regions adopted for the determination of the Galactic radial abundance gradient of He.  For the normalization, we have considered that oxygen per unit of mass represents $55 \pm 10$\,per cent of the metallicity $Z$ \citep{Peimbert2007}. The value of $Y_p$ considering $t ^ 2 = 0$ is $0.229 \pm 0.014$ while the one obtained considering $t ^ 2> 0$ is $0.236 \pm 0.012$. Since both values are formally consistent with the most precise determinations of $Y_p$ within the uncertainty bars, it gives further support to the reliability of our determination of the Galactic radial abundance gradient of He. 

\subsection{The radial gradients of heavy elements under the \texorpdfstring{$t^2$ paradigm}{TEXT}}
\label{sec:disc_metal_grad}

The resulting gradients of C, N, O, Ne, and S considering $t^{2} = 0.0$ are in excellent agreement with those reported by \citet{arellano2020,arellano2021} and therefore, the results and the discussion regarding these elements remain unchanged. However, we find some differences in the gradients of Cl and Ar. In this work we only consider those regions with estimates of both Ar$^{+2}$ and Ar$^{+3}$ since the ICFs constrained only with the Ar$^{+2}$ abundance are rather uncertain \citep[][]{Amayo2021}. Due to this methodological difference, our Ar gradient with $t^2=0.0$ is slightly flatter in comparison to what is found by \citet{arellano2020}, although both are consistent within the uncertainties. In the case of Cl, \citet{arellano2020} used the ICF scheme derived by \citet{Esteban2015}, obtaining a steeper gradient than our results, although they are also consistent within the uncertainties.

As mentioned in Sec~\ref{sec:chemical_abundances_grad}, it is usual to estimate the value of $t^2$ from the measured ADF(O$^{2+}$). However, there are fewer studies that also calculate the ADF for other ions such as C$^{2+}$, O$^{+}$, N$^{2+}$ or Ne$^{2+}$ --getting that the ADF can be different in each ion-- and essentially none considering other different ions. Therefore, the consistency between the solar abundances of O and the abundances of nearby H~{\sc ii} regions with $t^2>0$ is not surprising, since in these regions the RLs provide abundances closer to the solar ones. However, this does not necessarily imply that this consistency must be replicated in the abundances of other elements such as N, S, Cl or Ar, since by construction, the determination of $t ^ 2$ does not consider the ADF of any ion of these elements. Therefore, the discussion about the gradients of elements different than O and their comparison with the solar abundances is pertinent and it is not a circular argument. 

The definition of $t^2$ is not incompatible with the temperature stratification of the nebulae -- which explains that generally $T_{\rm e}$([N~{\sc ii}]) is higher than $T_{\rm e}$([O~{\sc iii}]) --; it just quantifies the level of temperature inhomogeneities within the emission volume (out of the expected stratification). Assuming a global value of $t ^ 2>0$ in each H~{\sc ii} region affects the determination of the abundances of low and high ionization ions in different proportions since, generally, the temperatures associated with each volume are different in addition to the different dependence of $t^2$ with the excitation energy of the emission lines \citep[see eq.(11) in][]{Peimbert04}. This means that the parameters of the linear fits to the gradients -- slope and intercept -- for $ t ^ 2> 0 $ are susceptible to change with respect to the case of assuming $ t ^ 2 = 0 $.

Fig.~\ref{fig:AbundGradients} shows the Galactic radial abundance gradients derived in this work by considering $t^2 = 0.038\pm 0.004$ and $t^2 = 0$. In the second panel of the right column, we show the O gradient. The blue circles represent the total O abundances derived from RLs for M~8, M~16, M~17, M~20, M~42, NGC~3576, NGC~3603, Sh~2-311 and NGC~2579 \citep[][]{Esteban05,Esteban:2013}. The slope with $t ^ 2> 0$ is slightly steeper with respect to the $t^2=0$ case. The resulting gradient is consistent with the recent O gradient of \citet{Luck2018}, based on an updated sample of $\sim$400 classical Cepheids in the Galactic disc. If the paradigm of temperature fluctuations is valid, then the agreement between the O gradient with $t^2>0$ and the solar abundances implies that the depletion of O into dust grains in the solar neighborhood regions is smaller than the uncertainty bars, in consistency with the results of \citet{Mesa-Delgado2009} and \citet{Peimbert2010} who estimated a depletion of O into dust of up to 0.11 dex.

\begin{figure}
\includegraphics[width=\columnwidth]{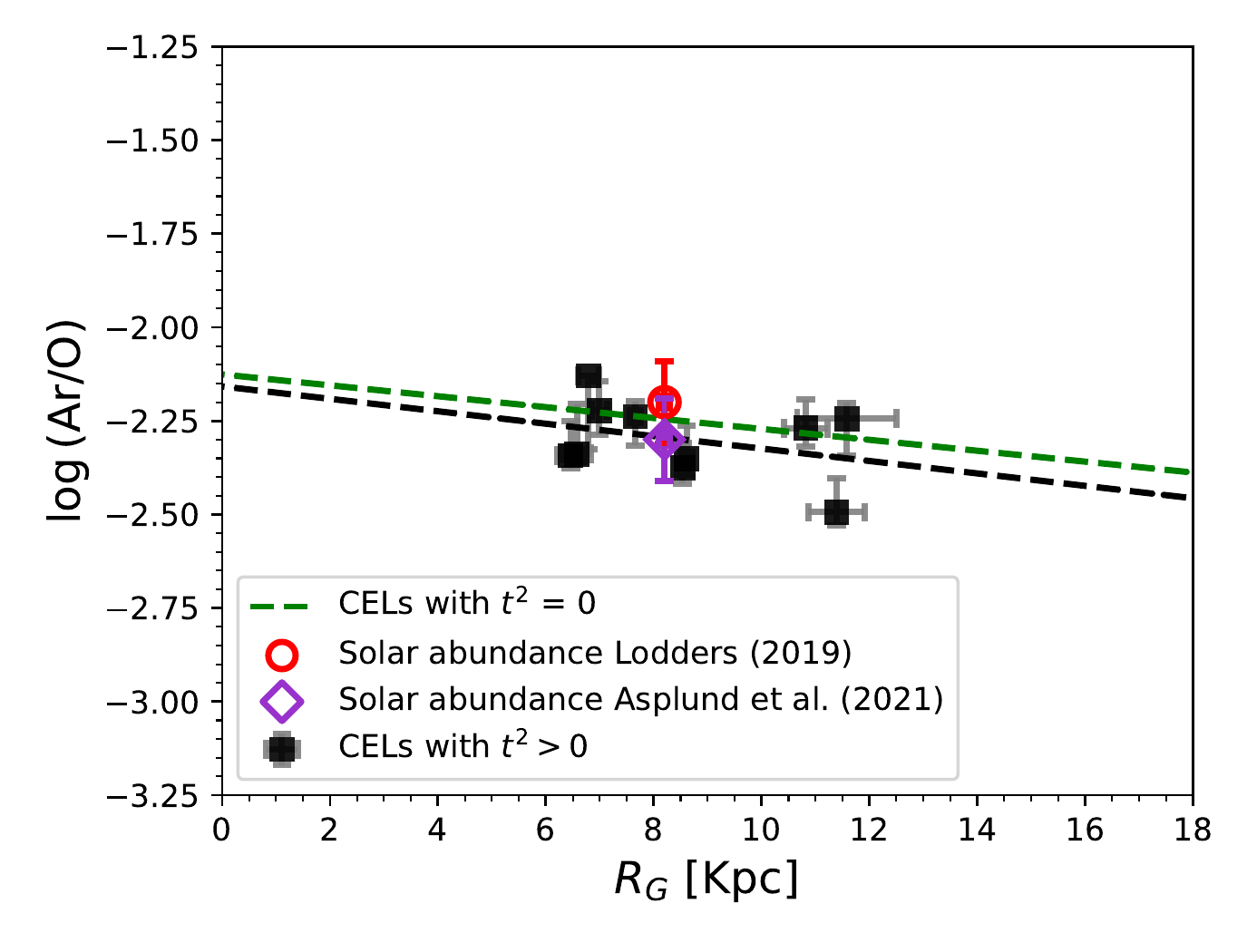}
\caption{Ar/O radial gradient with the same layout than Fig.~\ref{fig:AbundGradients}}
\label{fig:ArO}
\end{figure}

The radial distribution of the Ar/O abundance ratio, shown in Fig.~\ref{fig:ArO}, is almost flat, both in the case with $t ^ 2 = 0$ and $t ^ 2> 0$ (with slopes of $-0.011\pm0.010$ and $-0.014\pm0.010$ dex kpc$^{-1}$, respectively). Since both elements are produced by $\alpha$-particle captures in massive stars, their abundance ratio should be constant, although there may be a small fraction of Ar produced in Type-Ia supernovae, according to the Galactic chemical evolution models of \citealt{Kobayashi2020}, which would add an extra scattering factor in the Ar/O distribution, as discussed in \citet[][]{Amayo2021}.
The dispersion of the order of 0.1 dex in the distribution of Ar/O can serve as an upper limit to the possible depletion of O into dust grains, since there are several factors that contribute to the observed dispersion, such as uncertainties in the ionic abundances of both O and Ar as well as of the ICF of Ar and the possible production of Ar in supernovae.

\begin{figure}
\includegraphics[width=\columnwidth]{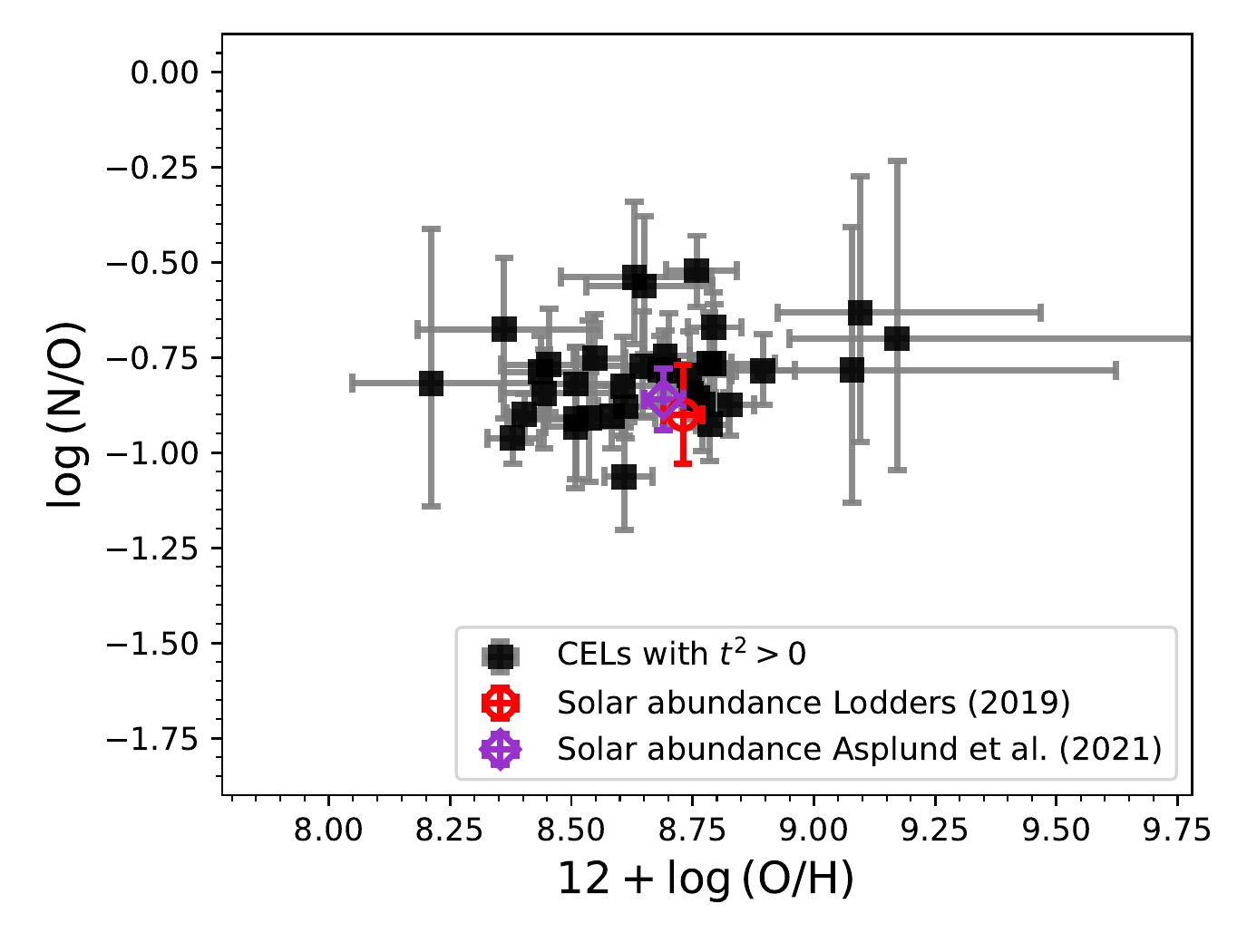}
\caption{N/O as a function of $12 + \log$(O/H) with the same layout than Fig.~\ref{fig:AbundGradients}.}
\label{fig:NO}
\end{figure}

The solar N/H given by \citet{Lodders2019} and \citet{Asplund2021} is between the values predicted by the radial gradient of N considering $t^2=0$ and $t^2>0.0$. The slope obtained with $t^2 > 0$, shown in Table~\ref{tab:gradients_table}, is slightly steeper when compared to some values in the literature \citep[e.g.,][]{Shaver1983, Afflerbach1996, Esteban:2018,arellano2021}. It has better consistency with the slope value of $-0.085\pm0.020$ dex kpc$^{-1}$ determined by \citet{Carigi2005}. Additionally, we determine the gradient of log(N/O) when $t^2>0$, since this ratio has important implications for the N production mechanisms. Taking into account temperature inhomogeneities, we obtain a radial gradient of $\log$(N/O) with a slope of $-0.018\pm0.015$ dex kpc$^{-1}$. This means that $t^2>0$ do not alter the N/O gradient, which is consistent with that found by \citet{arellano2020, arellano2021} considering $t^2=0$. When comparing N/O to the O/H, we do not observe a clear plateau as metallicity decreases, and all objects seems to follow a slightly positive trend instead. The distribution of N/O versus O/H, shown in Fig.~\ref{fig:NO} has a positive slope of $0.12 \pm 0.20$ when $t^{2} > 0$, a value that is maintained even when the most metallic and uncertain points are removed. This supports the contribution of some secondary N in the Galactic disc and is expected due to the metallicity range of our sample, since previous works in our Galaxy and other star forming galaxies suggest the primary N production mechanism dominates only at $12+\log$(O/H) $\lesssim 8.0$ \citep{VanZee1998, Henry2000, Vincenzo2016, Esteban2020}. However, a flat distribution cannot be ruled out considering the uncertainty of the slope of our N/O versus O/H distribution.

The radial gradients of Ne, S and Ar under the paradigm of $t^2 > 0$ are consistent with the solar values given by \citet{Lodders2019} and \citet{Asplund2021}, while those obtained with $t^2 = 0$ fail to reproduce both sets of solar abundances. The slopes obtained for these gradients, taking into account temperature inhomogeneities, are steeper than those derived by \citet{arellano2020} with $t^2 = 0$.  These gradients are also consistent with the O one within the uncertainties, as expected because they are $\alpha$-elements and are produced by the same population of massive stars \citep{Kunth83}. In the case of $t^2>0$, we find that the S abundances of Sh2~257, Sh2~271 and IC~5146 can be calculated without an ICF due to their low degree of ionization (following the criterion of \citet{arellano2020} of O$^{++}$/(O$^{+}$+O$^{++}$) $\leq 0.03$).


In the case of Cl, we do not use an ICF for M~20, M~17, M~42, NGC~2579, NGC~3576, and NGC~3603 when $t^2 > 0$, either because they have low degree of ionization (O$^{++}$/(O$^{+}$+O$^{++}$) $\leq 0.26$ \citep{arellano2020} or because they show [Cl~{\sc ii}], [Cl~{\sc iii}] and [Cl~{\sc iv}] emission lines in their spectra, allowing to compute the Cl abundance directly by summing up ionic abundances. In Fig.~\ref{fig:AbundGradients} it can be seen that the Cl gradient with $t^2>0$ obtained with the rest of objects is consistent with the solar values from \citet{Lodders2019} and \citet{Asplund2021}, while the homogeneous temperature scenario fails to reproduce these values \citep[a behaviour also found by][]{arellano2020}. Nevertheless, it also must be considered that the solar values of Cl/H are somewhat uncertain, since they can only be obtained from sunspot spectra because the solar spectrum does not include atomic features of Cl. This implies that its determination relies additionally on the molecular data used and on the atmosphere model to be suitable for a sunspot with unknown temperature, which makes their uncertainty probably larger than quoted \citep{Asplund2021}. 

As discussed by \citet{arellano2020}, we also expect a lockstep evolution of Cl and the rest of $\alpha-$elements (O among them), because they are produced in the same nucleosynthesis processes (being Cl an intermediate product of these). We find that the Cl gradient slope is consistent with those of all the $\alpha-$elements, except for O, both when $t^2 = 0$ and when $t^2 > 0$. Both in the case $t^2 = 0$ and $t^2> 0$, the slopes of the gradients of O and Cl differ by 50\%. This difference do not rely on the ICF used as it remains even when the ICF from \citet{Esteban2015} is used. Instead, this inconsistency seems to be originated in a statistical difference between the sample of regions used to derive the O and Cl gradients. The number of objects used to obtain the gradient of O is larger than that of Cl, sampling a wider range of Galactocentric distances. If we constrain our sample to those regions in the range of Galactocentric distances between $6.0 \text{ kpc} < R_G < 15.0 \text{ kpc}$ -- which covers the well-sampled area in the Cl gradient -- we obtain slopes of the O gradient of $-0.038\pm0.010$ dex kpc$^{-1}$ and $-0.051 \pm 0.010$ dex kpc$^{-1}$ for $t^2 = 0$ and $t^2 > 0$, respectively. These values are in better agreement with the slopes of the Cl gradients for each case, as expected under the mentioned scenario.

It is interesting to compare our gradients with $t^2 > 0$ with those gradients derived from CELs in the FIR. These lines, due to their origin, have a much lower dependence on temperature than optical CELs. Therefore, if chemical abundances based on optical CELs are underestimated in the case of $t^2=0$, the abundances derived with FIR should be less affected, being more similar to our predictions with $t^2 > 0$. \citet{Rudolph2006} determined the gradients of N, O and S using FIR CELs. 
For the O gradient, our slope is steeper and our intercept is higher in both  cases: $t^2 = 0$ and $t^2 > 0$. This may be due to the fact that the determination of the total O abundance in the work of \citet[][]{Rudolph2006} is not direct, requiring an ICF to correct for the lack of [O\thinspace II] lines in their FIR spectra. Errors in the ICF are especially critical for the innermost regions of the Galaxy where O$^+$ is usually the dominant ion. In the case of N, although our determinations and those of \citet{Rudolph2006} require the use of ICFs to estimate the contribution of N$^{2+}$ and N$^{+}$, respectively, both kinds of results are in agreement when the abundance derived with $t^2 > 0$ are considered in the optical data. In the case of S, both ions  S$^{+}$ and S$^{3+}$ may have relevant contributions to the total abundance. The slope of the radial gradient of S from \citet[][]{Rudolph2006} is consistent with ours in the case of $t^2 > 0$, this is not the case when comparing the intercept. 

In the case of the N/O ratio, the slopes of both gradients are consistent within the uncertainties. Moreover the N/O vs O/H distribution have a similar qualitative behavior: the observed trend in the optical with $t^2 = 0$ is practically flat \citep[In our case log(N/O) = -0.77 $\pm$ 0.08,][]{arellano2021}, while that observed in both FIR and optical CELs with $t ^ 2> 0$ has a positive slope. Both the FIR and the optical data with $t^2 > 0$ may support the contribution of some secondary N in the Galactic disc. However, the uncertainties are large enough to prevent us from being conclusive. The comparison between the optical data results assuming $t^2 = 0$ and the gradients obtained for H~{\sc ii} regions by other groups or other objects has been extensively discussed in \citet[][]{arellano2020,arellano2021}.

In our sample of Galactic H~{\sc ii} regions, we found no evidence of azimuthal variations in the chemical abundances and, essentially, their distribution depends only on the Galactocentric distance. However, our sample is concentrated in two Galactic quadrants, as shown in Fig.~\ref{fig:Gal_dis_fig}, whose characteristics may be different from the  H~{\sc ii} regions located in the rest of the Galaxy \citep[][]{Wenger19}. 

Recently, \citet[][]{DeCia2021} have claimed that there are variations of up to 1 dex in the metallicity of the neutral gas of the solar neighborhood. We definitively do not observe such variations in  H~{\sc ii} regions. The O abundances -- the proxy of metallicity in ionized nebulae-- present a rather uniform linear distribution with the Galactocentric radius, independently of considering $t^2 = 0$ or $t^2> 0$. In fact, the radial gradient of O is already perceptible even if we limit ourselves to heliocentric distances less than 3 kpc -- as the data of \citet[][]{DeCia2021} -- given the quality of our spectra and the detail of our analysis. As mentioned before and contrary to what happens with Fe and Ni atoms, the impact of O depletion onto dust grains is limited to a factor of the order of 0.1 dex at maximum, an order of magnitude below the variations acclaimed by \citet[][]{DeCia2021}. This implies that any hypothetical pristine gas inclusions in the solar neighborhood must be well mixed in  present-day Galactic H~{\sc ii} regions.

\section{Conclusions}
\label{sec:concl}

In this work, we determine the radial abundances gradients of He, C, N, O, Ne, S, Cl and Ar in the Milky Way assuming the temperature fluctuations paradigm, $t^2>0$, and the most recent parallax-based distances using Gaia EDR3 data. The analysis of the physical conditions and  chemical abundances of this work is complementary to the previous ones by \citet{Mendez2020} and \citet{arellano2020, arellano2021}, where  a homogeneous temperature structure, $t ^ 2 = 0$, is assumed. 

We calculate the parallax-based distances to the Galactic H~{\sc ii} regions through the identification of their ionizing sources and using the \citet{Bailer-Jones2021} statistical approach. The comparison of the Gaia EDR3 parallax-based distances with kinematic ones shows that the Galactic rotation curve of \citet{Reid2014, Reid2019}, calibrated with maser parallaxes obtained at radio wavelengths, provide more consistent results than the classic rotation curve of \citet{Brand93} for the regions of our sample.


To apply the Peimbert's $t^2$ formalism \citep[][]{Peimbert67}, we use the weighted mean of the $t^2$ values obtained for M~8, M~16, M~17, M~20, NGC~3576, NGC~3603, Sh~2-311 and NGC~2579 assuming that the ADF(O$^{++}$) is produced by temperature fluctuations \citep[][]{garciarojasyesteban07,Esteban:2013}. That sample of H~{\sc ii} regions covers $R_G$ values  between 6.5 and 11.6 kpc. We find that $t^2$ shows a rather constant distribution with $R_G$, which suggests that a $ t ^ 2 = 0.038 \pm 0.004$ may be representative for our sample. We apply the \citet{Peimbert67} formalism using the representative $t^2$ value and the ionic abundances determined by \citet{arellano2020, arellano2021}, obtained under the assumption of $ t ^ 2 = 0 $. To estimate the total abundances of C, N, Ne, S, Cl, and Ar, we use the ICFs from \citet{Amayo2021}, including the uncertainties of the ICFs in our calculations.

We find that total abundances determined from CELs of all elements at a given $R_G$ increase by up to 0.3 dex when $t^2 > 0$ is considered, becoming more consistent with the solar values recommended by \citet{Lodders2019} and \citet{Asplund2021}. The radial distributions of the $\alpha$-elements are consistent with a common slope --which is expected due to their common nucleosynthesis-- when using $t ^ 2> 0$. 
The $t^2$ formalism does not seem to introduce inconsistencies in the distributions of the chemical abundances studied. Therefore it can be invoked as a likely explanation of the ADF problem in Galactic H~{\sc ii} regions. However, the demonstration of the physical processes that may generate these temperature fluctuations is a matter of debate and there is still no definitive solution to this problem.



\section*{Acknowledgements}
We are very grateful to Dr. Laura Magrini for her useful comments to the first version of the paper. JEM-D is very grateful to Žofia Chrobáková for her suggestions and discussions about this work. JEM-D appreciates the friendly communication of Trey Wenger and the exchange of views regarding kinematic distances. JEM-D thanks the support of the Instituto de Astrof\'isica de Canarias under the Astrophysicist Resident Program and acknowledges support from the Mexican CONACyT (grant CVU 602402). JEM-D, CE and JG-R acknowledge support from (i) the Agencia Estatal de Investigaci\'on del Ministerio de Ciencia e Innovaci\'on (AEI-MCINN) under grant {\it Espectroscop\'\i a de campo integral de regiones \ion{H}{2} locales. Modelos para el estudio de regiones \ion{H}{2} extragal\'acticas} with reference 10.13039/501100011033 and (ii) the grant P/308614 financed by funds transferred from the Spanish Ministry of Science, Innovation and Universities (MCIU), charged to the General State Budgets and from the General Budgets of the Autonomous Community of the Canary Islands by the MCIU. AM-A thanks Grazyna Stasi\'nska for her suggestions and several conversations and CONACyT for her PhD. scholarship (CVU No. 825508). AM-A, GD-I, and LC thank support from PAPIIT (DGAPA-UNAM) grant no. IN$-$103820. 
LC acknowledges support from the grants IA-100420, IG-100622, and IN-100519
(DGAPA-PAPIIT,UNAM) and funding from the CONACYT grant CF19-39578.
JG-R acknowledges support from the Severo Ochoa excellence program CEX2019-000920-S.
This work has made use of data from the European Space Agency (ESA) mission {\it Gaia} (\url{https://www.cosmos.esa.int/Gaia}), processed by the {\it Gaia}
Data Processing and Analysis Consortium (DPAC,
\url{https://www.cosmos.esa.int/web/Gaia/dpac/consortium}). Funding for the DPAC
has been provided by national institutions, in particular the institutions
participating in the {\it Gaia} Multilateral Agreement.

\section*{Data availability}
This research is based on public data available in cited the references. Our results are entirely present in the tables of this work.



\bibliographystyle{mnras}
\bibliography{Mendez}

\newpage


\appendix
\setcounter{table}{0}
\renewcommand{\thetable}{A\arabic{table}}

\setcounter{figure}{0}
\renewcommand{\thefigure}{A\arabic{figure}}

\begin{table*}
\centering
\caption{Sample of Galactic H~{\sc ii} regions and ring nebulae for which distances have been analyzed in this work. The table contains the name of the ionizing/associated stars of each nebula, spectral type, Gaia DR2 ID and the geometric distances determined from their parallax in the heliocentric reference system ($d$). All objects have parallax accuracy better than 20\,per cent unless otherwise indicated.}
\label{tab:identi_stars}
\begin{tabular}{cccccccccc}
\hline 
Nebula & Ion./Assoc. Star & Spectral Type & Gaia DR2 ID  & $d$ DR2  & $d$ EDR3  & Ref. & Ref.\\
 &  &  &   & (kpc) & (kpc) &  I/A &  SpT \\

\hline
Sh~2-29 & HD~165921 & O7V+B0V & 4066278846098710016 & $1.07 ^{+0.07}_{-0.06}$ & $1.15 ^{+0.04}_{-0.03} $ & 1 & 32 \\

Sh~2-32 & ALS~17181 & B1V & 4066292830512043520 & $1.59 ^{+0.15}_{-0.13}$ & $1.46\pm 0.05$ & 1 & 33 \\
Sh~2-47 & BD-15~4913 & B0.5III & 4098093077571330816 & $1.66 ^{+0.14}_{-0.12}$ & $1.54\pm 0.05$ & 1 & 1 \\

Sh~2-48 & BD-14~5014 & O7.5V & 4098272263591120128 & $4.35 ^{+2.09}_{-1.22}$ $^{b}$ & $3.59^{+0.70}_{-0.61}$ $^{a}$ & 1 & 34 \\

\multirow{ 2}{*}{Sh~2-53} & TIC~413308470 & O6V-O5V & 4152566800618412928 & $4.36^{+1.70}_{-1.08}$ $^{b}$ & $3.38 \pm 0.18$ & 2 & 35 \\

 &TIC~333769778 & O8V-9V & 4152566628819783424 & $3.36^{+1.64}_{-0.91}$ $^{b}$ & $4.27^{+3.24}_{-1.20}$ $^{a}$ & 2 & 2 \\

\multirow{ 2}{*}{Sh~2-54} & HD~167971 & O8Ia-O4/5 & 4153657378750644864 & $1.98 ^{+0.64} _{-0.39}$ $^{a}$ & $1.34 ^{+0.15} _{-0.12}$ & 44 & 32\\
& WR~113 & WC8+O8/9 & 4153716198275554176 & $1.82^{+0.19}_{-0.16}$ & $1.91 \pm 0.05$ & 1 & 36 \\

Sh~2-57 & BD-08~4623 & B0.5III & 4156562838196105600 & $2.10^{+0.24}_{-0.20}$ & $2.10 ^{+0.06}_{-0.08}$ & 1 & 1 \\

Sh~2-61 & EM*~AS~310 & B1 & 4256460509022075008 & $2.39^{+0.30}_{-0.24}$ & $2.39 \pm 0.08$ & 3 & 37 \\

Sh~2-82 & HD~231616 & B0.5V & 4323127850373105536 & $0.85^{+0.10}_{-0.08}$ & $0.78 \pm 0.06$ & 1, 4 & 1, 4 \\


\multirow{ 3}{*}{Sh~2-88} & HD~338916 & O7.5V & 2020947043840246144 & $1.93^{+0.16}_{-0.14}$ & $2.06 ^{+0.08} _{-0.07}$ & 1 & 34 \\
 & HD~338926 & O8.5Ib & 2020923679216984064 & $2.08^{+0.19}_{-0.16}$ & $1.99 \pm 0.06$ & 1 & 1 \\
 & LS~II~+25~09 & B0.5V & 2020883959358967552 & $4.38^{+2.98}_{-1.89}$ $^{b}$ & N/A$^{\text{(I)}}$ & 1 & 1 \\

Sh~2-90 & TIC~287521091 & O8–O9MS & 2027904753426401152 & $6.47^{+2.74}_{-1.77}$ $^{b}$ & $3.36^{+0.39}_{-0.28}$ & 5 & 5 \\

Sh~2-93 & [F89b]~S93~1 & B2V & 2027545659769078016 & $3.30^{+2.65}_{-1.61}$ $^{b}$ & $3.54^{+1.85}_{-1.35}$ $^{b}$ & 4 & 38 \\

Sh~2-98 &  WR~130  & WN8 & 2030934212864924032  & $10.12 ^{+3.69} _{-2.71}$ $^{b}$  &  $11.24 ^{+2.59} _{-2.52}$ $^{b}$ & 1 & 43  \\

Sh~2-100 & s~D & O5 & 2058236705827295872 & $9.46^{+2.39}_{-1.70}$ $^{b}$ & $11.40^{+1.14}_{-1.01}$ & 6 & 6 \\

Sh~2-127 & WB~85~B & O8.5 & 2176011164377605888 & $9.45^{+2.59}_{-1.89}$ $^{b}$ & $12.29^{+5.19}_{-2.38}$ $^{b}$ & 7 & 7 \\

Sh~2-128 & ALS~19702 & O7 & 2177724310884924288 & $8.62^{+2.24}_{-1.60}$ $^{b}$ & $7.85^{+1.61}_{-1.11}$ $^{a}$ & 8 & 8 \\

Sh~2-132 & WR~153 & WN6o+O6I & 2006219978921880192 & $4.14 ^{+0.57}_{-0.45}$ & $4.53^{+0.29}_{-0.27}$ & 9 & 39 \\

Sh~2-152 & Sh2-152~4 & O8.5V & 2013435042930935936 & $4.96^{+2.14}_{-1.48}$ $^{b}$ & $4.45^{+1.78}_{-1.34}$ $^{b}$ & 10 & 10 \\

Sh~2-156 & Anon & O7 & 2013668306897124736 & $2.57 ^{+0.30}_{-0.25}$ & $2.56^{+0.22}_{-0.19}$ & 11 & 11 \\

Sh~2-158 & TYC~4279-1463-1 & O3.5V+O9.5V & 2014960718396556160 & $2.67 ^{+0.27} _{-0.22}$ & $2.84 ^{+0.14} _{-0.16}$ & 51 & 34 \\

Sh~2-175 & LS~I~+64~26 & B1.5V & 527233525086096768 & $2.12^{+0.14}_{-0.12}$ & $2.02 \pm 0.05$ & 1 & 10 \\

Sh~2-203 & LS~I~+55~47 & O9.5IV & 446947804498934272  & $2.55 ^{+0.25} _{-0.21}$  & $2.39 ^{+0.08} _{-0.09}$ & 48 & 49  \\

Sh~2-206 & BD+50~886  & O4V & 250748269579173248$^{\text{(II)}}$ & $3.14 ^{+0.82} _{-0.56}$ $^{a}$  & $2.96 ^{+0.17} _{-0.15}$  & 1 & 34 \\

Sh~2-207 & MFJ~SH~2-207~1 & O9.5V  & 275291258896477312  & $3.53 ^{+0.52} _{-0.40}$  &  $3.59 ^{+0.23} _{-0.26}$ & 13  &  50 \\

Sh~2-208 &  MFJ~SH~2-208~3  & B0V  & 275284696186500096  & $4.98 ^{+0.95} _{-0.70}$ $^{a}$  & $4.02 ^{+0.27} _{-0.25}$  &  13 & 13  \\

\multirow{ 3}{*}{Sh~2-209} & ALS~18696 & B1III & 271701112917796096 & $2.60^{+0.26}_{-0.22}$ & $2.61 \pm 0.14$ & 12 & 12 \\

&  ALS~18697  &  O9III & 271701009838634752 & $2.55 ^{+0.59} _{-0.41}$ & $3.12 ^{+0.61} _{-0.49}$ & 12 & 12 \\

&  [CW84]~S209-3   &  BI & 271701112917794176 & $2.41 ^{+0.82} _{-0.51}$ $^{a}$ & $3.18 ^{+0.86} _{-0.62}$ $^{a}$ & 12 & 12 \\

\multirow{ 4}{*}{Sh~2-212} & MFJ~SH~2-212~2 & O7 & 260166892342134528 & $0.78 ^{+0.53} _{-0.23}$ $^{a}$ & $7.95 ^{+4.94} _{-2.71}$ $^{b}$ & 13, 14 & 32 \\

 & MFJ~SH~2-212~5 & ? & 260166681885774208 & $5.39 ^{+1.62} _{-1.09}$ $^{b}$ & $4.30 ^{+0.65} _{-0.51}$ & 13, 14 & - \\

 & MFJ~SH~2-212~7 & ? & 260167270299248384 & $5.10 ^{+1.06} _{-0.77}$ $^{a}$ & $3.94 ^{+0.36} _{-0.35}$ & 13, 14 & - \\

 & MFJ~SH~2-212~11 & ? & 260167339018718208 & $5.98^{+2.17}_{-1.46}$ $^{b}$ & $4.90 ^{+1.22} _{-0.96}$ $^{b}$ & 13, 14 & - \\
 
Sh~2-219 & LS~V~+47~22 & B0.5V & 254926272030205440 & $3.84^{+0.66}_{-0.50}$ & $4.16 ^{+0.32} _{-0.28}$ & 1 & 1 \\

Sh~2-228 &  ALS~19710  & O8V  & 187250236123135360  & $3.23 ^{+0.52} _{-0.40}$  & $2.56 \pm 0.09$ & 12  & 12  \\

Sh~2-235 & BD+35~1201 & O9.5V & 3455729793008409088 & $1.65^{+0.12}_{-0.11}$ & $1.66 \pm 0.07$ & 15, 16 & 40 \\

Sh~2-237 & LS~V~+34~46 & B2V & 182584041919665024 & $2.16^{+0.35}_{-0.27}$ & $2.07 \pm 0.06$ & 1, 17 & 17 \\

Sh~2-255 & LS~19 & O9.5V & 3373362495052179456 & $2.38^{+0.43}_{-0.32}$ & $1.96 ^{+0.12} _{-0.09}$ & 18 & 18 \\

Sh~2-257 & HD~253327 & B0.5V & 3373408159144488320 & $0.29 ^{+0.07} _{-0.05}$ & $2.55 ^{+1.99} _{-1.03}$ $^{b}$ & 18 & 18 \\

Sh~2-266 & MWC~137 & sgB[e] & 3344973478481675264 & $5.52^{+1.73}_{-1.17}$ $^{b}$ & $4.60 ^{+0.53} _{-0.51}$ & 19 & 41 \\

Sh~2-270 & [NS84]~8 & B0.5V & 3344023436011171200 & $3.84^{+2.47}_{-1.54}$ $^{b}$ & $1.79 ^{+1.27} _{-0.42}$ $^{b}$ & 20 & 20 \\

Sh~2-271 & ALS~18672 & O9V & 3331937432404833792 & $3.64^{+0.58}_{-0.44}$ & $3.25 ^{+0.11} _{-0.12}$ & 13, 21 & 13 \\

Sh~2-283 & MFJ~SH~2-283~8  & B0V  & 3119828097371908992  & $6.79 ^{+1.75} _{-1.24}$ $^{b}$ & $5.37 ^{+0.58} _{-0.41}$  & 13 & 13  \\

Sh~2-285 & [L85]~S285~1 & B0V & 3112496794360485248 & $3.82^{+0.75}_{-0.55}$ & $4.38 ^{+0.42} _{-0.31}$ & 13, 22 & 22 \\

Sh~2-288 & RAFGL~5223 & B0V & 3107524257323390208 & $5.26^{+1.53}_{-1.06}$ $^{b}$ & $5.06 ^{+0.49} _{-0.47}$ & 3 & 42 \\

Sh~2-297 & HD~53623 & B0V & 3045713939855362944 & $1.06^{+0.08}_{-0.07}$ & $1.08 ^{+0.04} _{-0.05}$ & 23 & 23 \\

Sh~2-298 & WR~7 & WN4b & 3032940844556081408 & $4.32^{+0.90}_{-0.66}$ $^{a}$ & $4.12 ^{+0.54} _{-0.40}$ & 24 & 39 \\

Sh~2-301 &  MFJ~SH~2-301~1  & O6.5V &  2934257408213035264$^{\text{(III)}}$ & $4.45 ^{+1.04} _{-0.73}$ $^{a}$  & $3.21 \pm 0.14$  & 13  & 34 \\

Sh~2-308 & WR~6 & WN4b & 2922367976673391232 & $2.27^{+0.30}_{-0.24}$ & $1.51 ^{+0.10} _{-0.08}$ & 25 & 39 \\

\multirow{ 3}{*}{Sh~2-311} & HD~64315 & O5.5V+O7V & 5602025904044961536 & $7.51 ^{+2.56} _{-1.82}$ $^{b}$ & $6.62 ^{+2.38} _{-1.72}$ $^{b}$ & 1 & 32 \\
 & HD~64568 & O3V & 5602033390154015744 & $5.39 ^{+1.26} _{-0.90}$ $^{a}$ & $4.26 ^{+0.25} _{-0.34}$ & 1 & 32 \\

 & LSS~830 & O7V & 5602027755165929344 & $4.72 ^{+0.87} _{-0.65}$ & $5.28 ^{+0.40} _{-0.35}$ & 1 & 34 \\

\hline
\end{tabular}
\end{table*}

\begin{table*}
\ContinuedFloat
\centering
\caption{Continued.}
\begin{tabular}{cccccccccc}
\hline 
Nebula & Ion./Assoc. Star & Spectral Type & Gaia DR2 ID  & $d$ DR2  & $d$ EDR3  & Ref. & Ref.\\
 &  &  &   & (kpc) & (kpc) &  I/A &  SpT \\

\hline

IC~5146 & BD+46~3474 & B0V & 1974546106933956608 & $0.72 \pm 0.02$ & $0.76 \pm 0.01$ & 1 & 1 \\

RCW~52 & LS~1887 & O8V & 5350529730352416384 & $2.45 ^{+0.20} _{-0.17}$ & $2.33 \pm 0.07$ & 1 & 42 \\

RCW~58 & WR~40 & WN8h & 5240040631514998272 & $3.90^{+0.54}_{-0.43}$ & $2.70 ^{+0.13} _{-0.11}$ & 26 & 43 \\

G2.4+1.4 & WR~102 & WR & 4067059872964459648 & $2.67^{+0.22}_{-0.19}$ & $2.65 \pm 0.15$ & - & 42 \\

\multirow{ 2}{*}{NGC~2579} & VdBH~13a~B & O5V & 5542914544020602880 & $5.85 ^{+0.73} _{-0.59}$ & $5.19 ^{+0.48} _{-0.35}$ & 27 & 27 \\

 & VdBH~13b~A & O6.5V & 5542914715819276160$^{\text{(IV)}}$ & $2.99 ^{+2.04} _{-1.11}$ $^{b}$ & $2.75 ^{+1.33} _{-0.88}$ $^{a}$ & 27 & 27 \\

NGC~3576 & HD~97484 & O8+O8V & 5337239658437912960 & $2.76 ^{+0.39} _{-0.30}$ & $2.48 ^{+0.11} _{-0.10}$ & 28, 45 & 45 \\

NGC~3603 & - & - & - & $7.02 \pm 0.10$ & - & 29 & - \\

NGC~6888 & WR~136 & WN6 & 2061690233159124352 & $1.95^{+0.14}_{-0.12}$ & $1.67 \pm 0.04$ & - & 43 \\

NGC~7635 & BD+60~2522 & O6.5 & 2014149897293278848 & $2.50^{+0.20}_{-0.17}$ & $2.83 ^{+0.13} _{-0.10}$ & - & 32 \\

M~8 & HD~165052 & O7V+O7.5V & 4066064956700837248 & $1.24 ^{+0.08} _{-0.07}$ & $1.22 \pm 0.04$ & 30 & 30 \\

M~16 & - & - & - & $1.71 \pm {0.18}$ & - & 31 & - \\

M~17 & - & - & - & $1.82\pm {0.16}$ & - & 31 & - \\

M~20 & HD~164492 & O7V+B6V+A2Ia+aBe? & 4069268658691324672 & $1.53 ^{+0.26} _{-0.19}$ & $1.42 ^{+0.09} _{-0.08}$ & 31 & 46, 47 \\

M~42 & - & - & - & $0.41\pm {0.01}$ & - & 31 & - \\

\hline
\end{tabular}
\begin{description}
\item (1)\citet{Avedisova84}, (2)\citet{Paron13}, (3)\citet{Hunter90}, (4)\citet{Forbes89}, (5)\citet{Samal14}, (6)\citet{samal10}, (7)\citet{rudolph96}, (8)\citet{Bohigas03}, (9)\citet{Harten78}, (10)\citet{Russeil07}, (11)\citet{Lynds83}, (12)\citet{chini84}, (13)\citet{Moffat79}, (14)\citet{Deharveng08}, (15)\citet{Kirsanova08}, (16)\citet{camargo11}, (17)\citet{pandey13}, (18)\citet{Ojha11}, (19)\citet{Esteban98}, (20)\citet{Neckel84}, (21)\citet{Persi87}, (22)\citet{Rolleston94}, (23)\citet{Mallick12}, (24)\citet{Esteban89}, (25)\citet{Chu82}, (26)\citet{Chu822}, (27)\citet{Copetti07}, (28)\citet{Townsley11}, (29)\citet{Drew19}, (30)\citet{Damiani17}, (31)\citet{Binder2018}, (32)\citet{Sota14}, (33)\citet{Naze09}, (34)\citet{Maiz2016}, (35)\citet{Ji12}, (36)\citet{Skiff14}, (37)\citet{hernandez04}, (38)\citet{Hunter90}, (39)\citet{Smith96}, (40)\citet{Georgelin73}, (41)\citet{Mehner16}, (42)\citet{Reed03}, (43)\citet{Hamann06}, (44)\citet{DeBecker05}, (45)\citet{Povich17}, (46)\citet{Rho04}, (47)\citet{rho06}, (48)\citet{Churchwell73}, (49)\citet{Roman-Lopes19}, (50)\citet{Molina-Lera18}, (51)\citet{Lynds86},  \\
\item $^{\text{(I)}}$ The parallax could not be estimated in EDR3.
\item $^{\text{(II)}}$ In the EDR3 the new ID is 250748269580520320
\item$^{\text{(III)}}$ In the EDR3 the new ID is 2934257408221360384
\item$^{\text{(IV)}}$ In the EDR3 the new ID is 5542914715819272960
\item $^a$: $0.2\leq e_\text{plx}/\text{plx}<0.3$
\item $^b$: $e_\text{plx}/\text{plx} \geq 0.3$ or negative.
\end{description}
\end{table*}

\begin{table*}
\centering
\caption{Galactocentric distance estimates for the sample of Galactic H~{\sc ii} regions and ring nebulae. }
\label{tab:Galactocentric_dis}
\begin{tabular}{ccccccccc}
\hline 
Nebula & \multicolumn{6}{c}{$R_G$ (kpc)} \\
& \multicolumn{2}{c}{Geometric} & Photogeometric & 1/Plx & \multicolumn{2}{c}{Kinematic}  \\

 & DR2 & EDR3 & EDR3 &  EDR3 & W19 & R03-R07 \\

\hline
Sh~2-29 & $7.13 \pm 0.17$ & $7.06 \pm 0.14$ & $7.07 \pm 0.13$ & $7.07 \pm 0.13$ & - & $5.81 \pm 0.89$ \\
Sh~2-32 & $6.62 \pm 0.25$ & $6.75 \pm 0.15$ & $6.76 \pm 0.16$ & $6.75 \pm 0.15$ & - & $5.69 \pm 1.00$ \\
Sh~2-47 & $6.62 \pm 0.24$ & $6.73 \pm 0.15$ & $6.71 \pm 0.14$ & $6.72 \pm 0.14$ & - & $6.59 \pm 0.47$ \\
Sh~2-48 & $4.28 \pm 1.80$ & $4.86 \pm 0.69$ & - & - & - & $4.86 \pm 0.27$ \\
Sh~2-53 & $4.35 ^{+1.51} _{-1.17}$ & $5.10 \pm 0.25$ & $5.07 \pm 0.28$ & $5.10 \pm 0.27$ & $4.68 \pm 0.28$ & $5.11 \pm 0.25$ \\
Sh~2-54 & $6.52 \pm 0.26$ & $6.42 \pm 0.15$ & $6.43 \pm 0.16$ & $6.43 \pm 0.15$ & $6.08 \pm 0.27$ & $6.16 \pm 0.37$ \\
Sh~2-57 & $6.33 \pm 0.30$ & $6.32 \pm 0.17$ & $6.33 \pm 0.17$ & $6.33 \pm 0.16$ & - & $6.52 \pm 0.26$ \\
Sh~2-61 & $6.16 \pm 0.34$ & $6.15 \pm 0.16$ & $6.16 \pm 0.16$ & $6.16 \pm 0.15$ & - & $5.97 \pm 0.24$ \\
Sh~2-82 & $7.73 \pm 0.16$ & $7.76 \pm 0.13$ & $7.75 \pm 0.14$ & $7.77 \pm 0.13$ & - & $7.57 \pm 0.23$ \\
Sh~2-83 & - & - & - & - & $13.24 \pm 1.13$ & - \\
Sh~2-88 & $7.47 \pm 0.14$ & $7.44 \pm 0.12$ & $7.44 \pm 0.11$ & $7.44 \pm 0.11$ & $7.52 \pm 0.19$ & $10.94 ^{+12.83} _{-3.29}$ \\
Sh~2-90 & $7.92 \pm 1.34$ & $7.33 \pm 0.12$ & $7.33 \pm 0.12$ & - & $7.78 \pm 0.25$ & $11.38 ^{+13.36} _{-3.52}$ \\
Sh~2-93 & $7.56 ^{+0.54} _{-0.26}$ & $7.47 \pm 0.39$ & $7.58 ^{+0.70} _{-0.28}$ & - & $7.66 \pm 0.19$ & $8.48 ^{+4.44} _{-0.76}$ \\
Sh~2-98 & $10.33 ^{+2.81} _{-2.15}$ & $11.10 \pm 1.97$ & $11.49 ^{+1.72} _{-1.47}$ & - & $11.53 \pm 0.88$ & $8.53 \pm 0.26$ \\
Sh~2-100 & $10.13 ^{+1.74} _{-1.31}$ & $11.57 \pm 0.88$ & $11.08 \pm 1.27$ & - & $9.37 \pm 0.41$ & $9.83 \pm 0.29$ \\
Sh~2-127 & $13.08 \pm 2.15$ & $15.59 \pm 4.40$ & $14.25 \pm 2.32$ & - & $13.76 \pm 1.31$ & $15.06 \pm 0.39$ \\
Sh~2-128 & $12.64 \pm 1.87$ & $12.13 \pm 1.24$ & $11.52 \pm 0.99$ & - & $11.78 \pm 0.78$ & $13.02 \pm 0.46$ \\
Sh~2-132 & $9.98 \pm 0.44$ & $10.21 \pm 0.27$ & $10.23 \pm 0.26$ & $10.18 \pm 0.30$ & - & $10.98 \pm 0.33$ \\
Sh~2-152 & $10.77 ^{+1.75} _{-1.35}$ & $10.50 \pm 1.30$ & $10.15 \pm 1.10$ & - & $10.61 ^{+0.71} _{-0.43}$ & $9.55 \pm 0.11$ \\
Sh~2-156 & $9.40 \pm 0.26$ & $9.39 \pm 0.22$ & $9.38 \pm 0.22$ & $9.38 \pm 0.24$ & $10.75 \pm 0.58$ & $9.86 \pm 0.48$ \\
Sh~2-158 & $9.52 \pm 0.25$ & $9.62 \pm 0.19$ & $9.62 \pm 0.21$ & $9.60 \pm 0.19$ & $9.62 \pm 0.09$ & $9.57 \pm 0.09$ \\
Sh~2-175 & $9.45 \pm 0.19$ & $9.38 \pm 0.13$ & $9.38 \pm 0.13$ & $9.38 \pm 0.14$ & - & $9.09 \pm 0.14$ \\
Sh~2-203 & $10.35 \pm 0.32$ & $10.22 \pm 0.18$ & $10.21 \pm 0.17$ & $10.22 \pm 0.17$ & - & $9.91 \pm 0.28$ \\
Sh~2-206 & $11.09 \pm 0.78$ & $10.88 \pm 0.25$ & $10.88 \pm 0.26$ & $10.86 \pm 0.26$ & $10.77 \pm 0.59$ & $11.36 \pm 0.81$ \\
Sh~2-207 & $11.44 \pm 0.56$ & $11.50 \pm 0.33$ & $11.44 \pm 0.34$ & $11.41 \pm 0.34$ & - & $12.39 \pm 0.91$ \\
Sh~2-208 & $12.74 \pm 0.99$ & $11.88 \pm 0.34$ & $11.71 \pm 0.46$ & $11.91 \pm 0.45$ & - & $12.36 \pm 0.94$ \\
Sh~2-209 & $10.55 \pm 0.34$ & $10.57 \pm 0.23$ & $10.50 \pm 0.23$ & $10.53 \pm 0.24$ & $14.75 ^{+1.56} _{-1.10}$ & $16.43 \pm 1.92$ \\
Sh~2-212 & $13.84 \pm 2.20$ & $12.81 \pm 1.25$ & $12.02 \pm 1.02$ & - & $14.76 \pm 1.30$ & $14.27 ^{+1.38} _{-1.55}$ \\
Sh~2-219 & $11.86 \pm 0.74$ & $12.18 \pm 0.42$ & $12.16 \pm 0.40$ & $12.12 \pm 0.36$ & - & $12.97 \pm 0.81$ \\
Sh~2-228 & $11.42 \pm 0.57$ & $10.73 \pm 0.18$ & $10.71 \pm 0.20$ & $10.71 \pm 0.19$ & - & $10.83 ^{+1.72} _{-1.37}$ \\
Sh~2-235 & $9.84 \pm 0.23$ & $9.86 \pm 0.17$ & $9.85 \pm 0.15$ & $9.85 \pm 0.17$ & - & $103.95 ^{+103.77} _{-64.79}$ \\
Sh~2-237 & $10.34 \pm 0.43$ & $10.26 \pm 0.16$ & $10.22 \pm 0.18$ & $10.25 \pm 0.17$ & - & $11.85 ^{+3.46} _{-2.32}$ \\
Sh~2-255 & $10.53 \pm 0.50$ & $10.12 \pm 0.22$ & $10.14 \pm 0.20$ & $10.11 \pm 0.20$ & $9.89 \pm 0.08$ & $10.61 \pm 0.38$ \\
Sh~2-257 & $10.53 ^{+0.50} _{-0.52}$ & $10.13 \pm 0.22$ & $10.13 \pm 0.20$ & $10.10 \pm 0.20$ & $9.89 \pm 0.08$ & $10.61 \pm 0.36$ \\
Sh~2-266 & $13.69 \pm 1.79$ & $12.67 \pm 0.65$ & $12.87 \pm 0.77$ & - & - & $16.17 \pm 3.14$ \\
Sh~2-270 & $12.09 ^{+2.51} _{-2.15}$ & $10.00 \pm 1.25$ & $10.39 \pm 1.32$ & - & - & $15.18 \pm 2.21$ \\
Sh~2-271 & $11.72 \pm 0.64$ & $11.34 \pm 0.22$ & $11.37 \pm 0.23$ & $11.33 \pm 0.23$ & - & $11.89 \pm 1.49$ \\
Sh~2-283 & $14.39 \pm 1.89$ & $13.08 \pm 0.66$ & $13.20 \pm 0.63$ & - & - & $15.98 \pm 0.29$ \\
Sh~2-285 & $11.52 \pm 0.80$ & $12.09 ^{+0.46} _{-0.50}$ & $11.99 \pm 0.50$ & $12.10 \pm 0.44$ & - & $15.84 \pm 0.27$ \\
Sh~2-288 & $12.71 \pm 1.50$ & $12.54 ^{+0.50} _{-0.51}$ & $12.50 \pm 0.85$ & - & $12.99 \pm 0.86$ & $14.15 \pm 1.08$ \\
Sh~2-297 & $8.97 \pm 0.16$ & $8.99 \pm 0.14$ & $8.99 \pm 0.13$ & $8.99 \pm 0.14$ & $8.56 \pm 0.28$ & $9.24 \pm 0.35$ \\
Sh~2-298 & $11.59 \pm 0.83$ & $11.42 \pm 0.53$ & $11.38 \pm 0.58$ & - & $11.73 \pm 0.72$ & $13.06 \pm 0.67$ \\
Sh~2-301 & $11.48 \pm 0.91$ & $10.51 \pm 0.21$ & $10.55 \pm 0.20$ & $10.52 \pm 0.21$ & $11.60 \pm 0.73$ & $11.40 \pm 0.47$ \\
Sh~2-308 & $9.67 \pm 0.32$ & $9.14 \pm 0.17$ & $9.10 \pm 0.19$ & $9.13 \pm 0.16$ & - & $9.37 \pm 0.28$ \\
Sh~2-311 & $11.13 \pm 0.84$ & $11.60 \pm 0.39$ & $11.60 \pm 0.44$ & $11.55 \pm 0.45$ & $10.84 \pm 0.73$ & $11.51 \pm 0.40$ \\
IC~5146 & $8.29 \pm 0.10$ & $8.29 \pm 0.10$ & $8.29 \pm 0.10$ & $8.29 \pm 0.10$ & - & - \\
RCW~52 & $7.83 \pm 0.10$ & $7.83 \pm 0.10$ & $7.83 \pm 0.10$ & $7.83 \pm 0.10$ & - & $8.36 ^{+1.34} _{-0.22}$ \\
RCW~58 & $7.64 \pm 0.16$ & $7.60 \pm 0.10$ & $7.60 \pm 0.10$ & $7.60 \pm 0.10$ & - & $8.53 \pm 0.24$ \\
G2.4+1.4 & $5.53 \pm 0.32$ & $5.55 \pm 0.27$ & $5.60 \pm 0.26$ & $5.60 \pm 0.23$ & - & - \\
NGC~2579 & $11.26 \pm 0.61$ & $10.80 \pm 0.40$ & $10.74 \pm 0.39$ & $10.74 \pm 0.39$ & - & - \\
NGC~3576 & $7.65 \pm 0.11$ & $7.66 \pm 0.10$ & $7.66 \pm 0.10$ & $7.66 \pm 0.10$ & - & - \\
NGC~3603 & $8.61 \pm 0.11$ & - & - & - & - & $9.21 \pm 0.28$ \\
NGC~6888 & $7.94 \pm 0.10$ & $7.95 \pm 0.10$ & $7.95 \pm 0.10$ & $7.95 \pm 0.10$ & - & - \\
NGC~7635 & $9.42 \pm 0.22$ & $9.64 \pm 0.18$ & $9.61 \pm 0.17$ & $9.62 \pm 0.17$ & $10.28 ^{+0.66} _{-0.44}$ & - \\
M~8 & $6.97 \pm 0.18$ & $6.98 \pm 0.14$ & $7.00 \pm 0.14$ & $7.00 \pm 0.14$ & - & $5.74 \pm 0.98$ \\
M~16 & $6.58 \pm 0.27$ & - & - & - & - & $6.16 \pm 0.38$ \\
M~20 & $6.69 \pm 0.35$ & $6.79 \pm 0.19$ & $6.78 \pm 0.22$ & $6.83 \pm 0.19$ & - & $5.08 \pm 0.70$ \\
M~17 & $6.45 \pm 0.25$ & - & - & - & $6.49 \pm 0.09$ & $6.21 \pm 0.45$ \\
M~42 & $8.54 \pm 0.11$ & - & - & - & $8.70 \pm 0.01$ & $9.43 \pm 0.56$ \\

\hline
\end{tabular}
\begin{description}
\item W19: \citet{Wenger19}.
\item R03-R07: \citet{Russeil03}, \citet{Russeil07}.
\end{description}
\end{table*}

\begin{figure}
\includegraphics[width=\columnwidth]{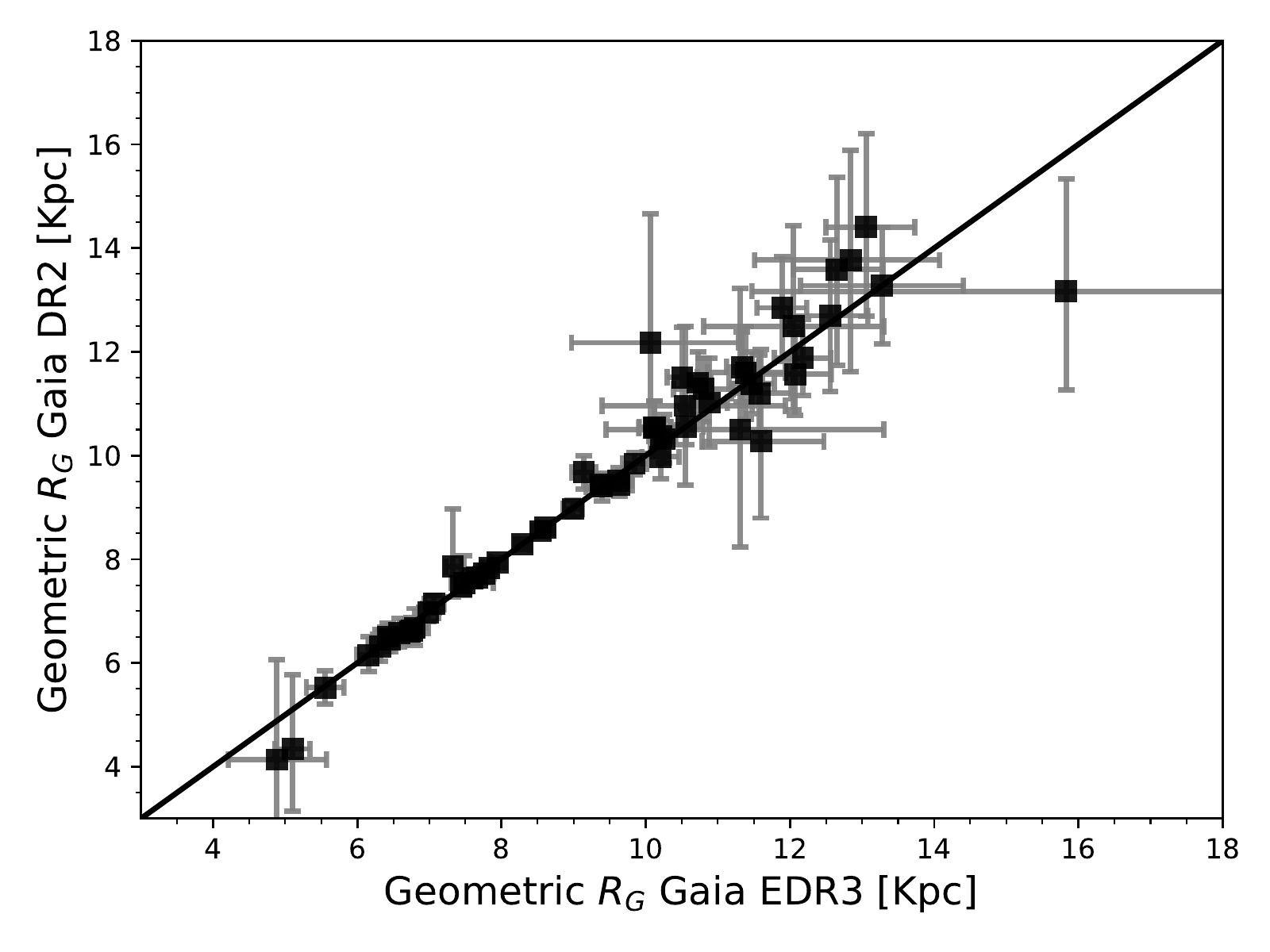}
\includegraphics[width=\columnwidth]{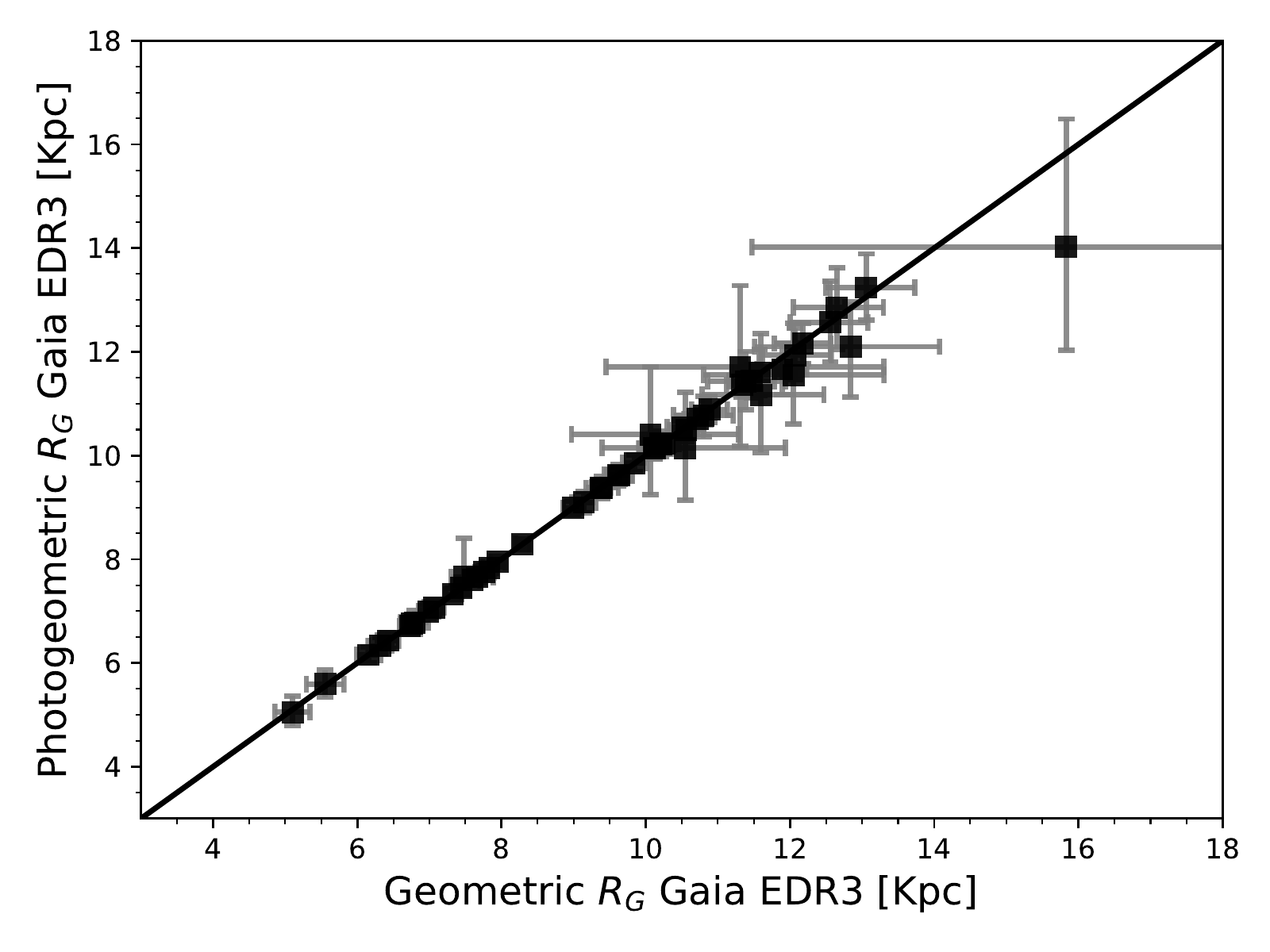}
\includegraphics[width=\columnwidth]{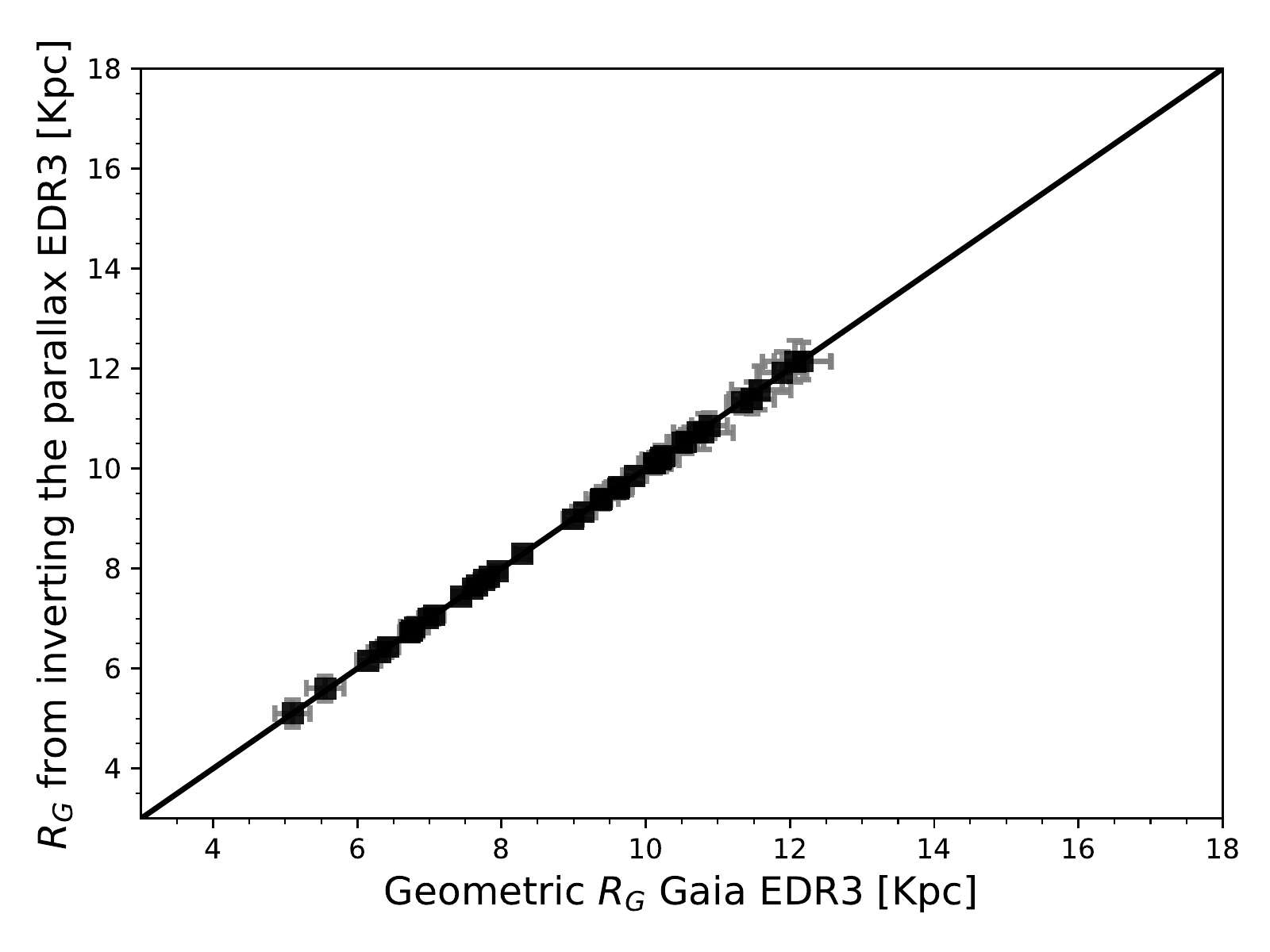}
\caption{Comparison of the Gaia EDR3 geometric Galactocentric distance with other determinations based on Gaia parallaxes. \textit{Top panel:} Gaia DR2 geometric distance. \textit{Middle panel:} Gaia EDR3 photogeometric distance. \textit{Bottom panel:} Distance obtained directly as the inverse of Gaia EDR3 parallax in a sub-sample with uncertainties in the parallax smaller than 10\,per cent.}
\label{fig:geo_EDR3vs_other_Gaia_distances}
\end{figure}

\begin{figure}
\includegraphics[width=\columnwidth]{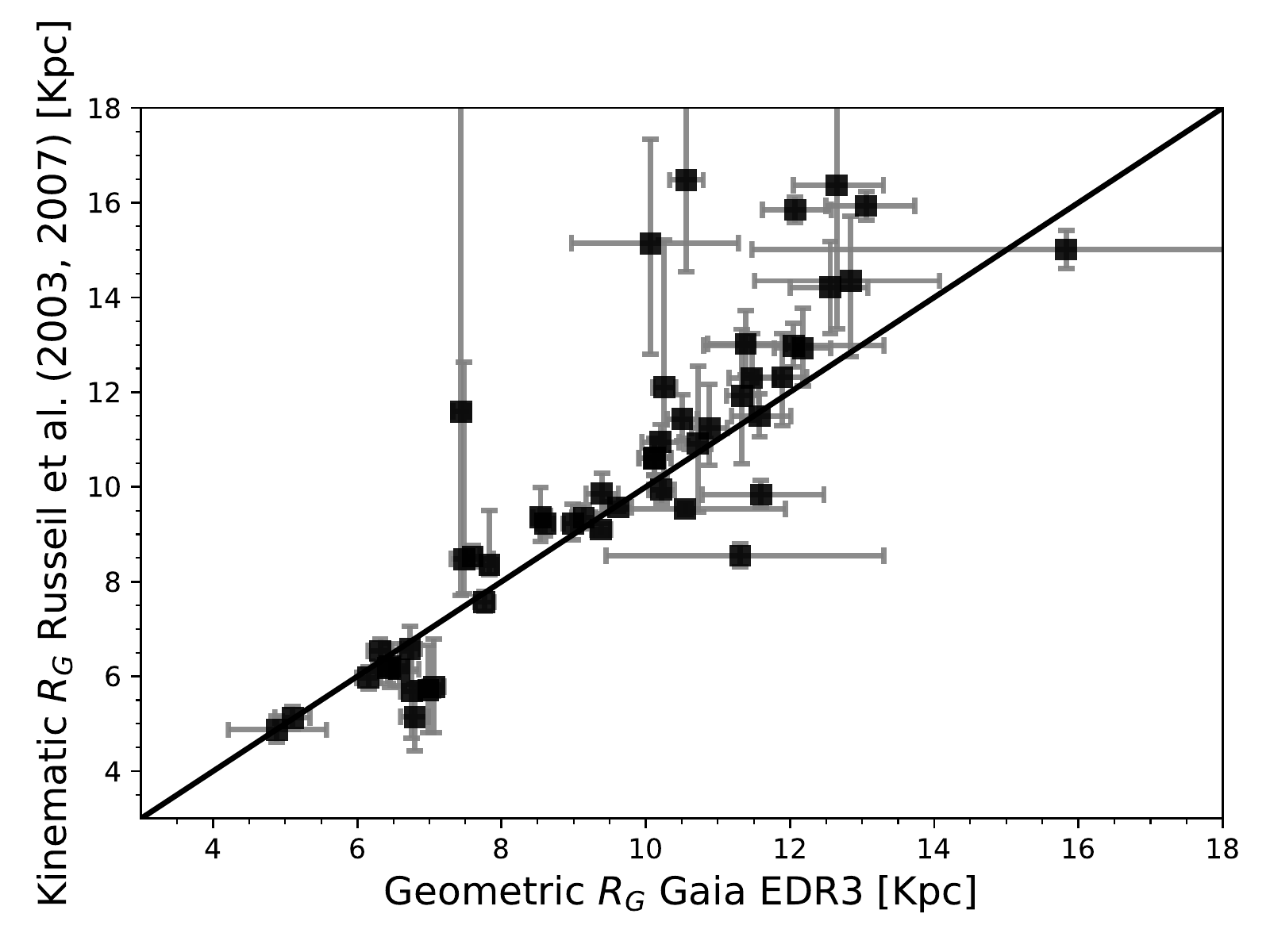}
\includegraphics[width=\columnwidth]{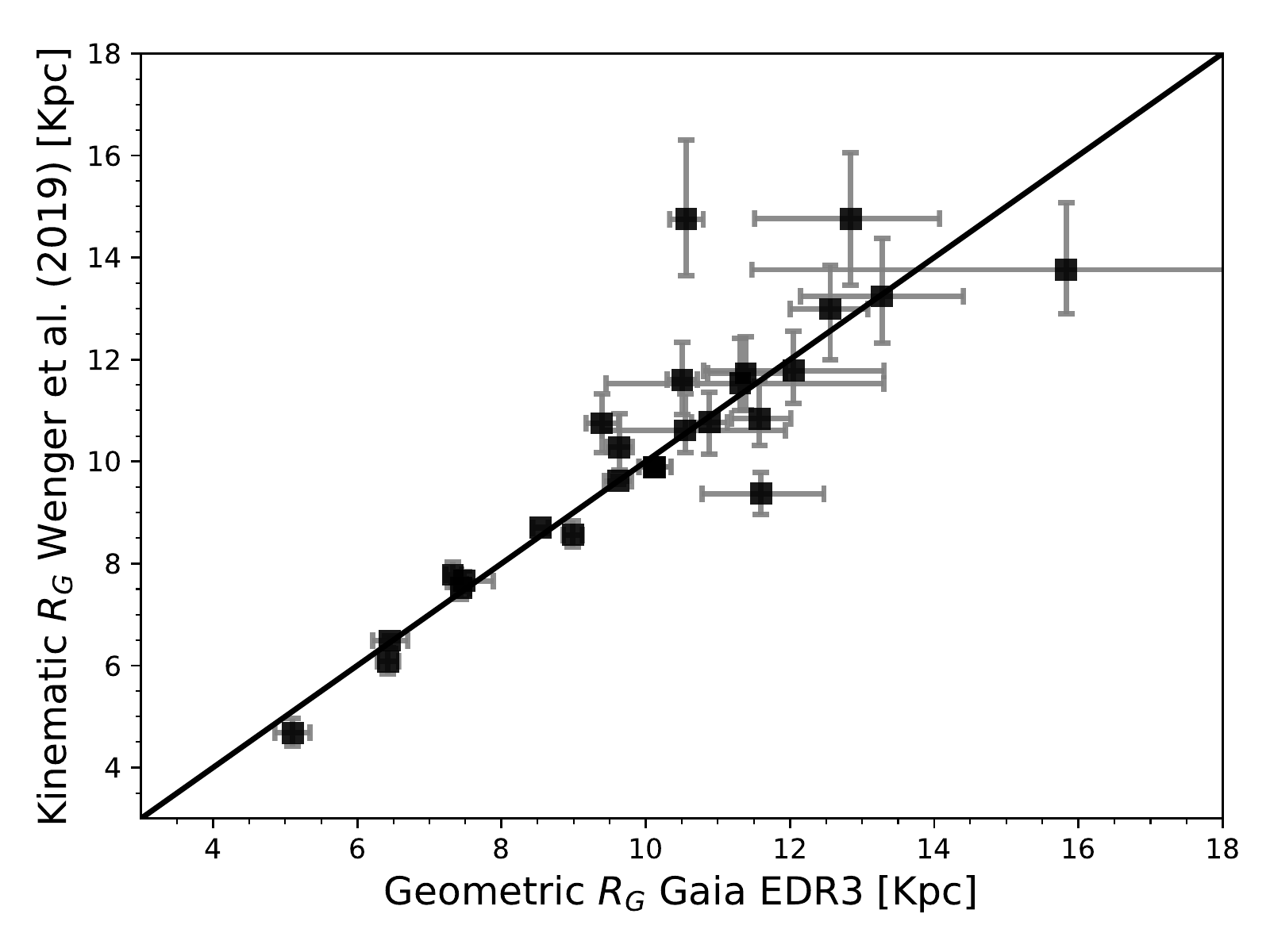}
\caption{Comparison of the Gaia EDR3 geometric Galactocentric distances with kinematic distance determinations. \textit{Top panel:} Determinations from \citet{Russeil03} and \citet{Russeil07} based on the rotation curve of \citet{Brand93}. \textit{Botton panel:} Determinations from \citet{Wenger19} based on the rotation curve of \citet{Reid2014} using a Monte Carlo approach. }
\label{fig:kin_compari}
\end{figure}

\begin{landscape}
\begin{table}
\centering
\caption{ Physical conditions, Ionic abundances derived assuming $t^2=0.038\pm 0.004$ in units of 12+log(X$^{i+}$/H$^+$) and references of spectroscopic data}
\label{tab:ionic_abundances_1_t2}
\scriptsize
\begin{tabular}{cccccccccccccc}
\hline
Region   & Reference &$n_{\rm e}$[\ion{O}{II}] & $n_{\rm e}$[\ion{S}{II}] & $n_{\rm e}$[\ion{Cl}{III}] & $T_{\rm e}$[\ion{N}{II}] & $T_{\rm e}$[\ion{O}{III}]   & He$^{+}$             & C$^{2+}$           & N$^{+}$                  & O$^{+}$                  & O$^{2+}$                 & O$^{2+}$        & O$^{2+}$/O\\                                                                          
         &           &(cm$^{-3}$)              & (cm$^{-3}$)              & (cm$^{-3}$)                & (K)                      & (K)                        & (RLs)                & (RLs)              &                          &                          &                          & (RLs)           & (CELs) \\                                                                    
\hline
Sh~2-29  & 1  &       -                        & $100 \pm 40$             & $700_{-600}^{3000}$        & $7500 \pm 200$           & $6300 \pm 200^{a}$         & $10.52 \pm 0.01$     &-                   &$8.05 ^{+0.06} _{-0.05}$  &$8.83 \pm 0.07$           &$7.98 ^{+0.15} _{-0.13}$  &-                &0.12  \\                                                                  
Sh~2-32  & 1  &       -                        & $100 \pm 50$             & -                          & $7100 \pm 200$           & $5800 \pm 300^{a}$         & -                    &-                   &$8.02 ^{+0.07} _{-0.08}$  &$8.78 \pm 0.11$           &$7.52 ^{+0.25} _{-0.17}$  &-                &0.05  \\                                                                  
Sh~2-47  & 1  &       -                        & $100 \pm 70$             & -                          & $7000 \pm 500$           & $5600_{-700}^{+600}$ $^{a}$   & $9.92 \pm 0.03$      &-                   &$8.06^{+0.18} _{-0.19}$  &$8.93 ^{+0.34} _{-0.32}$  &$8.53 ^{+0.76} _{-0.57}$  &-                &0.29  \\
Sh~2-48  & 1  &       -                        & $100 \pm 60$             & -                          & $6800 \pm 500$           & $5300 \pm 700^{a}$         & $10.87 \pm 0.07$     &-                   &$7.99 ^{+0.20} _{-0.19}$  &$8.77 ^{+0.37} _{-0.34}$  &$8.99 ^{+0.82} _{-0.50}$  &-                &0.62  \\                                                                  
Sh~2-53  & 1  &       -                        & $300 \pm 100$            & -                          & $7000 \pm 400$           & $5600 \pm 600^{a}$         & $10.59 \pm 0.04$     &-                   &$8.29 ^{+0.16} _{-0.15}$  &$8.98 \pm 0.28$           &$8.52 ^{+0.50} _{-0.42}$  &-                &0.26  \\                                                                  
Sh~2-54  & 1  &       -                        & $200 \pm 50$             & $200_{-610}^{900}$         & $7500 \pm 200$           & $8800_{-1000}^{+800}$      & $10.95 \pm 0.03$     &$8.49 \pm 0.11$     &$7.76 ^{+0.06} _{-0.07}$  &$8.55 ^{+0.13} _{-0.12}$  &$7.91 ^{+0.36} _{-0.35}$  &-                &0.19  \\
Sh~2-61  & 2  &       -                        & $1100 \pm 200$           & -                          & $7500 \pm 400$           & $6300 \pm 500^{a}$         & -                    &-                   &$8.09 ^{+0.10} _{-0.09}$  &$8.64 ^{+0.14} _{-0.15}$  &$6.34 ^{+0.35} _{-0.25}$  &-                &0.01  \\                                                                  
Sh~2-82  & 1  &       -                        & $100 \pm 70$             & -                          & $7000 \pm 900$           & $5600 \pm 1200^{a}$        & $10.14 \pm 0.02$     &-                   &$8.06 ^{+0.29} _{-0.30}$  &$8.89 ^{+0.44} _{-0.49}$  &-                         &-                &$\sim 0.0$  \\                                                                  
Sh~2-83  & 3  &       -                        & $300 \pm 100$            & -                          & $11900 \pm 600$          & $10400 \pm 500$            & $10.899 \pm 0.007$   &-                   &$6.40 \pm 0.08$           &$7.24 ^{+0.13} _{-0.14}$  &$8.43 \pm 0.09$           &-                &0.94  \\                                                                  
Sh~2-88  & 1  &       -                        & $900 \pm 100$            & $< 100$                    & $8300 \pm 200$           & $7300 \pm 300^{a}$         & $10.67 \pm 0.04$     &-                   &$7.83 \pm 0.05$           &$8.64 \pm 0.09$           &$8.03 ^{+0.14} _{-0.12}$  &-                &0.20  \\                                                                  
Sh~2-90  & 2  &       -                        & $200 \pm 100$            & -                          & $8200 \pm 200$           & $7200 \pm 300^{a}$         & $10.92 \pm 0.01$     &-                   &$7.92 ^{+0.06} _{-0.05}$  &$8.53 \pm 0.08$           &$8.37 ^{+0.15} _{-0.12}$  &-                &0.41  \\                                                                  
Sh~2-93  & 1  &       -                        & $500 \pm 50$             & -                          & $7900 \pm 300$           & $7000 \pm 400^{a}$         & $10.47 \pm 0.02$     &-                   &$8.04 \pm 0.08$           &$8.61 ^{+0.14} _{-0.15}$  &$7.68 ^{+0.23} _{-0.22}$  &-                &0.11  \\                                                                  
Sh~2-100 & 3  &       -                        & $400 \pm 200$            & -                          & $8600 \pm 300$           & $8200 \pm 100$             & $10.984 \pm 0.015$   &$8.36 \pm 0.08$     &$6.93 ^{+0.07} _{-0.08}$  &$7.90 ^{+0.11} _{-0.10}$  &$8.70 ^{+0.06} _{-0.05}$  &$8.52 \pm 0.06$  &0.87  \\                                                                                
Sh~2-127 & 3  &       -                        & $600 \pm 100$            & -                          & $9800 \pm 200$           & $9500 \pm 200^{a}$         & $10.799 \pm 0.008$   &-                   &$7.41 \pm 0.03$           &$8.33 \pm 0.05$           &$7.61 \pm 0.06$           &-                &0.16  \\                                                                  
Sh~2-128 & 3  &       -                        & $500 \pm 100$            & -                          & $10600 \pm 200$          & $9900 \pm 300$             & $10.928 \pm 0.015$   &-                   &$6.91 ^{+0.05} _{-0.04}$  &$7.94 \pm 0.06$           &$8.18 ^{+0.08} _{-0.07}$  &-                &0.63  \\                                                                  
Sh~2-132 & 4  &       -                        & $300 \pm 100$            & -                          & $9300 \pm 500$           & $8800 \pm 600^{a}$         & $10.85 \pm 0.06$     &-                   &$7.60 \pm 0.09$           &$8.50 \pm 0.15$           &$7.38 \pm 0.20$           &-                &0.07  \\                                                                  
Sh~2-152 & 2  &       -                        & $700 \pm 100$            & -                          & $8200 \pm 100$           & $7300 \pm 100^{a}$         & $10.802 \pm 0.007$   &$7.97 \pm 0.14$     &$7.74 \pm 0.03$           &$8.63 ^{+0.04} _{-0.03}$  &$8.11 \pm 0.07$           &-                &0.23  \\                                                                  
Sh~2-156 & 4  &       -                        & $900 \pm 100$            & -                          & $9400 \pm 400$           & $9000 \pm 400$             & $10.90 \pm 0.05$     &-                   &$7.41 \pm 0.07$           &$8.37 \pm 0.13$           &$7.93 ^{+0.10} _{-0.11}$  &-                &0.26  \\                                                                  
Sh~2-175 & 2  &       -                        & $< 100$                  & -                          & $7000 \pm 1300$          & $5600 \pm 400^{a}$         & -                    &-                   &$7.98 ^{+0.18} _{-0.12}$  &$8.77 ^{+0.23} _{-0.15}$  &-                         &-                &$\sim 0.0$  \\                                                                  
Sh~2-209 & 3  &       -                        & $300 \pm 300$            & -                          & $10600 \pm 800$          & $10700_{-1200}^{+1000}$    & $10.893 \pm 0.046$   &-                   &$6.84 \pm 0.14$           &$7.74 \pm 0.32$           &$8.05 ^{+0.24} _{-0.23}$  &-                &0.67  \\                                                                  
Sh~2-212 & 3  &       -                        & $< 100$                  & -                          & $8300 \pm 700$           & $11100 \pm 1000$           & $11.019 \pm 0.019$   &-                   &$6.78 ^{+0.17} _{-0.16}$  &$8.35 \pm 0.31$           &$7.98 ^{+0.20} _{-0.18}$  &-                &0.30  \\                                                                  
Sh~2-219 & 2  &       -                        & $< 100$                  & -                          & $8600 \pm 300$           & $7800 \pm 500^{a}$         & $10.35\pm 0.04$      &-                   &$7.60 ^{+0.07} _{-0.06}$  &$8.45 ^{+0.11} _{-0.12}$  &$6.38 ^{+0.22} _{-0.19}$  &-                &0.01  \\
Sh~2-235 & 2  &       -                        & $< 100$                  & -                          & $8100 \pm 200$           & $7100 \pm 200^{a}$         & $10.59 \pm 0.01$     &-                   &$7.73 ^{+0.05} _{-0.04}$  &$8.60 ^{+0.08} _{-0.07}$  &$7.31 ^{+0.12} _{-0.11}$  &-                &0.05  \\
Sh~2-237 & 2  &       -                        & $400 \pm 100$            & -                          & $8700 \pm 300$           & $8000 \pm 400^{a}$         & $9.67 \pm 0.07$      &-                   &$7.69 \pm 0.05$           &$8.50 \pm 0.09$           &$6.71 \pm 0.13$           &-                &0.02  \\
Sh~2-257 & 2  &       -                        & $100 \pm 100$            & -                          & $7900 \pm 200$           & $6900 \pm 300^{a}$         & $10.24 \pm 0.02$     &-                   &$7.78 \pm 0.07$           &$8.60 \pm 0.10$           &$7.00 ^{+0.17} _{-0.16}$  &-                &0.02  \\                                                                  
Sh~2-266 & 2  &       -                        & $300 \pm 200$            & -                          & $8300 \pm 400$           & $7400 \pm 600^{a}$         & $10.32 \pm 0.07$     &-                   &$7.69 ^{+0.11} _{-0.12}$  &$8.36 ^{+0.18} _{-0.16}$  &$6.44 ^{+0.27} _{-0.26}$  &-                &0.01  \\                                                                  
Sh~2-270 & 2  &       -                        & $400 \pm 100$            & -                          & $9300 \pm 1000$          & $8800 \pm 1300^{a}$        & -                    &-                   &$7.38 ^{+0.20} _{-0.18}$  &$8.24 ^{+0.28} _{-0.30}$  &-                         &-                &$\sim 0.0$  \\                                                                  
Sh~2-271 & 2  &       -                        & $< 100$                  & -                          & $8700 \pm 200$           & $7900 \pm 300^{a}$         & $10.22 \pm 0.03$     &-                   &$7.65 \pm 0.06$           &$8.44 \pm 0.08$           &$6.16 \pm 0.12$           &-                &0.01  \\                                                                  
Sh~2-285 & 2  &       -                        & $< 100$                  & -                          & $8500 \pm 300$           & $7700 \pm 40^{a}$          & -                    &-                   &$7.59 \pm 0.07$           &$8.38 \pm 0.11$           &-                         &-                &$\sim 0.0$  \\                                                                  
Sh~2-288 & 3  &       -                        & $400 \pm 300$            & -                          & $9400 \pm 300$           & $9200 \pm 500$             & $10.792 \pm 0.010$   &-                   &$7.34 \pm 0.07$           &$8.35 \pm 0.10$           &$7.99 ^{+0.15} _{-0.13}$  &-                &0.30  \\                                                                  
Sh~2-297 & 2  &       -                        & $< 100$                  & -                          & $7800 \pm 200$           & $6700 \pm 300^{a}$         & $10.17 \pm 0.04$     &-                   &$7.89 ^{+0.06} _{-0.05}$  &$8.67 \pm 0.08$           &$7.17 ^{+0.16} _{-0.14}$  &-                &0.03  \\                                                                  
Sh~2-298 & 2  &       -                        & $< 100$                  & -                          & $11700 \pm 500$          & $11700 \pm 200$            & $10.971 \pm 0.017$   &-                   &$7.41 \pm 0.07$           &$8.24 \pm 0.11$           &$8.26 ^{+0.04} _{-0.03}$  &-                &0.51  \\                                                                                         
Sh~2-311 & 5  &       $300 \pm 100$            & $300 \pm 100$            & -                          & $9300 \pm 200$           & $8900 \pm 100$             & $10.914 \pm 0.006$   &$8.01 \pm 0.05$     &$7.44 ^{+0.04} _{-0.05}$  &$8.42 \pm 0.06$           &$8.08 \pm 0.04$           &$8.04 \pm 0.04$  &0.32  \\                                                                  
IC~5146  & 6  &       $< 100$                  & $< 100$                  & -                          & $7100 \pm 100$           & $5700 \pm 200^{a}$         & -                    &-                   &$8.03 ^{+0.06} _{-0.05}$  &$8.80 ^{+0.11} _{-0.10}$  &-                         &-                &$\sim 0.0$  \\                                                                  
NGC~2579 & 7  &       $1200 \pm 300$           & $800 \pm 200$            & -                          & $8600 \pm 300^{a}$       & $9300 \pm 200$             & $10.934 \pm 0.026$   &$8.18 \pm 0.05$     &$6.99 \pm 0.06$           &$8.13 ^{+0.10} _{-0.11}$  &$8.44 \pm 0.05$           &$8.46 \pm 0.03$  &0.67  \\                                                                                         
NGC~3576 & 8  &       $1700 \pm 200$           & $1100 \pm 300$           & -                          & $8800 \pm 200$           & $8400 \pm 100$             & $10.941 \pm 0.015$   &$8.44 \pm 0.02$     &$7.21 ^{+0.06} _{-0.05}$  &$8.21 \pm 0.06$           &$8.65 \pm 0.04$           &$8.62 \pm 0.05$  &0.74  \\                                                                                
NGC~3603 & 9  &       $2700_{-500}^{+800}$     & $3100_{-700}^{+1300}$    & -                          & $11200 \pm 500$          & $9000 \pm 100$             & $10.996 \pm 0.027$   &$8.49 \pm 0.07$     &$6.57 ^{+0.07} _{-0.08}$  &$7.44 \pm 0.14$           &$8.67 \pm 0.05$           &$8.69 \pm 0.05$  &0.94  \\                                                                                
M~8      & 10 &       $1600 \pm 600$           & $1300 \pm 200$           & -                          & $8400 \pm 100$           & $8000 \pm 100$             & $10.838 \pm 0.009$   &$8.31 \pm 0.02$     &$7.70 \pm 0.03$           &$8.52 \pm 0.06$           &$8.20 ^{+0.05} _{-0.04}$  &$8.23 \pm 0.03$  &0.32  \\                                                                                         
M~16     & 9  &       $1200 \pm 200$           & $1100 \pm 200$           & -                          & $8300 \pm 100$           & $7600 \pm 200$             & $10.892 \pm 0.009$   &$8.39 \pm 0.04$     &$7.89 ^{+0.04} _{-0.05}$  &$8.64 ^{+0.07} _{-0.06}$  &$8.26 ^{+0.10} _{-0.07}$  &$8.30 \pm 0.04$  &0.30  \\                                                                                         
M~17     & 10 &       $500 \pm 100$            & $400 \pm 100$            & -                          & $8900 \pm 200$           & $8000 \pm 100$             & $10.967 \pm 0.013$   &$8.73 \pm 0.03$     &$6.98 ^{+0.06} _{-0.05}$  &$7.95 ^{+0.08} _{-0.07}$  &$8.77 ^{+0.06} _{-0.05}$  &$8.68 \pm 0.02$  &0.87  \\                                                                                         
M~20     & 9  &       $300 \pm 100$            & $300 \pm 100$            & -                          & $8300 \pm 200$           & $7800 \pm 300$             & $10.854 \pm 0.009$   &$8.19 \pm 0.05$     &$7.76 ^{+0.05} _{-0.04}$  &$8.66 ^{+0.06} _{-0.05}$  &$8.06 ^{+0.11} _{-0.10}$  &$8.00 \pm 0.22$  &0.20  \\                                                                                         
M~42     & 11 &       $6900_{-1500}^{+2400}$   & $4900_{-1800}^{+2900}$    & -                          & $10200 \pm 200$          & $8300 \pm 100$             & $10.946 \pm 0.014$   &$8.34 \pm 0.02$     &$7.00 \pm 0.06$           &$7.85 \pm 0.12$           &$8.73 \pm 0.04$           &$8.57 \pm 0.01$  &0.88  \\
\hline
\end{tabular}
\begin{description}
\item (1) \citet{arellano2021}, (2) \citet{Esteban:2018}, (3) \citet{Esteban:2017}, (4) \citet{Fernandez-Martin:2017}, (5) \citet{Garcia-Rojas:2005}, (6) \citet{Garcia-Rojas:2014}, (7) \citet{Esteban:2013}, (8) \citet{Garcia-Rojas:2004}, (9) \citet{Garcia-Rojas:2006}, (10) \citet{Garcia-Rojas:2007}, (11) \citet{Esteban:2004}.
\item $^{a}$ $T_{\rm e}$ estimated using the temperature relation from \citet{Esteban2009}.
\end{description}
\end{table}
\end{landscape}

\begin{table*}
\centering
\caption{Ionic abundances derived assuming $t^2=0.038\pm 0.004$ in units of 12+log(X$^{i+}$/H$^+$).}
\label{tab:ionic_abundances_2_t2}
\begin{tabular}{ccccccccc}
\hline 
Nebula & Ne$^{2+}$ & S$^{+}$ & S$^{2+}$ &Cl$^{+}$ &Cl$^{2+}$ & Cl$^{3+}$ & Ar$^{2+}$  & Ar$^{3+}$ \\
\hline

Sh~2-29 & - &$6.96 ^{+0.06} _{-0.05}$ &$7.06 \pm 0.20$ &- &$4.90 ^{+0.13} _{-0.12}$ &- &$6.09 \pm 0.10$ &- \\
Sh~2-32 & - &$6.96 \pm 0.08$ &- &- &- &- &$5.37 \pm 0.17$ &- \\
Sh~2-47 & - &$6.99 ^{+0.18} _{-0.17}$ &- &- &- &- &$5.42 ^{+0.28} _{-0.25}$ &- \\
Sh~2-48 & - &$6.73 ^{+0.18} _{-0.16}$ &- &- &- &- &$6.59 \pm 0.36$ &- \\
Sh~2-53 & - &$6.79 ^{+0.20} _{-0.17}$ &- &- &- &- &$6.28 ^{+0.19} _{-0.20}$ &- \\
Sh~2-54 & - &$6.36 \pm 0.08$ &$6.97 ^{+0.30} _{-0.26}$ &- &$5.26 ^{+0.10} _{-0.09}$ &- &$6.23 ^{+0.16} _{-0.13}$ &- \\
Sh~2-61 & - &$6.98 \pm 0.10$ &- &- &- &- &- &- \\
Sh~2-82 & - &$6.94 ^{+0.35} _{-0.29}$ &- &- &- &- &$5.83 ^{+0.51} _{-0.47}$ &- \\
Sh~2-83 & $7.90 ^{+0.13} _{-0.12}$ &$5.28 \pm 0.07$ &$6.39 ^{+0.10} _{-0.12}$ &- &$4.68 \pm 0.09$ &- &$5.81 \pm 0.06$ &- \\
Sh~2-88 & - &$6.39 ^{+0.06} _{-0.05}$ &$7.19 ^{+0.16} _{-0.15}$ &- &$5.04 ^{+0.12} _{-0.14}$ &- &$6.24 ^{+0.09} _{-0.08}$ &- \\
Sh~2-90 & - &$6.50 ^{+0.05} _{-0.06}$ &- &- &$5.29 ^{+0.09} _{-0.10}$ &- &$6.49 ^{+0.11} _{-0.09}$ &- \\
Sh~2-93 & - &$6.71 ^{+0.09} _{-0.08}$ &$7.07 ^{+0.28} _{-0.26}$ &- &- &- &$6.14 \pm 0.15$ &- \\
Sh~2-100 & $8.15 ^{+0.07} _{-0.06}$ &$5.68 \pm 0.07$ &$7.09 ^{+0.13} _{-0.11}$ &- &$5.15 ^{+0.08} _{-0.07}$ &- &$6.48 \pm 0.06$ &$5.09 \pm 0.11$ \\
Sh~2-127 & - &$6.03 ^{+0.04} _{-0.03}$ &$6.87 ^{+0.07} _{-0.06}$ &- &$4.86 \pm 0.05$ &- &$5.82 \pm 0.05$ &- \\
Sh~2-128 & $7.35 ^{+0.12} _{-0.11}$ &$5.72 \pm 0.03$ &$6.60 ^{+0.07} _{-0.06}$ &- &$4.82 \pm 0.05$ &- &$5.97 \pm 0.03$ &- \\
Sh~2-132 & - &$6.51 ^{+0.10} _{-0.09}$ &$6.73 \pm 0.05$ &- &- &- &$6.24 ^{+0.11} _{-0.10}$ &- \\
Sh~2-152 & - &$6.32 \pm 0.03$ &$7.21 ^{+0.06} _{-0.05}$ &- &$5.17 \pm 0.04$ &- &$6.34 \pm 0.04$ &- \\
Sh~2-156 & $6.74 ^{+0.14} _{-0.13}$ &$6.00 ^{+0.09} _{-0.08}$ &$6.98 \pm 0.06$ &- &$4.99 \pm 0.09$ &- &$6.27 \pm 0.04$ &- \\
Sh~2-175 & - &$6.94 ^{+0.16} _{-0.10}$ &- &- &- &- &- &- \\
Sh~2-209 & - &$5.61 ^{+0.11} _{-0.12}$ &$6.44 ^{+0.31} _{-0.29}$ &- &- &- &- &- \\
Sh~2-212 & $7.06 ^{+0.24} _{-0.22}$ &$5.32 \pm 0.17$ &$6.75 \pm 0.29$ &- &$5.06 \pm 0.19$ &- &$6.16 ^{+0.19} _{-0.18}$ &- \\
Sh~2-219 & - &$6.46 \pm 0.06$ &- &- &$4.96 ^{+0.29} _{-0.30}$ &- &$5.42 \pm 0.08$ &- \\
Sh~2-235 & - &$7.09 \pm 0.08$ &$6.94 ^{+0.09} _{-0.08}$ &- &$4.93 \pm 0.07$ &- &$6.11 ^{+0.06} _{-0.05}$ &- \\
Sh~2-237 & - &$6.57 ^{+0.07} _{-0.06}$ &- &- &$5.01 \pm 0.22$ &- &$5.29 ^{+0.12} _{-0.11}$ &- \\
Sh~2-257 & - &$7.19 ^{+0.18} _{-0.17}$ &$7.00 ^{+0.19} _{-0.20}$ &- &$4.83 ^{+0.12} _{-0.11}$ &- &$5.61 ^{+0.15} _{-0.14}$ &- \\
Sh~2-266 & - &$6.77 ^{+0.10} _{-0.09}$ &- &- &- &- &- &- \\
Sh~2-270 & - &$6.42 ^{+0.18} _{-0.17}$ &- &- &- &- &- &- \\
Sh~2-271 & - &$6.84 \pm 0.09$ &$6.68 \pm 0.09$ &- &$4.87 \pm 0.10$ &- &$5.25 ^{+0.09} _{-0.11}$ &- \\
Sh~2-285 & - &$5.64 ^{+0.07} _{-0.08}$ &- &- &- &- &- &- \\
Sh~2-288 & - &$6.06 \pm 0.07$ &$6.81 ^{+0.11} _{-0.12}$ &- &$4.87 \pm 0.07$ &- &$5.96 ^{+0.07} _{-0.06}$ &- \\
Sh~2-297 & - &$6.65 ^{+0.06} _{-0.05}$ &- &- &$4.91 \pm 0.11$ &- &$5.71 ^{+0.09} _{-0.10}$ &- \\
Sh~2-298 & $8.01 ^{+0.05} _{-0.04}$ &$6.56 \pm 0.06$ &$6.61 \pm 0.06$ &- &$4.93 ^{+0.10} _{-0.09}$ &- &$6.05 \pm 0.03$ &$4.48 ^{+0.08} _{-0.10}$ \\
Sh~2-311 & $7.33 \pm 0.05$ &$6.34 \pm 0.04$ &$6.81 \pm 0.04$ &$4.41 \pm 0.04$ &$4.95 ^{+0.05} _{-0.04}$ &- &$6.24 \pm 0.04$ &- \\
IC~5146 & - &$7.00 \pm 0.06$ &$6.52 ^{+0.07} _{-0.06}$ &- &- &- &- &- \\
NGC~2579 & $7.58 \pm 0.06$ &$5.66 \pm 0.07$ &$6.80 ^{+0.05} _{-0.06}$ &$3.67 \pm 0.06$ &$5.05 \pm 0.07$ &$2.97 ^{+0.07} _{-0.06}$ &$6.34 \pm 0.04$ &$4.19 ^{+0.14} _{-0.12}$ \\
NGC~3576 & $8.09 \pm 0.04$ &$5.94 \pm 0.05$ &$7.00 ^{+0.06} _{-0.05}$ &$3.93 \pm 0.05$ &$5.04 \pm 0.04$ &$3.45 ^{+0.06} _{-0.05}$ &$6.51 ^{+0.04} _{-0.03}$ &$4.71 \pm 0.06$ \\
NGC~3603 & $8.15 \pm 0.06$ &$5.16 ^{+0.09} _{-0.08}$ &$6.81 \pm 0.08$ &$3.26 \pm 0.07$ &$4.86 \pm 0.08$ &$4.07 \pm 0.05$ &$6.36 \pm 0.05$ &$5.06 \pm 0.10$ \\
M~8 & $7.47 \pm 0.06$ &$6.22 ^{+0.05} _{-0.04}$ &$7.04 \pm 0.04$ &$4.36 \pm 0.03$ &$5.14 ^{+0.04} _{-0.03}$ &- &$6.41 \pm 0.03$ &$4.24 ^{+0.12} _{-0.11}$ \\
M~16 & $7.61 ^{+0.11} _{-0.10}$ &$6.60 ^{+0.04} _{-0.05}$ &$6.96 \pm 0.05$ &$4.57 ^{+0.06} _{-0.05}$ &$5.13 \pm 0.04$ &- &$6.45 ^{+0.04} _{-0.03}$ &$4.49 ^{+0.23} _{-0.21}$ \\
M~17 & $8.19 ^{+0.06} _{-0.05}$ &$5.67 \pm 0.06$ &$6.99 ^{+0.04} _{-0.05}$ &$3.73 ^{+0.10} _{-0.11}$ &$5.05 \pm 0.06$ &$3.41 ^{+0.20} _{-0.22}$ &$6.50 \pm 0.03$ &$4.57 ^{+0.16} _{-0.14}$ \\
M~20 & $7.13 \pm 0.11$ &$6.47 ^{+0.04} _{-0.03}$ &$6.96 \pm 0.04$ &$4.60 ^{+0.04} _{-0.05}$ &$5.12 \pm 0.05$ &- &$6.46 \pm 0.05$ &$4.59 ^{+0.27} _{-0.24}$ \\
M~42 & $8.17 \pm 0.04$ &$5.57 ^{+0.18} _{-0.16}$ &$6.79 \pm 0.06$ &$3.61 \pm 0.08$ &$4.92 ^{+0.06} _{-0.05}$ &$3.90 ^{+0.07} _{-0.06}$ &$6.41 ^{+0.05} _{-0.04}$ &$4.95 ^{+0.07} _{-0.06}$ \\

\hline
\end{tabular}
\end{table*}

\begin{landscape}
\begin{table}
\centering
\caption{Total abundances of C, N, O, Ne, S, Cl, and Ar derived assuming $t^{2}=0.0$ and $t^2=0.038\pm 0.004$ in units of 12+log(X/H).}
\label{tab:total_abundances}
\scriptsize
\begin{tabular}{ccccccccccccccc}
\hline
Region   & \multicolumn{2}{c}{C/H}                                   & \multicolumn{2}{c}{N/H}                                   & \multicolumn{2}{c}{O/H}                                   & \multicolumn{2}{c}{Ne/H}                                  & \multicolumn{2}{c}{S/H}                                   & \multicolumn{2}{c}{Cl/H}                                  & \multicolumn{2}{c}{Ar/H} \\
         & $t^2=0.0$                   & $t^2>0.0$                   & $t^2=0.0$                   & $t^2>0.0$                   & $t^2=0.0$                   & $t^2>0.0$                   & $t^2=0.0$                   & $t^2>0.0$                   & $t^2=0.0$                   & $t^2>0.0$                   & $t^2=0.0$                   & $t^2>0.0$                   & $t^2=0.0$                   & $t^2>0.0$\\
\hline
Sh~2-29  &                        --- &                       --- &     $7.87_{-0.05}^{+0.06}$ &     $8.11_{-0.05}^{+0.07}$ &               $8.65\pm0.07$ &     $8.89_{-0.06}^{+0.07}$ &                        --- &                        --- &     $6.99_{-0.08}^{+0.09}$ &     $7.29_{-0.09}^{+0.15}$ &     $5.07_{-0.13}^{+0.15}$ &               $5.17\pm0.11$ &                        --- &                        --- \\
Sh~2-32  &                        --- &                       --- &     $7.80_{-0.07}^{+0.08}$ &     $8.03_{-0.07}^{+0.09}$ &               $8.54\pm0.11$ &               $8.79\pm0.11$ &                        --- &                        --- &                        --- &                        --- &                        --- &                        --- &                        --- &                        --- \\
Sh~2-47  &                        --- &                       --- &     $7.92_{-0.09}^{+0.30}$ &     $8.30_{-0.21}^{+0.59}$ &     $8.75_{-0.13}^{+0.32}$ &     $9.08_{-0.19}^{+0.54}$ &                        --- &                        --- &                        --- &                        --- &                        --- &                        --- &                        --- &                        --- \\
Sh~2-48  &                        --- &                       --- &     $8.02_{-0.21}^{+0.34}$ &     $8.47_{-0.27}^{+0.86}$ &     $8.70_{-0.14}^{+0.38}$ &     $9.17_{-0.22}^{+0.68}$ &                        --- &                        --- &                        --- &                        --- &                        --- &                        --- &                        --- &                        --- \\
Sh~2-53  &                        --- &                       --- &     $8.11_{-0.14}^{+0.26}$ &     $8.46_{-0.18}^{+0.35}$ &     $8.78_{-0.20}^{+0.25}$ &     $9.10_{-0.17}^{+0.37}$ &                        --- &                        --- &                        --- &                        --- &                        --- &                        --- &                        --- &                        --- \\
Sh~2-54  &                        --- &                       --- &     $7.66_{-0.08}^{+0.16}$ &     $7.87_{-0.10}^{+0.15}$ &     $8.42_{-0.08}^{+0.13}$ &     $8.65_{-0.10}^{+0.13}$ &                        --- &                        --- &               $6.80\pm0.18$ &               $7.05\pm0.20$ &     $5.29_{-0.13}^{+0.09}$ &               $5.49\pm0.12$ &                        --- &                        --- \\
Sh~2-61  &                        --- &                       --- &               $7.89\pm0.12$ &               $8.09\pm0.12$ &               $8.41\pm0.17$ &               $8.63\pm0.15$ &                        --- &                        --- &                        --- &                        --- &                        --- &                        --- &                        --- &                        --- \\
Sh~2-82  &                        --- &                       --- &     $7.81_{-0.18}^{+0.32}$ &               $8.05\pm0.31$ &     $8.60_{-0.31}^{+0.49}$ &               $8.88\pm0.45$ &                        --- &                        --- &                        --- &                        --- &                        --- &                        --- &                        --- &                        --- \\
Sh~2-83  &                        --- &                       --- &     $7.53_{-0.16}^{+0.15}$ &     $7.68_{-0.14}^{+0.17}$ &     $8.27_{-0.08}^{+0.09}$ &     $8.45_{-0.10}^{+0.09}$ &               $7.76\pm0.13$ &               $7.95\pm0.11$ &               $6.44\pm0.11$ &               $6.59\pm0.12$ &     $4.72_{-0.09}^{+0.10}$ &               $4.81\pm0.10$ &                        --- &                        --- \\
Sh~2-88  &                        --- &                       --- &     $7.73_{-0.05}^{+0.06}$ &     $7.95_{-0.06}^{+0.09}$ &     $8.53_{-0.07}^{+0.08}$ &     $8.74_{-0.06}^{+0.09}$ &                        --- &                        --- &     $6.98_{-0.11}^{+0.09}$ &               $7.24\pm0.12$ &               $5.14\pm0.13$ &               $5.27\pm0.13$ &                        --- &                        --- \\
Sh~2-90  &                        --- &                       --- &               $7.98\pm0.07$ &     $8.24_{-0.08}^{+0.09}$ &     $8.51_{-0.06}^{+0.07}$ &     $8.76_{-0.06}^{+0.08}$ &                        --- &                        --- &                        --- &                        --- &               $5.27\pm0.10$ &     $5.41_{-0.09}^{+0.12}$ &                        --- &                        --- \\
Sh~2-93  &                        --- &                       --- &     $7.88_{-0.07}^{+0.10}$ &               $8.09\pm0.10$ &               $8.44\pm0.13$ &               $8.65\pm0.12$ &                        --- &                        --- &     $6.92_{-0.17}^{+0.13}$ &     $7.19_{-0.18}^{+0.25}$ &                        --- &                        --- &                        --- &                        --- \\
Sh~2-100 &                        --- &                       --- &               $7.63\pm0.12$ &               $7.90\pm0.12$ &     $8.49_{-0.03}^{+0.04}$ &     $8.77_{-0.04}^{+0.06}$ &               $7.94\pm0.05$ &     $8.24_{-0.06}^{+0.08}$ &               $6.94\pm0.10$ &               $7.20\pm0.11$ &     $5.06_{-0.06}^{+0.08}$ &     $5.24_{-0.07}^{+0.08}$ &               $6.31\pm0.05$ &     $6.50_{-0.06}^{+0.05}$ \\
Sh~2-127 &                        --- &                       --- &               $7.36\pm0.03$ &               $7.50\pm0.04$ &               $8.26\pm0.04$ &     $8.40_{-0.04}^{+0.05}$ &                        --- &                        --- &     $6.74_{-0.05}^{+0.06}$ &               $6.92\pm0.05$ &               $5.00\pm0.05$ &               $5.10\pm0.05$ &                        --- &                        --- \\
Sh~2-128 &                        --- &                       --- &               $7.26\pm0.06$ &     $7.42_{-0.07}^{+0.08}$ &               $8.21\pm0.05$ &               $8.38\pm0.05$ &               $7.44\pm0.11$ &               $7.61\pm0.11$ &     $6.52_{-0.05}^{+0.06}$ &               $6.67\pm0.06$ &     $4.80_{-0.05}^{+0.04}$ &     $4.90_{-0.04}^{+0.05}$ &                        --- &                        --- \\
Sh~2-132 &                        --- &                       --- &     $7.49_{-0.07}^{+0.10}$ &     $7.63_{-0.08}^{+0.15}$ &     $8.37_{-0.15}^{+0.14}$ &     $8.54_{-0.17}^{+0.14}$ &                        --- &                        --- &              $6.75\pm0.051$ &               $6.90\pm0.06$ &                        --- &                        --- &                        --- &                        --- \\
Sh~2-152 &                        --- &                       --- &               $7.66\pm0.03$ &     $7.90_{-0.04}^{+0.05}$ &               $8.53\pm0.03$ &     $8.75_{-0.02}^{+0.04}$ &                        --- &                        --- &     $6.98_{-0.03}^{+0.04}$ &               $7.26\pm0.05$ &               $5.24\pm0.03$ &               $5.37\pm0.04$ &                        --- &                        --- \\
Sh~2-156 &                        --- &                       --- &               $7.43\pm0.10$ &     $7.60_{-0.09}^{+0.11}$ &               $8.34\pm0.10$ &               $8.51\pm0.11$ &               $7.14\pm0.14$ &     $7.38_{-0.17}^{+0.14}$ &               $6.88\pm0.05$ &               $7.02\pm0.05$ &               $5.06\pm0.10$ &     $5.17_{-0.09}^{+0.11}$ &                        --- &                        --- \\
Sh~2-175 &                        --- &                       --- &     $7.74_{-0.09}^{+0.11}$ &     $7.99_{-0.11}^{+0.17}$ &               $8.51\pm0.15$ &               $8.76\pm0.16$ &                        --- &                        --- &                        --- &                        --- &                        --- &                        --- &                        --- &                        --- \\
Sh~2-209 &                        --- &                       --- &     $7.23_{-0.22}^{+0.30}$ &     $7.39_{-0.22}^{+0.38}$ &     $8.08_{-0.12}^{+0.26}$ &               $8.21\pm0.16$ &                        --- &                        --- &                        --- &     $6.53_{-0.19}^{+0.32}$ &                        --- &                        --- &                        --- &                        --- \\
Sh~2-212 &                        --- &                       --- &               $6.85\pm0.22$ &     $7.00_{-0.22}^{+0.26}$ &     $8.34_{-0.13}^{+0.26}$ &     $8.51_{-0.20}^{+0.28}$ &               $7.50\pm0.28$ &               $7.70\pm0.32$ &               $6.57\pm0.25$ &               $6.74\pm0.31$ &     $5.05_{-0.19}^{+0.26}$ &               $5.23\pm0.22$ &                        --- &                        --- \\
Sh~2-219 &                        --- &                       --- &               $7.45\pm0.06$ &               $7.60\pm0.07$ &               $8.28\pm0.12$ &     $8.44_{-0.09}^{+0.14}$ &                        --- &                        --- &                        --- &                        --- &     $5.96_{-0.38}^{+0.23}$ &     $5.98_{-0.33}^{+0.40}$ &                        --- &                        --- \\
Sh~2-235 &                        --- &                       --- &               $7.56\pm0.04$ &     $7.73_{-0.05}^{+0.04}$ &               $8.42\pm0.07$ &     $8.61_{-0.06}^{+0.08}$ &                        --- &                        --- &               $7.11\pm0.06$ &     $7.27_{-0.06}^{+0.07}$ &               $5.31\pm0.11$ &     $5.36_{-0.08}^{+0.10}$ &                        --- &                        --- \\
Sh~2-237 &                        --- &                       --- &               $7.54\pm0.05$ &     $7.69_{-0.06}^{+0.05}$ &     $8.34_{-0.09}^{+0.08}$ &               $8.51\pm0.09$ &                        --- &                        --- &                        --- &                        --- &               $5.74\pm0.25$ &     $5.79_{-0.26}^{+0.22}$ &                        --- &                        --- \\
Sh~2-257 &                        --- &                       --- &               $7.60\pm0.06$ &               $7.78\pm0.07$ &               $8.41\pm0.10$ &               $8.61\pm0.10$ &                        --- &                        --- &               $7.18\pm0.11$ &               $7.40\pm0.11$ &               $5.44\pm0.17$ &               $5.45\pm0.17$ &                        --- &                        --- \\
Sh~2-266 &                        --- &                       --- &               $7.52\pm0.11$ &     $7.69_{-0.11}^{+0.09}$ &               $8.19\pm0.16$ &               $8.36\pm0.18$ &                        --- &                        --- &                        --- &                        --- &                        --- &                        --- &                        --- &                        --- \\
Sh~2-270 &                        --- &                       --- &               $7.26\pm0.19$ &               $7.40\pm0.17$ &               $8.09\pm0.26$ &               $8.23\pm0.31$ &                        --- &                        --- &                        --- &                        --- &                        --- &                        --- &                        --- &                        --- \\
Sh~2-271 &                        --- &                       --- &               $7.50\pm0.05$ &               $7.65\pm0.05$ &               $8.27\pm0.09$ &     $8.44_{-0.08}^{+0.09}$ &                        --- &                        --- &     $6.89_{-0.05}^{+0.08}$ &     $7.08_{-0.06}^{+0.07}$ &               $6.04\pm0.18$ &               $6.05\pm0.18$ &                        --- &                        --- \\
Sh~2-285 &                        --- &                       --- &     $7.44_{-0.08}^{+0.06}$ &     $7.59_{-0.06}^{+0.07}$ &               $8.21\pm0.11$ &               $8.38\pm0.10$ &                        --- &                        --- &                        --- &                        --- &                        --- &                        --- &                        --- &                        --- \\
Sh~2-288 &                        --- &                       --- &     $7.40_{-0.09}^{+0.10}$ &     $7.58_{-0.11}^{+0.09}$ &               $8.34\pm0.08$ &     $8.51_{-0.06}^{+0.10}$ &                        --- &                        --- &     $6.68_{-0.09}^{+0.08}$ &               $6.87\pm0.10$ &               $4.92\pm0.07$ &               $5.03\pm0.09$ &                        --- &                        --- \\
Sh~2-297 &                        --- &                       --- &               $7.71\pm0.05$ &     $7.90_{-0.05}^{+0.06}$ &     $8.48_{-0.07}^{+0.09}$ &     $8.68_{-0.07}^{+0.08}$ &                        --- &                        --- &                        --- &                        --- &     $5.45_{-0.15}^{+0.14}$ &               $5.47\pm0.15$ &                        --- &                        --- \\
Sh~2-298 &                        --- &                       --- &               $7.68\pm0.08$ &     $7.80_{-0.09}^{+0.08}$ &     $8.43_{-0.05}^{+0.06}$ &     $8.55_{-0.04}^{+0.06}$ &     $8.27_{-0.07}^{+0.08}$ &     $8.40_{-0.08}^{+0.09}$ &               $6.80\pm0.04$ &     $6.89_{-0.04}^{+0.03}$ &     $4.94_{-0.08}^{+0.10}$ &     $5.02_{-0.09}^{+0.10}$ &               $6.00\pm0.03$ &     $6.09_{-0.03}^{+0.04}$ \\
Sh~2-311 &               $8.21\pm0.10$ &              $8.26\pm0.09$ &               $7.50\pm0.05$ &               $7.68\pm0.06$ &     $8.41_{-0.04}^{+0.05}$ &     $8.58_{-0.04}^{+0.05}$ &               $7.70\pm0.05$ &     $7.93_{-0.07}^{+0.05}$ &     $6.78_{-0.02}^{+0.03}$ &               $6.93\pm0.03$ &               $4.93\pm0.03$ &               $5.11\pm0.05$ &                        --- &                        --- \\
IC~5146  &                        --- &                       --- &               $7.81\pm0.05$ &     $8.04_{-0.05}^{+0.06}$ &               $8.57\pm0.09$ &     $8.81_{-0.09}^{+0.10}$ &                        --- &                        --- &               $6.89\pm0.04$ &     $7.13_{-0.04}^{+0.05}$ &                        --- &                        --- &                        --- &                        --- \\
NGC~2579 &     $8.18_{-0.06}^{+0.05}$ &    $8.20_{-0.06}^{+0.05}$ &               $7.36\pm0.09$ &               $7.55\pm0.10$ &     $8.41_{-0.05}^{+0.06}$ &     $8.61_{-0.04}^{+0.06}$ &     $7.60_{-0.08}^{+0.09}$ &     $7.79_{-0.07}^{+0.10}$ &               $6.71\pm0.05$ &               $6.85\pm0.05$ &               $4.94\pm0.07$ &     $5.07_{-0.07}^{+0.06}$ &               $6.21\pm0.03$ &     $6.36_{-0.03}^{+0.04}$ \\
NGC~3576 &               $8.48\pm0.03$ &    $8.50_{-0.03}^{+0.04}$ &               $7.62\pm0.06$ &     $7.86_{-0.08}^{+0.09}$ &               $8.54\pm0.02$ &     $8.79_{-0.03}^{+0.04}$ &               $8.01\pm0.03$ &     $8.26_{-0.04}^{+0.06}$ &     $6.90_{-0.04}^{+0.05}$ &     $7.07_{-0.05}^{+0.06}$ &     $4.92_{-0.04}^{+0.03}$ &     $5.08_{-0.04}^{+0.05}$ &               $6.35\pm0.03$ &               $6.53\pm0.04$ \\
NGC~3603 &     $8.53_{-0.07}^{+0.08}$ &              $8.54\pm0.07$ &               $7.69\pm0.15$ &               $7.91\pm0.14$ &     $8.46_{-0.03}^{+0.04}$ &     $8.70_{-0.04}^{+0.05}$ &     $7.93_{-0.05}^{+0.06}$ &     $8.18_{-0.06}^{+0.07}$ &     $6.85_{-0.09}^{+0.08}$ &     $7.01_{-0.09}^{+0.08}$ &               $4.82\pm0.07$ &     $4.93_{-0.06}^{+0.07}$ &     $6.24_{-0.04}^{+0.06}$ &               $6.37\pm0.05$ \\
M~8      &               $8.47\pm0.06$ &              $8.53\pm0.06$ &               $7.73\pm0.04$ &               $7.95\pm0.05$ &               $8.48\pm0.04$ &     $8.69_{-0.04}^{+0.05}$ &     $7.79_{-0.04}^{+0.05}$ &     $8.07_{-0.07}^{+0.06}$ &               $6.92\pm0.03$ &     $7.10_{-0.03}^{+0.04}$ &               $5.04\pm0.02$ &               $5.30\pm0.05$ &               $6.28\pm0.02$ &               $6.48\pm0.03$ \\
M~16     &     $8.53_{-0.10}^{+0.09}$ &    $8.62_{-0.10}^{+0.08}$ &     $7.89_{-0.05}^{+0.06}$ &               $8.12\pm0.07$ &     $8.57_{-0.04}^{+0.05}$ &     $8.79_{-0.05}^{+0.06}$ &     $7.91_{-0.09}^{+0.11}$ &               $8.23\pm0.13$ &               $6.93\pm0.03$ &     $7.11_{-0.03}^{+0.04}$ &               $5.06\pm0.04$ &               $5.29\pm0.05$ &               $6.30\pm0.02$ &               $6.52\pm0.03$ \\
M~17     &     $8.79_{-0.03}^{+0.04}$ &    $8.80_{-0.04}^{+0.03}$ &     $7.67_{-0.07}^{+0.09}$ &     $7.95_{-0.09}^{+0.10}$ &               $8.54\pm0.04$ &               $8.83\pm0.05$ &               $7.96\pm0.04$ &     $8.27_{-0.05}^{+0.06}$ &     $6.90_{-0.04}^{+0.05}$ &               $7.10\pm0.05$ &     $4.93_{-0.05}^{+0.06}$ &     $5.08_{-0.05}^{+0.06}$ &               $6.32\pm0.02$ &               $6.51\pm0.03$ \\
M~20     &               $8.44\pm0.20$ &              $8.52\pm0.19$ &               $7.69\pm0.05$ &     $7.90_{-0.05}^{+0.06}$ &     $8.55_{-0.05}^{+0.04}$ &               $8.76\pm0.05$ &               $7.50\pm0.11$ &               $7.82\pm0.10$ &     $6.88_{-0.02}^{+0.03}$ &     $7.07_{-0.03}^{+0.04}$ &     $5.06_{-0.03}^{+0.04}$ &     $5.23_{-0.03}^{+0.04}$ &               $6.31\pm0.04$ &               $6.53\pm0.05$ \\
M~42     &     $8.46_{-0.05}^{+0.04}$ &    $8.47_{-0.05}^{+0.03}$ &               $7.77\pm0.13$ &               $8.02\pm0.14$ &               $8.52\pm0.02$ &     $8.78_{-0.03}^{+0.05}$ &               $7.96\pm0.03$ &     $8.25_{-0.04}^{+0.06}$ &     $6.74_{-0.06}^{+0.07}$ &     $6.92_{-0.06}^{+0.08}$ &     $4.86_{-0.04}^{+0.05}$ &               $4.98\pm0.05$ &               $6.27\pm0.04$ &               $6.43\pm0.04$ \\
\hline
\end{tabular}
\end{table}
\end{landscape}

\bsp	
\label{lastpage}
\end{document}